\documentclass[12pt,reqno]{amsart}

\usepackage[letterpaper,hmargin=1in,top=1in,bottom=1.2in,footskip=0.7in]{geometry}
\usepackage{placeins}
\usepackage[small]{caption}
\usepackage{subcaption}
\usepackage{amssymb}
\usepackage{graphicx}
\usepackage{tabularx}
\usepackage{booktabs}
\usepackage[authoryear,round]{natbib}
\usepackage{xcolor}
\definecolor{vdarkred}{rgb}{0.6,0,0.2}
\definecolor{vdarkblue}{rgb}{0,0.2,0.6}
\usepackage[colorlinks,linkcolor=vdarkblue,citecolor=vdarkred]{hyperref}
\usepackage{comment}

\renewcommand{\Im}{\mathfrak{Im}}
\renewcommand{\Re}{\mathfrak{Re}}
\renewcommand{\P}{\mathbb{P}}
\newcommand{\E}{\operatorname{\mathbb{E}}}
\newcommand{\R}{\mathbb{R}}
\newcommand{\C}{\mathbb{C}}
\newcommand{\one}{\mathbf{1}}
\newcommand{\diag}{\operatorname{diag}}
\newcommand{\supp}{\operatorname{supp}}
\newcommand{\Tr}{\operatorname{Tr}}
\newcommand{\dist}{\operatorname{dist}}
\newcommand{\rank}{\operatorname{rank}}
\newcommand{\spec}{\mathrm{spec}}
\newcommand{\TW}{\mathrm{TW}}
\newcommand{\la}{\label}
\newcommand{\be}{\begin{equation}}
\newcommand{\ee}{\end{equation}}
\newcommand{\eps}{\varepsilon}
\newcommand{\ii}{\mathbf i}

\newtheorem{Th}{Theorem}[section]
\newtheorem{propo}[Th]{Proposition}
\newtheorem{coro}[Th]{Corollary}
\newtheorem{lem}[Th]{Lemma}
\theoremstyle{definition}
\newtheorem{rmk}[Th]{Remark}
\newtheorem{Def}[Th]{Definition}
\newtheorem{Ass}[Th]{Assumption}

\newcommand{\bTh}{\begin{Th}}
\newcommand{\eTh}{\end{Th}}
\newcommand{\bpropo}{\begin{propo}}
\newcommand{\epropo}{\end{propo}}
\newcommand{\blem}{\begin{lem}}
\newcommand{\elem}{\end{lem}}
\newcommand{\brem}[1][]{%
    \begin{rmk}[#1]%
    \renewcommand{\qedsymbol}{\ensuremath{\diamond}}%
    \pushQED{\qed}%
}

\newcommand{\erem}{%
    \popQED%
    \end{rmk}%
}
\newcommand{\bDef}{\begin{Def}}
\newcommand{\eDef}{\end{Def}}
\newcommand{\bAss}{\begin{Ass}}
\newcommand{\eAss}{\end{Ass}}
\newcommand{\bpr}{\begin{proof}}
\newcommand{\epr}{\end{proof}}

\begin{document}
\author{Anna Bykhovskaya}
\address[Anna Bykhovskaya]{Department of Economics, Duke University, Durham, NC, 27708, }
\email{anna.bykhovskaya@duke.edu}

\author{Tomas Espana}
\address[Tomas Espana]{Department of Operations Research and Financial Engineering, Princeton University, Princeton, NJ, 08540, USA}
\email{tomas.espana@princeton.edu}

\author{Jianqing Fan}
\address[Jianqing Fan]{Department of Operations Research and Financial Engineering, Princeton University, Princeton, NJ, 08540, USA}
\email{jqfan@princeton.edu}

\title{Skew-Symmetric Factor Models}

\begin{abstract}
The paper introduces and analyzes a factor model for (real) skew-symmetric data, which arise naturally in directed flows and in pairwise comparisons used for rankings. Working within a signal-plus-noise framework, we analyze spectral inference and establish a BBP-type phase transition in signal strength. Under Gaussian noise, we derive joint eigenvalue fluctuation limits. Above the separation threshold, the estimated signal strength is asymptotically normal whenever the population strength has multiplicity one. For a general multiplicity $k$, the joint limiting fluctuations of the corresponding sample eigenvalues above the threshold are given by the eigenvalues of a scaled $k\times k$ Gaussian unitary ensemble matrix. The skew-symmetric structure induces paired eigenspaces, and we characterize the canonical angles between the population and sample eigenspaces. Below the threshold, the corresponding largest sample eigenvalue converges to the edge of the semicircle law, with fluctuations governed by the Tracy--Widom $F_2$ distribution, in contrast to the $F_1$ law familiar from real symmetric models. The approach is illustrated through financial and demographic applications.
\end{abstract}

\date{\today}
\maketitle
\section{Introduction}
\subsection{Motivation}

Skew-symmetric matrices capture a feature shared by many kinds of data: reversing the order of two entities reverses the sign of the measurement. Bilateral trade balances, net migration flows, margins of victory between teams, and net preferences between alternatives all have this property. These examples motivate the statistical study of skew-symmetric matrices. One way to describe systematic patterns in such data is through a low-rank model.

For example, differences between latent scores form a skew-symmetric matrix of rank at most two and provide a basic model for ranking and pairwise comparisons \citep{jiang2011statistical,d2021ranking}. More general low-rank signals can be represented as sums of rank-two skew-symmetric components, each formed from a pair of latent score vectors. A natural approach to estimating such low-rank structure from noisy observations uses the leading singular values and their associated subspaces. Much of the asymptotic theory for these methods concerns sample covariance and correlation matrices, including work on high-dimensional PCA \citep{johnstone2001distribution} and its applications to factor estimation, forecasting, and selection of the number of factors in large panels \citep{stock2002forecasting,bai2002determining,bai2003inferential}.

In this paper, we introduce a new factor model based on skew-symmetric spiked random matrices.
We assume that the observed matrix has the form
\begin{equation}\label{eq:intro_model}
\sum_{j=1}^r \theta_j\bigl(u_jv_j^{\top}-v_ju_j^{\top}\bigr)
+\frac{1}{\sqrt n}\mathcal{E}_n,
\end{equation}
where ${}^{\top}$ denotes matrix transposition, $(u_1,v_1,\ldots,u_r,v_r)$ is a deterministic orthonormal family in \(\mathbb R^n\), \(\theta_1\geq\cdots\geq\theta_r>0\) are the factor strengths and $r$ is the number of factors. Thus, the sum, which represents the signal, has rank $2r$. The noise matrix \(\mathcal{E}_n\) is real skew-symmetric, with i.i.d.\ centered entries of variance \(\sigma^2>0\) above the main diagonal. The form \eqref{eq:intro_model} is the skew-symmetric analogue of classical spiked random matrices and factor models. We let \(n\to\infty\) while $r$ and the factor strengths remain fixed, without imposing sparsity or localization assumptions on the factor directions. The normalization by $\sqrt{n}$ places the noise spectrum and the signal on the same scale. 

The goal is to estimate the signal in \eqref{eq:intro_model}, that is, the strengths $\theta_j$ and the directions $u_j,v_j$. We do this using spectral methods. Multiplying a real skew-symmetric matrix by $\ii$ gives a Hermitian matrix with real eigenvalues symmetric about zero. Spectral methods based on this transformation have been used, for example, for directed-network clustering \citep{cucuringu2020hermitian}. In the present model, we establish eigenvalue limits and fluctuation laws. A striking feature is that, under Gaussian noise, the limiting fluctuations differ from those in the corresponding real symmetric models, despite the observations being real. We also study estimation of the two-dimensional factor planes spanned by the pairs of signal directions. Simultaneous rotation of both directions within their plane leaves the factor $ \theta_j\left(u_jv_j^{\top}-v_ju_j^{\top}\right)$ unchanged, so the individual axes are not identified.

We illustrate the results using pairwise stock outperformance and migration imbalances between metropolitan areas. In each application, we estimate the strengths and planes of the leading factors and use the asymptotic results to assess the accuracy of the plane estimates.

\subsection{Signal strength}
The empirical spectral distribution of the observed matrix multiplied by $\ii$ converges to the semicircle law on $[-2\sigma,2\sigma]$ \citep{bai-silver-book}, while the finite-rank signal may produce outliers outside this bulk.
We show that a factor with strength $\theta_j>\sigma$ produces a positive outlier that converges almost surely to $\theta_j+\sigma^2/\theta_j$. For $\theta_j\leq\sigma$, the corresponding extreme eigenvalue converges almost surely to the noise edge $2\sigma$. This is the analogue of the BBP phase transition in high-dimensional PCA \citep{bbp-1,baik2006eigenvalues,paul2007asymptotics}. The separation threshold and outlier locations agree with those for additive real symmetric deformations \citep{capitaine-1}.

Under Gaussian noise, we establish joint fluctuation limits for the positive outliers. For a supercritical strength $\theta>\sigma$ of multiplicity $k$, the corresponding $k$ outliers, centered at their common limiting location and scaled by $n^{1/2}$, converge jointly to the ordered eigenvalues of a standard $k\times k$ Gaussian unitary ensemble (GUE) matrix multiplied by $\sigma\sqrt{1-\sigma^2/\theta^2}$. The limits associated with distinct strengths are independent. For $k=1$, the limit is Gaussian with variance $\sigma^2(1-\sigma^2/\theta^2)$. This agrees with the corresponding complex Hermitian spiked Wigner variance and is one half of the real symmetric spiked Wigner variance for the same signal strength and off-diagonal noise variance \citep{capitaine-1,capitaine-fluctuations}. Thus, the first-order agreement with a real symmetric model does not extend to the second-order behavior of the spectral estimates.

We also study the largest non-outlier eigenvalue. Under Gaussian noise, when no factor strength equals $\sigma$, this eigenvalue, centered at $2\sigma$ and scaled by $n^{2/3}/\sigma$, converges in distribution to the Tracy--Widom $F_2$ distribution. Its fluctuations are asymptotically independent of those of the supercritical outliers, including those associated with repeated strengths. This contrasts with the $F_1$ distribution familiar from real Gaussian PCA and symmetric Wigner models \citep{johnstone2001distribution,feral2007largest}.

Our outlier analysis uses a determinant equation involving the factor strengths and the noise resolvent projected onto the signal space. The limiting equation determines the outlier locations, while linearization around these locations reduces the eigenvalue fluctuations to fluctuations of resolvent quadratic forms. Under Gaussian noise, orthogonal invariance and the complex Gaussian structure of these forms yield the GUE limits. For the edge result, we use the fact that the squared positive eigenvalues of the undeformed Gaussian ensemble, after rescaling, have the joint eigenvalue distribution of a Laguerre unitary ensemble \citep[Exercise~1.3.5]{forrester2010log}. We show that the resulting $F_2$ edge limit persists under fixed skew-symmetric perturbations whose strengths differ from $\sigma$.



These results provide a basis for statistical inference on the presence and strength of factors. The Tracy--Widom $F_2$ limit gives asymptotic critical values under the null that all factor strengths are strictly below $\sigma$, while the outlier limits imply consistency against fixed alternatives containing a supercritical factor. Related spectral methods have been used for factor-number testing in large dynamic factor models \citep{onatski2009testing} and for factor-number estimation based on adjusted eigenvalues of the sample correlation matrix \citep{fan2022estimating}. For known $\sigma$, inverting the limiting outlier location $\theta+\sigma^2/\theta$ gives a consistent estimator of each supercritical strength. When the strength has multiplicity one, the Gaussian outlier limit also yields asymptotic normality of this estimator. An analogous inverse map for spiked covariance models is discussed by \citet[Section~1.1]{donoho2018optimal}.

\subsection{Signal directions}

In PCA, distinct population eigenvalues specify separate principal directions, each determined up to sign. In contrast, a distinctive geometric feature of our model is that each skew-symmetric factor $u_j v_j^{\top} - v_j u_j^{\top}$ is invariant under simultaneous rotation of $u_j$ and $v_j$ within their plane. 
When a factor strength has multiplicity one, the population signal therefore identifies a two-dimensional factor plane rather than a one-dimensional principal direction, with
no distinguished pair of axes within that plane.

We measure estimation accuracy by the canonical angles between the estimated and population factor planes; their cosines are also known as the canonical correlations. For each supercritical strength of multiplicity one, we show that both squared canonical correlations converge almost surely to $1-\sigma^2/\theta_j^2$. For every fixed $\theta_j>\sigma$, both limiting angles are positive. A factor strength can therefore be consistently estimated while the estimated plane retains a nonvanishing directional error. The common limiting angle decreases to zero as $\theta_j/\sigma$ tends to infinity.

Persistent directional error also occurs in high-dimensional PCA and weak-factor models \citep{paul2007asymptotics,onatski2012asymptotics,benaych2012singular,wang2017asymptotics} and additive deformations \citep{BENAYCHGEORGES2011494}. Nonvanishing angles between population and sample canonical variables also arise in high-dimensional canonical correlation analysis \citep{bykhovskaya2023high}. These fixed-strength asymptotics differ from strong-factor regimes, where increasing information about pervasive factors can yield consistent estimation \citep{bai2003inferential}.


\subsection{Relation to existing theory}
Multiplying the observed matrix by $\ii$ gives a finite-rank additive
perturbation of a complex Hermitian random matrix. Several standard results, however, impose assumptions that the skew-symmetric ensemble does not satisfy. For example, the invariant Hermitian framework of \citet{BENAYCHGEORGES2011494} requires unitary conjugation invariance of the noise or perturbation, whereas our Gaussian noise is invariant under real orthogonal conjugation but not under arbitrary unitary conjugation. The complex Hermitian formulations of \citet{knowles2013isotropic,knowles-deformed-wigner} assume that the nonconjugated second moments of the off-diagonal noise entries vanish. The off-diagonal entries of the transformed noise matrix are purely imaginary, so their nonconjugated second moments are strictly negative. More general local semicircle laws and isotropic resolvent estimates remain applicable \citep{erdos2013local,erdos-cov} and provide the basis for our first-order analysis; see also \citet[Section 3]{sosoe2016spectrum} for the application of local semicircle laws to skew-symmetric matrices. The fluctuation results require additional arguments that account for the skew-symmetric structure.

Another connection is with PCA, which arises from the uncentered sample covariance matrix, obtained by multiplying the data matrix by its transpose. For both the signal and the observed data, each nonzero conjugate eigenvalue pair gives a repeated eigenvalue of the corresponding covariance matrix, with real eigenvectors obtained from the real and imaginary parts of a corresponding eigenvector of the original matrix. Repeated spikes are allowed in the sample eigenvalue results of \citet{bai-yao} and the eigenspace results of \citet{bloemendal2016principal,bao2022statistical}. These results, however, assume independent sample vectors, whereas skew-symmetry makes the rows of the data matrix dependent. Thus, the cited theorems do not apply. Moreover, in real Gaussian PCA, a supercritical population spike of multiplicity two produces two sample outliers whose joint fluctuations are described by the eigenvalues of a scaled GOE matrix of size two. For a supercritical strength of multiplicity one in our Gaussian model, the two corresponding sample covariance eigenvalues remain exactly equal and have a common Gaussian limit after centering and scaling.

\subsection{Outline of the paper}

Section~\ref{sec:setting} introduces the model and notation. Section~\ref{sec:asy_eigenvalue} derives the asymptotic distributions of the largest eigenvalues, while Section~\ref{sec:asy_eigenvectors} analyzes the corresponding eigenvectors. Section~\ref{sec:empirical} presents two empirical illustrations of our results. Finally, Section~\ref{sec:conclusion} concludes. All proofs are given in the Appendix.

\smallskip

\noindent\textit{Notation.}
For an integer $n\geq1$, let $[n]=\{1,\ldots,n\}$. We use ${}^{\top}$ for the transpose and ${}^*$ for the conjugate transpose of a vector or matrix, and $e_j$ for the $j$-th canonical basis vector of $\R^n$. For complex scalars and vectors, $\Re$ and $\Im$ denote the real and imaginary parts, taken entrywise for vectors. We write 
$\Rightarrow$ for convergence in distribution, $\mathcal N(a,\Sigma)$ for a Gaussian random variable with mean $a$ and covariance $\Sigma$, and $\TW_2$ for a random variable with the Tracy--Widom $F_2$ distribution (see Definition~\ref{def:tracy_widom}).

\section{Setting}\label{sec:setting}

Let $X_n$ be an observed $n\times n$ real skew-symmetric matrix. For example, $(X_n)_{k\ell}$ may record the net migration flow from
location $k$ to location $\ell$, defined as the difference between
the flows in the two directions. We assume that
\begin{equation}\label{eq:dgp}
    X_n=S_n+\frac1{\sqrt{n}}\mathcal{E}_n,\qquad S_n=\sum_{j=1}^r \theta_j\bigl(u_jv_j^{\top}-v_ju_j^{\top}\bigr),
\end{equation}
where $r$ is the number of signals, $\theta_j\in\mathbb{R}$ are their strengths, $u_j,v_j\in\mathbb{R}^n$ are the corresponding directions, and $\mathcal E_n$ is the noise matrix. We study the limit $n\to\infty$. Under the assumptions below, the normalized noise $n^{-1/2}\mathcal{E}_n$ has spectral norm of order one, as does $S_n$.


\bAss\la{ass:ass1}
Let $\sigma>0$ be fixed, and let
$\{\eps_{k\ell}:1\leq k<\ell<\infty\}$ be an infinite i.i.d.\ array
with mean zero, variance $\sigma^2$ and a finite fourth moment.
For each $n$, define the real skew-symmetric noise matrix
$\mathcal E_n$ by $(\mathcal E_n)_{k\ell}=\eps_{k\ell}$ for
$k<\ell\leq n$.
\eAss
\bAss\la{ass:ass2}
The number of signals $r$ and their strengths
$\theta_1\geq\cdots\geq\theta_r>0$ are fixed and deterministic.
For each $n\geq2r$, the vectors $(u_1,v_1,\ldots,u_r,v_r)$ form
a deterministic orthonormal family in $\mathbb{R}^n$. These
vectors may depend on $n$; we omit this dependence from the
notation.
\eAss

\brem\la{rmk:random_factor_directions}
All results of the paper remain valid if the sequence of orthonormal families $(u_1,v_1,\ldots,u_r,v_r)$ is random and independent of the noise sequence $(\mathcal E_n)_n$, with almost sure convergence replaced by convergence in probability unless a suitable coupling across $n$ is imposed. Conditional on the factor directions, the deterministic results apply, and the resulting limits do not depend on their realization.
\erem

The signal matrix $S_n$ has rank $2r$. Its $j$th summand has rank two, spectral norm $\theta_j$, and nonzero eigenvalues $\pm \ii\theta_j$, with corresponding eigenvectors $v_j\mp \ii u_j$. These rank-two components are the skew-symmetric counterparts of the rank-one components used in symmetric spiked models. 
Every real skew-symmetric matrix of rank $2r$ admits such a decomposition with orthonormal directions, so orthonormality in Assumption~\ref{ass:ass2} is without loss of generality.

Real skew-symmetric matrices have purely imaginary eigenvalues
and become Hermitian after multiplication by $\ii$. We therefore set
$$    
W_n=\frac{\ii\mathcal{E}_n}{\sqrt n},    \qquad H_n=\ii X_n=W_n+\ii S_n.
$$
The transformed signal has the spectral decomposition
\begin{equation*}
    \ii S_n=
\sum_{j=1}^r
\theta_j\left(w_{j,+}w_{j,+}^*-w_{j,-}w_{j,-}^*\right),\qquad 
w_{j,+}
=
\frac{u_j-\ii v_j}{\sqrt2},
\qquad
w_{j,-}
=
\frac{u_j+\ii v_j}{\sqrt2}.
\end{equation*}
Thus, each rank-two skew-symmetric factor gives two rank-one
Hermitian spikes with eigenvalues $\theta_j$ and $-\theta_j$.
The corresponding eigenvectors $w_{j,+}$ and $w_{j,-}$ are
complex conjugates, and their real and imaginary parts span
the same real factor plane. Moreover, the family $(w_{1,+},\dots,w_{r,+},w_{1,-},\dots,w_{r,-})$ is orthonormal in $\C^n.$

The transformed noise matrix $W_n$ is Hermitian, but its invariance properties differ from those of the Gaussian unitary ensemble (GUE), even when $\eps_{k\ell}\thicksim\mathcal{N}(0,\sigma^2)$. Under Gaussian noise, the distribution of $W_n$ is invariant under real orthogonal conjugation: $OW_nO^\top\stackrel{d}{=}W_n$ for every deterministic real orthogonal matrix $O$. Invariance under arbitrary unitary conjugation, however, fails even in dimension two. Indeed, write
$$
W_2=\frac{\eps}{\sqrt2}
\begin{pmatrix}
0&\ii\\
-\ii&0
\end{pmatrix},
\qquad
\eps\sim\mathcal N(0,\sigma^2).
$$
The deterministic unitary matrix
$$
U=\frac1{\sqrt2}
\begin{pmatrix}
1&\ii\\
1&-\ii
\end{pmatrix}
$$
diagonalizes $W_2$:
$$
UW_2U^*
=\frac{\eps}{\sqrt2}
\begin{pmatrix}
1&0\\
0&-1
\end{pmatrix}.
$$
The diagonal entries of $W_2$ are identically zero, while those of $UW_2U^*$ are nondegenerate Gaussian random variables. Hence, the two matrices have different distributions.

\medskip

Let $\lambda_1\geq\cdots\geq\lambda_n$ denote the eigenvalues
of $H_n$. We suppress their dependence on $n$ in the notation.
The spectrum is symmetric around zero:
\[
    \lambda_{n-j+1}=-\lambda_j,\qquad j=1,\ldots,n.
\]
It, therefore, suffices to consider the positive eigenvalues.
The leading eigenvalues will be used to infer the signal
strengths $\theta_j$. Section \ref{sec:asy_eigenvalue} establishes their
limiting locations and fluctuation laws. Eigenvectors
corresponding to opposite nonzero eigenvalues can be chosen
as complex conjugates. Section~\ref{sec:asy_eigenvectors} studies their asymptotic behaviour and their use in estimating the real two-dimensional factor planes spanned by $u_j$ and $v_j$.

\section{Eigenvalue asymptotics}\label{sec:asy_eigenvalue}

This section studies the eigenvalues of $H_n$. We first establish
the limiting bulk distribution and the locations of the extreme
eigenvalues, and then derive their fluctuation laws.

\subsection{First-order asymptotics}

The first-order behaviour of the eigenvalues has two main features.
First, the limiting bulk distribution is determined by the
noise matrix $W_n$ and is the semicircle law on
$[-2\sigma,2\sigma]$. Second, a signal with strength
$\theta_j>\sigma$ produces an outlier above the right edge
of the semicircle, whereas for $\theta_j\leq\sigma$ the
corresponding eigenvalue converges to that edge.

\begin{propo}[Semicircle law; {\citet[Theorem 2.5]{bai-silver-book}}]
\label{prop:semicircle}
Suppose Assumption~\ref{ass:ass1} holds. Then, as $n\to\infty$,
the empirical spectral distribution of $W_n$ converges weakly, almost surely, to the semicircle law with variance $\sigma^2$
and density
\[
\frac{1}{2\pi\sigma^2}\sqrt{4\sigma^2-x^2}\,
\mathbf{1}_{[-2\sigma,\,2\sigma]}(x).
\]
If Assumption~\ref{ass:ass2} also holds, the empirical spectral distribution of $H_n$ converges weakly, almost surely, to the same law.
\end{propo}

The assertion for $H_n$ follows from the rank inequality in
Lemma~\ref{lem:rank_ineq}, since $H_n-W_n=\ii S_n$ has rank $2r$.
Thus, the signal does not change the limiting empirical spectral
distribution. Its effect on individual extreme eigenvalues is
described by the following theorem, which shows that positive sample eigenvalues overestimate population ones.

\begin{figure}[t]
    \centering
    \begin{subfigure}[t]{0.49\textwidth}
        \centering
        \includegraphics[width=\linewidth]{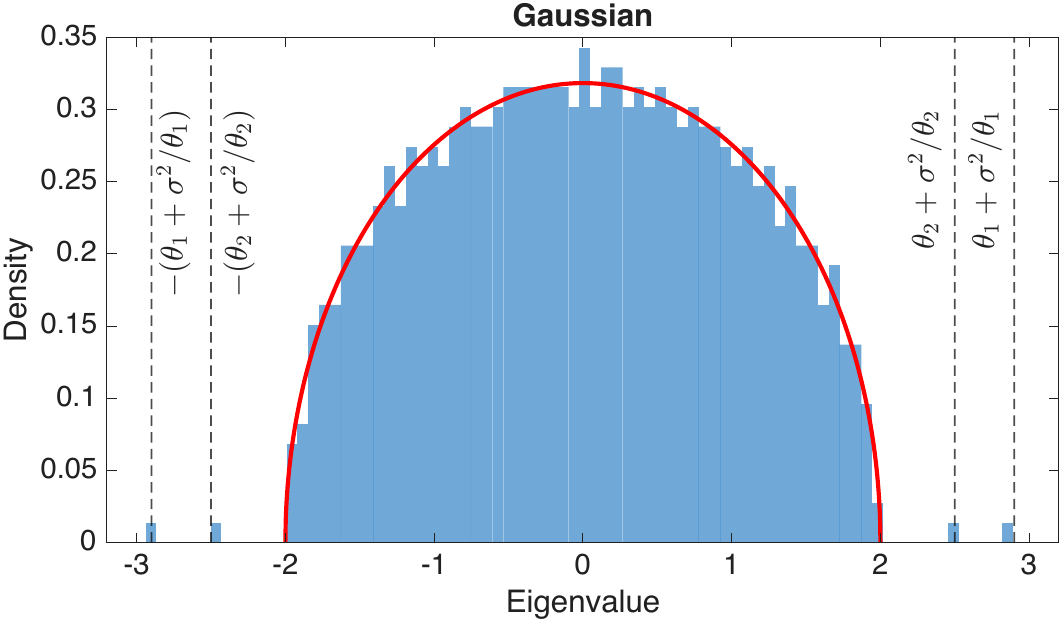}
        \caption{Gaussian noise.}
        \label{fig:lsd_and_spikes_gaussian}
    \end{subfigure}
    \hfill
    \begin{subfigure}[t]{0.49\textwidth}
        \centering
        \includegraphics[width=\linewidth]{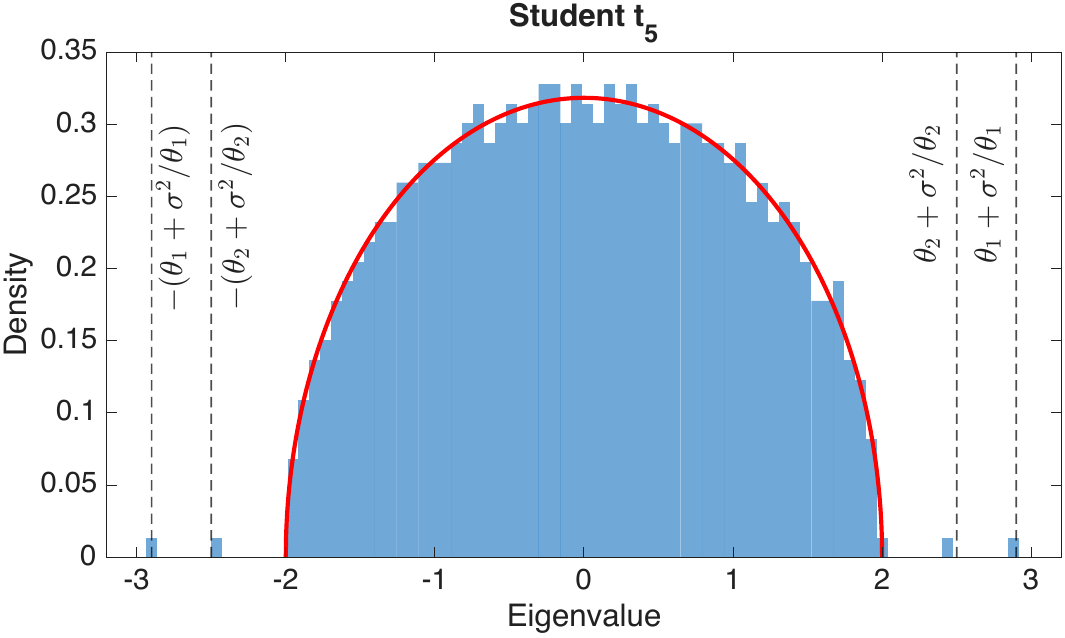}
        \caption{Standardized Student $t_5$ noise.}
        \label{fig:lsd_and_spikes_student}
    \end{subfigure}

    \caption{Empirical spectra of $H_n$ for $n=1000$ and $\sigma=1$.
    The signal is $S_n=\sum_{j=1}^3\theta_j(u_jv_j^\top-v_ju_j^\top)$,
    where $u_j=e_{2j-1}$, $v_j=e_{2j}$, and
    $(\theta_1,\theta_2,\theta_3)=(2.5,2,0.5)$.
    Blue histograms show the eigenvalue distributions, red curves show the semicircle density, and dashed lines mark the theoretical outlier locations for the two supercritical spikes.
    The third spike is subcritical.}
    \label{fig:lsd_and_spikes}
\end{figure}

\bTh\la{thm:general_random_skew_outliers}
Suppose Assumptions~\ref{ass:ass1} and~\ref{ass:ass2} hold. Set $m = \#\{j: \theta_j > \sigma\}$. Then, almost surely,
\be \la{eq:first_order_limss}
\lambda_j \longrightarrow\theta_j+\frac{\sigma^2}{\theta_j} \quad (1\leq j \leq m), \qquad \qquad \lambda_{m+k} \longrightarrow  2\sigma \quad \text{for every fixed } k\geq1.
\ee
The corresponding lower-edge limits follow from $\lambda_{n-j+1} = - \lambda_j$.
\eTh
Theorem~\ref{thm:general_random_skew_outliers} gives the
skew-symmetric analogue of the Baik--Ben Arous--P\'ech\'e
(BBP) phase transition \citep{bbp-1,feral2007largest}.
Signals with $\theta_j>\sigma$ are supercritical and produce
outliers. Critical signals, with $\theta_j=\sigma$, and
subcritical signals, with $\theta_j<\sigma$, do not separate
from the bulk. Since the eigenvalues of $H_n$ occur in opposite
pairs, separation takes place simultaneously at both spectral
edges; see Figure~\ref{fig:lsd_and_spikes}.

For known $\sigma$, Theorem~\ref{thm:general_random_skew_outliers} yields strongly consistent estimators of the supercritical signal strengths. Inverting the limiting outlier location gives 
\begin{equation}\label{eq:theta_hat}
\widehat{\theta}_j
=\frac{\lambda_j+\sqrt{\lambda_j^2-4\sigma^2}}{2}
\overset{\text{a.s.}}{\longrightarrow} \theta_j,
\qquad \theta_j>\sigma.
\end{equation}
The estimator is well defined for all sufficiently large \(n\) almost surely, since the limiting outlier location exceeds \(2\sigma\).

If $\sigma$ is unknown, it can be estimated from the spectral
bulk. Since the second moment of the semicircle law equals
$\sigma^2$, we consider the empirical second moment, allowing
for trimming of extreme eigenvalue pairs.

\begin{coro}
\label{cor:variance_estimation}
Suppose Assumptions \ref{ass:ass1} and \ref{ass:ass2} hold. Let $k_n$ be possibly data-dependent nonnegative integers such that $2k_n<n$ and $k_n/n^{1/3}\to0$ almost surely. Define the trimmed second-moment estimator
\[
\widehat{\sigma}_n^2
=
\frac{1}{n-2k_n}
\sum_{j=k_n+1}^{n-k_n}\lambda_j^2.
\]
Then, as $n\to\infty$,
\[
\widehat{\sigma}_n^2
\overset{\text{a.s.}}{\longrightarrow} \sigma^2,
\qquad
\mathbb{E}\bigl[\widehat{\sigma}_n^2\bigr]
\longrightarrow \sigma^2,
\qquad
\hat{\sigma}^2_n-\sigma^2=o_\P(n^{-2/3}).
\]
Thus, $\widehat{\sigma}_n^2$ is strongly consistent and asymptotically unbiased.
\end{coro}

The choice $k_n=0$ gives a strongly consistent untrimmed estimator.
In applications, $k_n$ may be chosen by visual
inspection of the spectrum to discard eigenvalue pairs
that appear separated from the bulk. Any such choice of $k_n$ bounded by a fixed integer satisfies the trimming condition.

Replacing $\sigma^2$ by $\widehat{\sigma}_n^2$ in the inverse
outlier map gives a feasible estimator of each supercritical
signal strength:
$$
\widehat{\theta}_j
=\frac{\lambda_j+
\sqrt{\lambda_j^2-4\widehat{\sigma}_n^2}}{2}
\overset{\text{a.s.}}{\longrightarrow}\theta_j,
\qquad \theta_j>\sigma.
$$

\subsection{Second-order asymptotics}

We refine the preceding first-order results. We now assume Gaussian noise and study the joint fluctuations of the positive outliers and the largest non-outlier eigenvalue. We first consider the case in which the supercritical strengths are distinct.


\FloatBarrier

\bTh\la{thm:eigenvalue_fluctuations}
Suppose that Assumptions \ref{ass:ass1} and \ref{ass:ass2} hold with $\eps_{k\ell}\thicksim\mathcal{N}(0,\sigma^2)$. Let $m\in\{0,\ldots,r\}$ and suppose that 
\[
\theta_1>\cdots>\theta_m>\sigma>\theta_{m+1}.
\]
For $m=0$, this condition means $\sigma>\theta_1$;
for $m=r$, the final inequality is omitted.

Then, as $n\to\infty$, the following convergences hold jointly,
with mutually independent limiting random variables:
\be\la{eq:joint_conv_eig_thm}
\begin{aligned}
\sqrt n\left(\lambda_j-\theta_j-\frac{\sigma^2}{\theta_j}\right)
&\Longrightarrow
\mathcal N\left(0,\sigma^2\left(1-\frac{\sigma^2}{\theta_j^2}\right)\right),
&& j=1,\ldots,m,\\
\frac{n^{2/3}}{\sigma}\left(\lambda_{m+1}-2\sigma\right)
&\Longrightarrow \TW_2.
\end{aligned}
\ee
The corresponding lower-edge fluctuations are the negatives
of those above, since ${\lambda_{n-j+1}=-\lambda_j}$.
\eTh

Several aspects of Theorem~\ref{thm:eigenvalue_fluctuations} merit further discussion.

\paragraph*{\textit{Outlier fluctuations.}}
Although the observed matrix $X_n$ is real, the limiting fluctuations
in Theorem~\ref{thm:eigenvalue_fluctuations} agree with those
in the corresponding complex Hermitian Gaussian models. For the same signal strength and off-diagonal noise variance, the outlier variance agrees with that for a finite-rank deformation of the Gaussian unitary ensemble (GUE) and is half that for its real
symmetric counterpart, the Gaussian orthogonal ensemble
(GOE); see \citet[Theorems~2.2--2.3]{capitaine-1} and
\citet[Theorem~3.4]{capitaine-fluctuations}. In our proof, two independent vectors of real Gaussian noise
entries combine into a complex Gaussian vector for each
factor plane. These complex Gaussian vectors enter the quadratic forms used to derive the outlier fluctuations. 

The limiting distributions in \eqref{eq:joint_conv_eig_thm}
depend on the signal strengths but not on the factor
directions. Orthogonal conjugation preserves the eigenvalues
and the distribution of Gaussian skew-symmetric noise,
so we may take the factor directions to be coordinate axes. For non-Gaussian noise, the choice of directions can matter, as we illustrate at the end of this section.

\paragraph*{\textit{Estimation of signal strengths.}}
The Gaussian limits also give asymptotic distributions
for the estimators of distinct supercritical signal strengths.
For known $\sigma$, applying the delta method to the
inverse estimator \eqref{eq:theta_hat} gives, for every such index $j$,
\[
\sqrt n(\widehat{\theta}_j-\theta_j)
\Longrightarrow
\mathcal N\left(
0,\frac{\sigma^2}{1-\sigma^2/\theta_j^2}
\right).
\]
These convergences hold jointly, with mutually independent
Gaussian limits, and translate into confidence
intervals for the supercritical strengths.

\paragraph*{\textit{Edge fluctuations and testing.}}
Tracy--Widom distributions describe the limiting fluctuations of the largest eigenvalues in a broad class of large random matrix models. The cumulative distribution function of $\TW_2$  is expressed as a Fredholm determinant in Definition~\ref{def:tracy_widom}, and its selected quantiles are tabulated in \citet{Bejan}. In the present model, the $\TW_2$ limit follows from a connection with the Laguerre unitary ensemble: the squared positive eigenvalues of $W_n$, after rescaling, have the same joint distribution as the eigenvalues of this ensemble \citep[Exercise~1.3.5]{forrester2010log}. We show that adding fixed signals preserves this edge limit for $\lambda_{m+1}$, provided no signal strength equals $\sigma$.

The asymptotic results in Theorem~\ref{thm:eigenvalue_fluctuations}
can be used to construct a test for the presence of a
supercritical signal. Consider
\be\la{eq:hyp_test_tw}
H_0:\theta_1<\sigma
\qquad\text{versus}\qquad
H_1:\theta_1>\sigma,
\ee
and reject $H_0$ when $\frac{n^{2/3}}{\sigma}(\lambda_1-2\sigma)$ exceeds the $(1-\alpha)$-quantile of $\TW_2$. By Corollary~\ref{cor:variance_estimation}, the estimator $\widehat\sigma_n$ satisfies $\widehat\sigma_n-\sigma=o_\P(n^{-2/3})$, so, under Gaussian noise, replacing $\sigma$ by $\widehat\sigma_n$ in the test statistic preserves its Tracy--Widom limit under $H_0$ and hence the asymptotic test level. For fixed signal strengths, the rejection probability
converges to $\alpha$ under $H_0$. Under $H_1$, the largest
eigenvalue converges to a limit strictly above $2\sigma$,
so the rejection probability converges to one. At the critical value $\theta_1=\sigma$, the limiting distribution is different. We expect that inference in this case can be developed along the lines of
\citet{bykhovskaya2025weak}.

\subsection{Relaxing assumptions} The preceding fluctuation theorem relies on two main assumptions: distinct spikes ($\theta_i\neq\theta_j$ for $i\neq j$) and Gaussian noise. In this subsection, we discuss how these assumptions can be relaxed. We first present a more general fluctuation theorem that allows for spikes of multiplicity greater than one. We then use simulations to examine how deviations from Gaussianity affect the results.

\subsubsection{Spikes with multiplicities}

We now give an extended version of Theorem \ref{thm:eigenvalue_fluctuations} that allows for multiplicities among the supercritical spikes.

Throughout, a standard $\mathrm{GUE}(k)$ matrix is a $k\times k$ Hermitian matrix with mutually independent entries on and above the diagonal. The diagonal entries have distribution $\mathcal N(0,1)$, while each entry above the diagonal has independent real and imaginary parts, both with distribution $\mathcal N(0,1/2)$.

\bTh\la{thm:eigenvalue_fluctuations_ext}
Suppose that Assumptions~\ref{ass:ass1} and~\ref{ass:ass2} hold with $\eps_{k\ell}\thicksim\mathcal{N}(0,\sigma^2)$. Let
$\vartheta_1>\cdots>\vartheta_s>\sigma$ be the distinct supercritical
strengths, with multiplicities $k_1,\ldots,k_s$, and set
\[
m=\sum_{\ell=1}^s k_\ell,\qquad
p_\ell=\sum_{h<\ell}k_h.
\]
For $m\geq1$, let $\mathcal G_1,\ldots,\mathcal G_s$ be independent
standard GUE matrices of sizes $k_1,\ldots,k_s$, with ordered
eigenvalues $\gamma_{\ell,1}\geq\cdots\geq\gamma_{\ell,k_\ell}$, $\ell=1,\ldots,s$.
Then, as $n\to\infty$, jointly,
\be\la{eq:joint_conv_eig_thm_ext}
\sqrt n\left(
\lambda_{p_\ell+b}-\vartheta_\ell-\frac{\sigma^2}{\vartheta_\ell}
\right)
\Longrightarrow
\sigma\sqrt{1-\frac{\sigma^2}{\vartheta_\ell^2}}\,\gamma_{\ell,b},
\qquad b=1,\ldots,k_\ell,\quad \ell=1,\ldots,s.
\ee

If, moreover, either $m=r$ or $\theta_{m+1}<\sigma$, then
\[
\frac{n^{2/3}}{\sigma}(\lambda_{m+1}-2\sigma)
\Longrightarrow \TW_2,
\]
jointly and independently with the preceding convergences.
When $m=0$ and $\theta_1<\sigma$, only the edge convergence remains. The corresponding negative-outlier and lower-edge limits follow
from spectral symmetry, $\lambda_{n-j+1}=-\lambda_j$.
\eTh

The matrices governing the outlier limits in
Theorem~\ref{thm:eigenvalue_fluctuations_ext} belong to the
ordinary GUE: they are Hermitian but generally not
skew-symmetric. In contrast, the transformed noise $W_n$
is a rescaled antisymmetric Gaussian unitary ensemble
(a-GUE) matrix. It is both Hermitian and skew-symmetric,
and its eigenvalues occur in opposite pairs.

A positive signal strength of multiplicity $k$ corresponds
to a $k$-dimensional complex eigenspace of $\ii S_n$.
Noise can mix the directions within this space.
To describe this interaction, we apply $W_n$ to an
orthonormal basis of the eigenspace and project the results
back onto it. In this basis, the resulting operator is
represented by a $k\times k$ Hermitian matrix.
Since the basis is complex, this matrix need not inherit
the skew-symmetry of $W_n$. Instead, skew-symmetry relates
the eigenspaces for opposite strengths. 

For $k=1$, a standard $\mathrm{GUE}(1)$ matrix is simply
a standard Gaussian random variable. Thus, strengths of
multiplicity one give the Gaussian outlier limits stated
in Theorem~\ref{thm:eigenvalue_fluctuations}.

\subsubsection{Gaussianity}
We examine departures from the Gaussianity assumed in
Theorems~\ref{thm:eigenvalue_fluctuations}
and~\ref{thm:eigenvalue_fluctuations_ext} in simulations with
$n=1000$, $r=\sigma=1$, and $\theta=1.5$ (supercritical signal) or $\theta=0.5$ (subcritical signal).
We compare dense directions,
$$
u=\frac{1}{\sqrt n}(1,\ldots,1)^\top,
\qquad
v=\frac{1}{\sqrt n}(1,-1,\ldots,1,-1)^\top,
$$
with sparse directions $u=e_1$ and $v=e_2$. The upper-triangular noise entries are independent and follow one of four distributions:
a standard Gaussian law, the uniform law on
$[-\sqrt3,\sqrt3]$, a Student $t_{4}$ law rescaled to have unit
variance, or a Student $t_{8}$ law rescaled to have unit variance. 

\begin{figure}[t]
    \centering
    \begin{subfigure}[t]{0.48\textwidth}
        \centering
        \includegraphics[width=\linewidth]{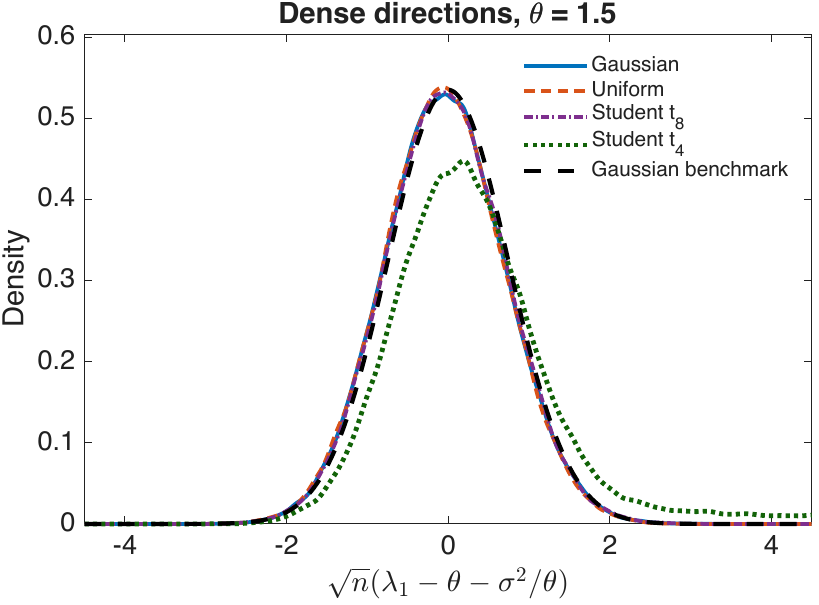}
        \caption{Dense signal directions.}
        \label{fig:non_gaussian_supercritical_dense}
    \end{subfigure}\hfill
    \begin{subfigure}[t]{0.48\textwidth}
        \centering
        \includegraphics[width=\linewidth]{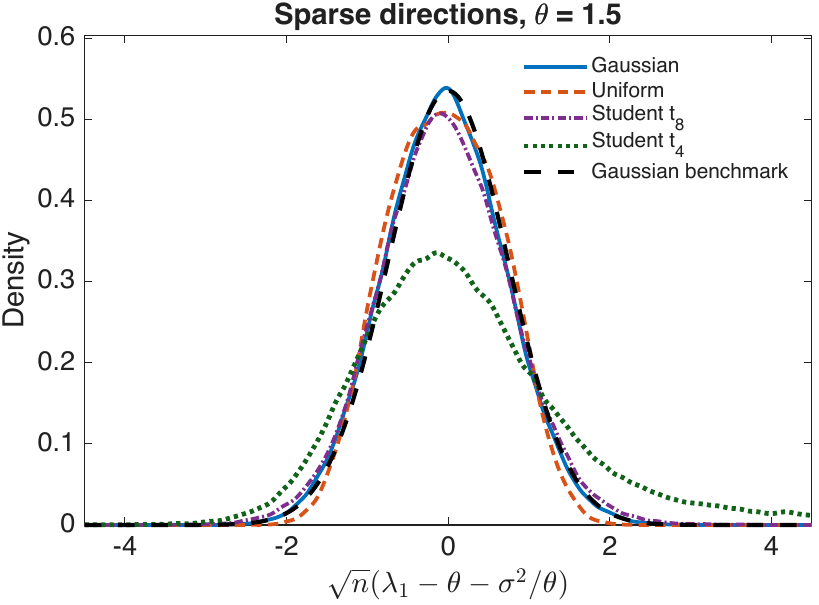}
        \caption{Sparse signal directions.}
        \label{fig:non_gaussian_supercritical_sparse}
    \end{subfigure}
    \caption{Supercritical fluctuations with $\theta=1.5$,
    $n=1000$, and $\sigma=1$.
    Colored curves show kernel density estimates of
    $\sqrt n(\lambda_1-\theta-\sigma^2/\theta)$
    from $10^5$ replications under Gaussian, uniform,
    and standardized Student $t_{8}$ and $t_{4}$ noise.
    The black dashed curve is the Gaussian benchmark
    $\mathcal N(0,5/9)$ from
    Theorem~\ref{thm:eigenvalue_fluctuations}.}
    \label{fig:non_gaussian_supercritical}
\end{figure}

For the supercritical strength $\theta=1.5$,
Figure~\ref{fig:non_gaussian_supercritical} shows close
agreement with the Gaussian benchmark under Gaussian,
uniform, and Student $t_{8}$ noise when the directions
are dense. For sparse directions, the estimated densities depend visibly on the noise law. This suggests a distinction similar to that established
for simple outliers in deformed Wigner matrices:  under appropriate assumptions, delocalized directions give universal Gaussian fluctuations,
whereas localized directions can yield limits that depend
on the entry distribution
\citep[Theorems~3.2--3.4]{capitaine-fluctuations}. 

\begin{figure}[t]
    \centering
    \begin{subfigure}[t]{0.48\textwidth}
        \centering
        \includegraphics[width=\linewidth]{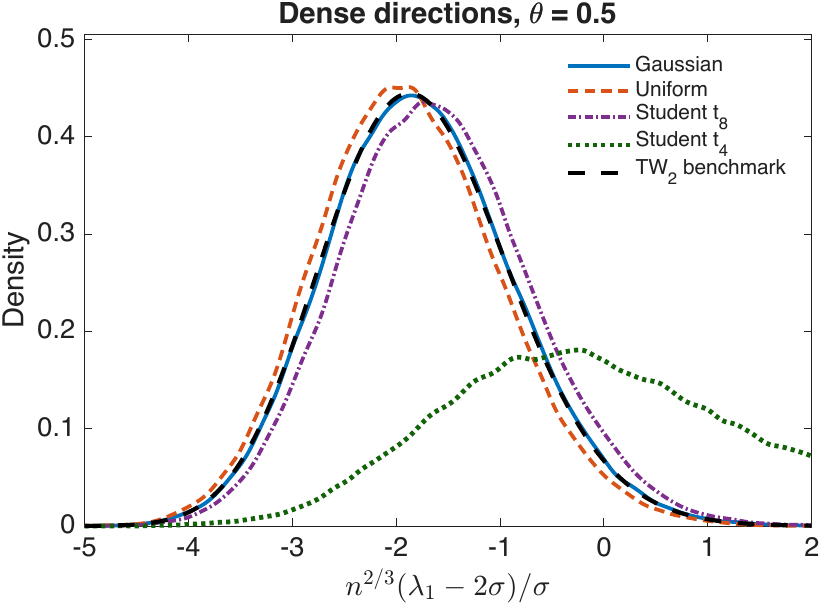}
        \caption{Dense signal directions.}
        \label{fig:non_gaussian_subcritical_dense}
    \end{subfigure}\hfill
    \begin{subfigure}[t]{0.48\textwidth}
        \centering
        \includegraphics[width=\linewidth]{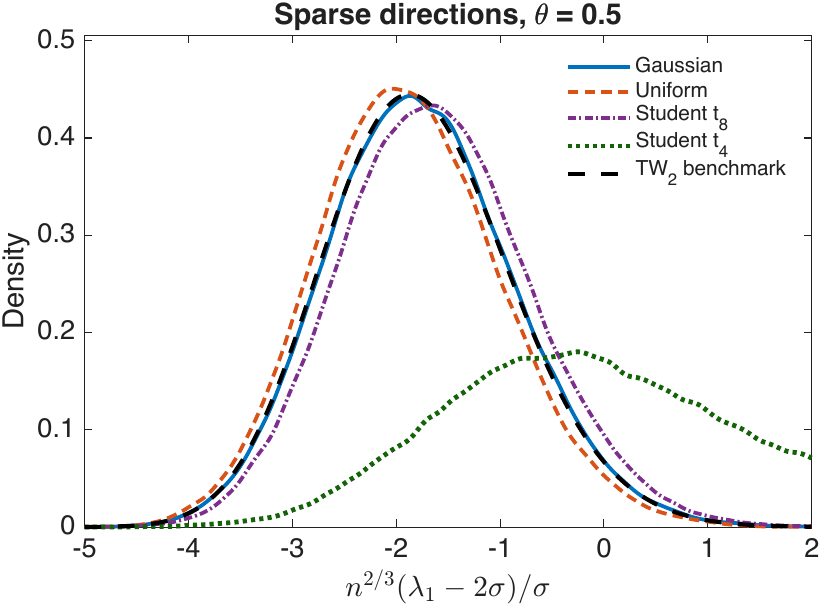}
        \caption{Sparse signal directions.}
        \label{fig:non_gaussian_subcritical_sparse}
    \end{subfigure}
    \caption{Subcritical fluctuations with $\theta=0.5$,
    $n=1000$, and $\sigma=1$.
    Colored curves show kernel density estimates of
    $n^{2/3}(\lambda_1-2\sigma)/\sigma$
    from $10^5$ replications under the same four noise laws.
    The black dashed curve is a numerical approximation
    to the $\TW_2$ density.}
    \label{fig:non_gaussian_subcritical}
\end{figure}

For the subcritical strength $\theta=0.5$,
Figure~\ref{fig:non_gaussian_subcritical} shows close
agreement with the $\TW_2$ benchmark under Gaussian,
uniform, and Student $t_{8}$ noise for both choices of
directions. This parallels the edge universality established
for deformed Wigner matrices without localization
assumptions on the signal directions
\citep[Theorem~2.7 and Remark~2.11]{knowles2013isotropic},
although those results do not apply directly to the
present setting. 
We expect the $\beta=2$ edge limit to hold for non-Gaussian
noise when the upper-triangular entries are i.i.d.$\,(0,\sigma^2)$ with finite $(4+\delta)$th moment for some $\delta>0$, with the signal assumptions unchanged.

Student $t_4$ noise shows substantial departures from
both benchmarks, for both choices of signal directions.
Unlike the other three noise laws, this distribution has
an infinite fourth moment and falls outside
Assumption~\ref{ass:ass1}. The fourth-moment condition
controls the contribution of unusually large noise entries,
which can otherwise affect even the first-order behavior
of the largest eigenvalues. In both Figures~\ref{fig:non_gaussian_supercritical} and \ref{fig:non_gaussian_subcritical}, the fluctuations under $t_4$ noise are nonetheless of the same order of magnitude as those under the other three distributions.

\FloatBarrier
\section{Eigenvector asymptotics}\label{sec:asy_eigenvectors}

In this section, we study the asymptotic behavior of the leading eigenvectors of $H_n$. A supercritical signal strength can be consistently estimated from its outlier eigenvalue. The associated directions $u_j,v_j$, however, are identified only up to simultaneous
rotation: rotating this pair leaves $u_jv_j^\top-v_ju_j^\top$ unchanged. For a strength of multiplicity one, the population signal therefore identifies the plane $\operatorname{Span}_{\mathbb R}(u_j,v_j)$, but does not select a unique pair of axes within it. We estimate this plane by the span of the real and imaginary parts of an eigenvector of $H_n$ associated with the corresponding positive outlier.

For each supercritical strength $\theta_j$ of multiplicity one, the corresponding eigenvalue $\lambda_j$ is positive and simple almost surely, for all sufficiently large $n$, by Theorem~\ref{thm:general_random_skew_outliers}. On this event, let $\widehat w_{j,n}$ be a unit eigenvector of $H_n$ associated with $\lambda_j$, and define
$$
\widehat u_{j,n}=\sqrt2\,\Re\widehat w_{j,n},
\qquad
\widehat v_{j,n}=-\sqrt2\,\Im\widehat w_{j,n}.
$$
Then $\widehat u_{j,n},\widehat v_{j,n}$ form an orthonormal pair
in $\mathbb R^n$. On the complementary event, set
$\widehat u_{j,n}=u_j$ and $\widehat v_{j,n}=v_j$.
This convention does not affect the following limiting results.

To measure estimation accuracy, we use the squared canonical
correlations between the population and estimated planes,
which are the cosines squared of their two canonical angles.
Equivalently, these are the eigenvalues of the operator
that projects from the population plane onto the estimated
plane and then back onto the population plane; see \citet[Chapter~11.3]{muirhead2009aspects} and
\citet[Section~1.3]{bykhovskaya2024canonical}.
The following theorem gives their almost-sure limits.

\bTh\la{thm:canonical_angles_supercritical}
Suppose Assumptions~\ref{ass:ass1} and~\ref{ass:ass2} hold.
Then, for each $j\in\{1,\ldots,r\}$ such that
$\theta_j>\sigma$ and $\theta_j$ has multiplicity one,
$$
    \begin{pmatrix}
        u_{j} & v_{j}
    \end{pmatrix}^{\top}\begin{pmatrix}
        \widehat u_{j,n} & \widehat v_{j,n}
    \end{pmatrix}\begin{pmatrix}
        \widehat u_{j,n} & \widehat v_{j,n}
    \end{pmatrix}^{\top}\begin{pmatrix}
        u_{j} & v_{j}
    \end{pmatrix}
    \overset{\mathrm{a.s.}}{\longrightarrow}
    \left(1-\frac{\sigma^2}{\theta_j^2}\right)I_2.
$$
In particular, both squared canonical correlations between the population plane $\operatorname{Span}_{\mathbb R}(u_{j},v_{j})$ and the estimated plane $\operatorname{Span}_{\mathbb R}(\widehat u_{j,n},\widehat v_{j,n})$ converge almost surely to $1-\sigma^2\theta_j^{-2}$.
\eTh

For each fixed supercritical strength $\theta_j$ of multiplicity one, both canonical angles converge almost surely to $\arcsin(\sigma/\theta_j)>0$. Thus, although the signal strength can be consistently estimated, its factor plane retains a nonvanishing estimation error.

\begin{figure}[t]
    \centering
    \includegraphics[width=0.5\textwidth]{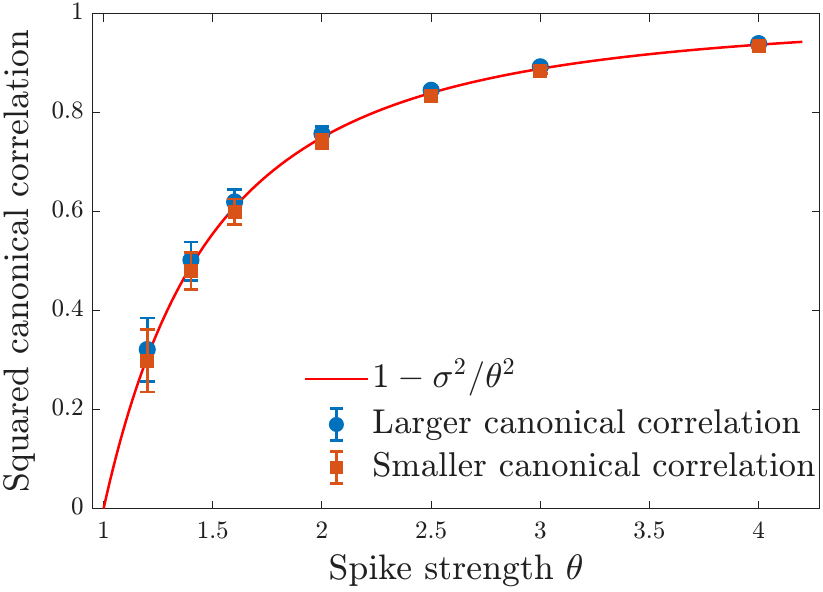}
    \caption{Squared canonical correlations between the population and estimated factor planes as functions of the signal strength $\theta$. We use Gaussian noise with $n=1000$ and $\sigma=1$, and a single signal $S_n=\theta(u_1v_1^\top-v_1u_1^\top)$ with $u_1=e_1$ and $v_1=e_2$. For each $\theta>\sigma$, the blue circles and orange squares show the means of the larger and smaller squared canonical correlations, respectively, over $500$ Monte Carlo replications. The vertical bars span the empirical $10$th to $90$th percentiles. The red curve shows the theoretical limit $1-\sigma^2/\theta^2$ from Theorem~\ref{thm:canonical_angles_supercritical}.}
    \label{fig:cca}
\end{figure}

The limiting angles decrease as $\theta_j$ increases or
$\sigma$ decreases, and tend to zero as the signal-to-noise ratio $\theta_j/\sigma$ tends to infinity. Conversely, they approach $\pi/2$ as $\theta_j$ decreases to $\sigma$. A signal just above the separation threshold can therefore produce an outlier while its estimated plane remains nearly orthogonal to the population plane. This pattern is illustrated in Figure~\ref{fig:cca}. The simulations also show that the variance of the canonical correlations decreases as $\theta$ increases.

In applications, we suggest estimating the limiting canonical angles as a measure of the accuracy of the estimated factor planes. Using the strongly consistent estimators of signal strength and noise variance introduced above, we obtain
\be\la{eq:cca_estimate}
\widehat c_j^{\,2}=1-\frac{\widehat\sigma_n^2}{\widehat\theta_j^2}.
\ee
For a fixed simple supercritical strength, this quantity is a strongly consistent estimator of the common limit of the two squared canonical correlations. When $\sigma^2$ is known, it replaces $\widehat\sigma^2_n$. Smaller estimated angles indicate more accurate estimation of the factor plane. Confidence regions for the factor planes would require further fluctuation results.


\FloatBarrier
\section{Empirical illustration}\label{sec:empirical}

We present two examples illustrating the application of our theory. The first examines the stock outperformance matrix, while the second focuses on metropolitan migration imbalances.

\subsection{S\&P 500}
\label{sec:sp500-empirical}

We illustrate our methods through an application to the relative performance of stocks in the S\&P 500. For each pair of stocks, we record how often one stock's daily return exceeds the other's. Our data come from Yahoo Finance and consist of daily closing prices adjusted for stock splits and dividends, from January 2, 2014 to August 21, 2026. We retain $n=447$ stocks with complete histories over this period, yielding $T=3177$ daily log return observations.\footnote{We exclude GOOG, FOX, and NWS to avoid duplicate share classes; we keep class A shares GOOGL, FOXA, and NWSA. We also exclude SW, FERG, and AMCR, whose early price histories are back-filled.} Let $r_{t\ell}=\log(P_{t\ell}/P_{t-1,\ell})$, where $P_{t\ell}$ is the adjusted price of stock $\ell$. We construct the skew-symmetric outperformance matrix\footnote{A natural alternative is the matrix of average differences of returns, $(\bar r_k-\bar r_\ell)_{k,\ell}$, where $\bar r_\ell=T^{-1}\sum_{t=1}^T r_{t\ell}$. However, this matrix equals $\bar r\mathbf1^\top-\mathbf1\bar r^\top$, so it has rank at most two and range contained in $\operatorname{span}\{\mathbf1,\bar r\}$. It therefore cannot recover multiple factor planes.}
\[
 (X_n)_{k \ell}=\frac1T\sum_{t=1}^{T}
 \left[\mathbf1\{r_{tk}>r_{t\ell}\}-\mathbf1\{r_{tk}<r_{t\ell}\}\right].
\]
Thus, a positive entry indicates that stock $k$ outperforms stock $\ell$ more often than the reverse, with a tie contributing zero. Each entry is an affine transformation of the fraction of days the trade ``buy $k$, sell $\ell$'' makes money. The matrix records the frequency of relative performance, rather than the magnitude of return differences. The construction is also invariant to common additive
shifts and positive rescalings of daily returns, even
when these shifts and scales vary over time.

\begin{figure}[t]
    \centering
    \begin{subfigure}[t]{0.48\linewidth}
        \centering
        \includegraphics[width=\linewidth]{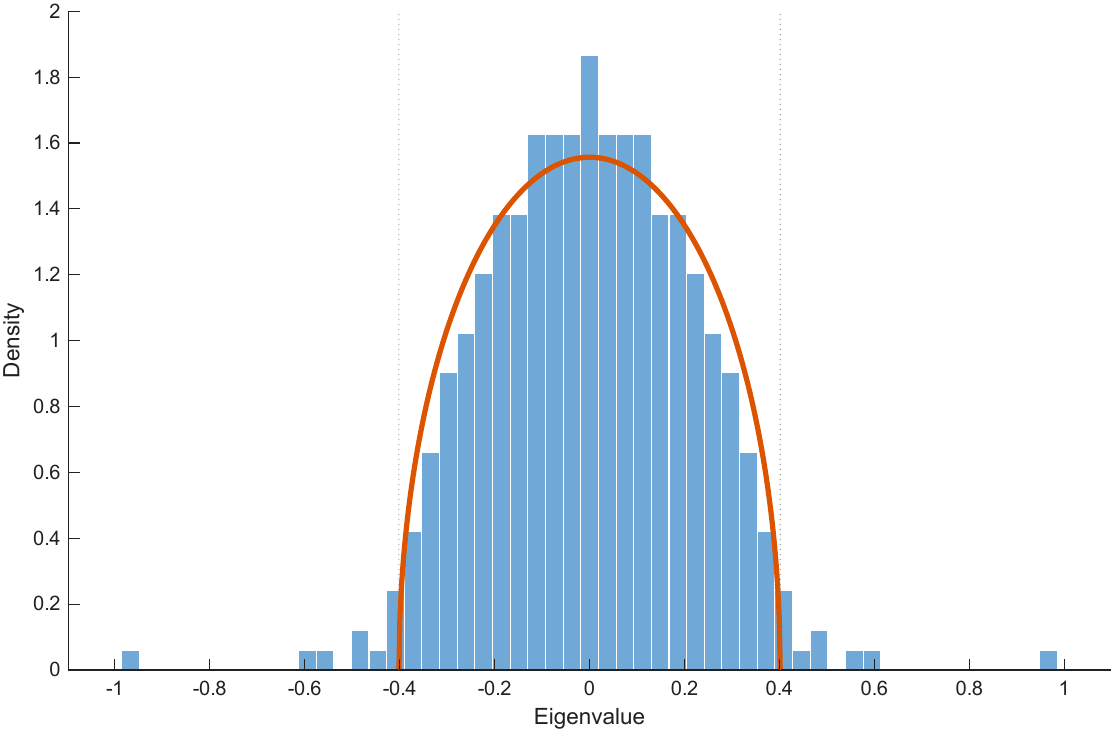}
        \caption{Full spectrum except $\lambda_1,\lambda_n$.}
        \label{fig:sp500_bulk}
    \end{subfigure}
    \hfill
    \begin{subfigure}[t]{0.48\linewidth}
        \centering
        \includegraphics[width=\linewidth]{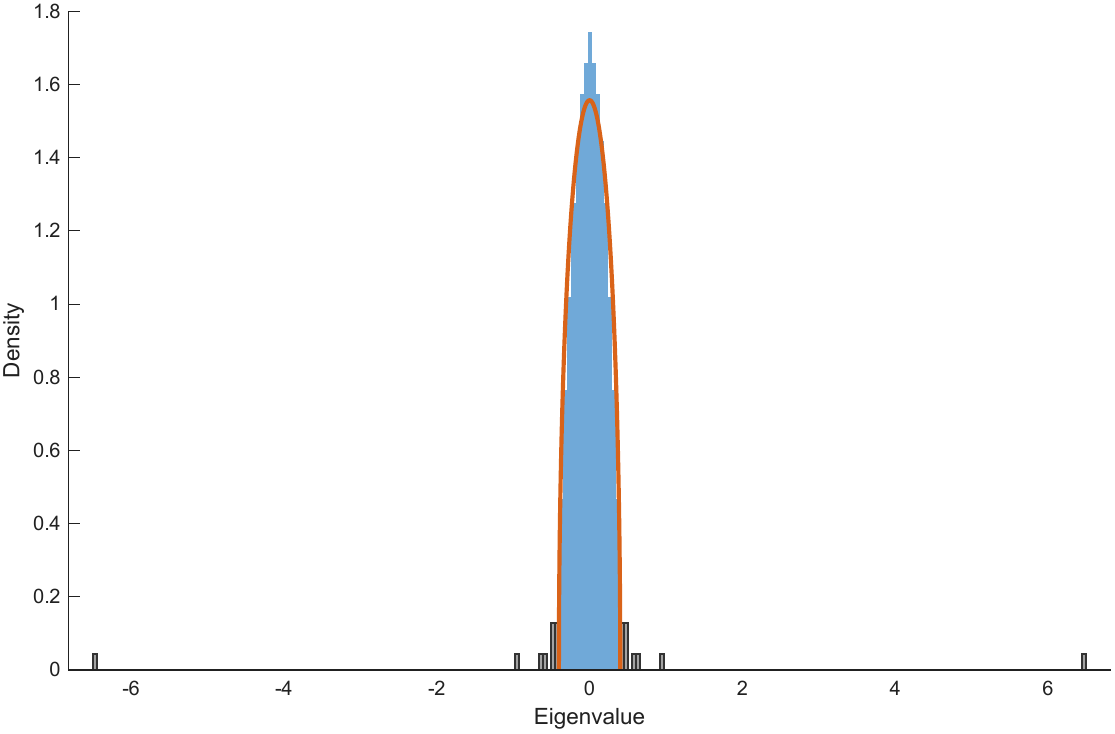}
        \caption{Full spectrum.}
        \label{fig:sp500_full_spectrum}
    \end{subfigure}
    \caption{S\&P~500: spectrum of $\ii X_n$ for S\&P~500 stock returns and fitted semicircle.}
    \label{fig:sp500_spectrum}
\end{figure}

Figure~\ref{fig:sp500_spectrum} shows the eigenvalues of $\ii X_n$. As suggested in Corollary~\ref{cor:variance_estimation}, we estimate $\sigma^2$ by the mean squared positive eigenvalue after omitting the four largest positive eigenvalues. This gives $\widehat\sigma\approx0.2$ and a fitted semicircle edge of $2\widehat\sigma\approx0.4$. 
There are ten positive eigenvalues to the right of this edge, each accompanied by its negative counterpart. They are highlighted in black in Figure~\ref{fig:sp500_full_spectrum}. For all ten, a sequential Tracy--Widom $F_2$ test rejects the null $\theta_j<\sigma$ at the $5\%$ level. The tenth, $\lambda_{10}=0.402$, is very close to the cutoff and may correspond to a critical signal, which the test cannot distinguish from a supercritical one. We therefore report results only for the first nine signals in Table~\ref{tab:sp500_angles}, which lists the corresponding positive eigenvalues, signal-strength estimates, and factor-plane accuracy implied by Theorem~\ref{thm:canonical_angles_supercritical}.

As is common for financial data sets, the leading eigenvalue dominates the spectrum: $\lambda_1$ is very close to the estimated signal strength $\widehat\theta_1$, and the estimated canonical angles are only $1.8^\circ$. These estimates are close to the strong-signal limit, in which the eigenvalue consistently estimates strength and the canonical angles vanish. The second eigenplane is also recovered relatively accurately, with estimated canonical angles of $12.8^\circ$. Accuracy decreases for the subsequent planes: the angles are $22.0^\circ$ and $25.4^\circ$ for the third and fourth planes, respectively, and range from $30.9^\circ$ to $50.4^\circ$ for the fifth through ninth. Thus, under the fitted model, these additional eigenvalues indicate further signals, yet their associated directions are estimated less precisely.

\begin{table}[t]
 \centering
 \small
 \begin{tabular}{@{}crrrr@{}}
 \toprule
 $j$ & $\lambda_j$ & $\widehat\theta_j$ & $\widehat{\cos^2\phi_j}$
 & $\widehat\phi_j$ (degrees) \\
 \midrule
 1 & 6.489 & 6.483 & 0.999 & 1.8 \\
 2 & 0.954 & 0.909 & 0.951 & 12.8 \\
 3 & 0.611 & 0.536 & 0.860 & 22.0 \\
 4 & 0.554 & 0.468 & 0.816 & 25.4 \\
 5 & 0.494 & 0.391 & 0.737 & 30.9 \\
 6 & 0.466 & 0.351 & 0.673 & 34.9 \\
 7 & 0.449 & 0.324 & 0.616 & 38.3 \\
 8 & 0.424 & 0.280 & 0.484 & 45.9 \\
 9 & 0.415 & 0.261 & 0.406 & 50.4 \\
 \bottomrule
 \end{tabular}
 \caption{S\&P~500: signal strengths and factor-plane accuracy.
    Each angular estimate applies to the common limit of both
    canonical angles of Theorem~\ref{thm:canonical_angles_supercritical}.\label{tab:sp500_angles}}
\end{table}

\begin{figure}[t]
 \centering
 \includegraphics[width=0.8\linewidth]{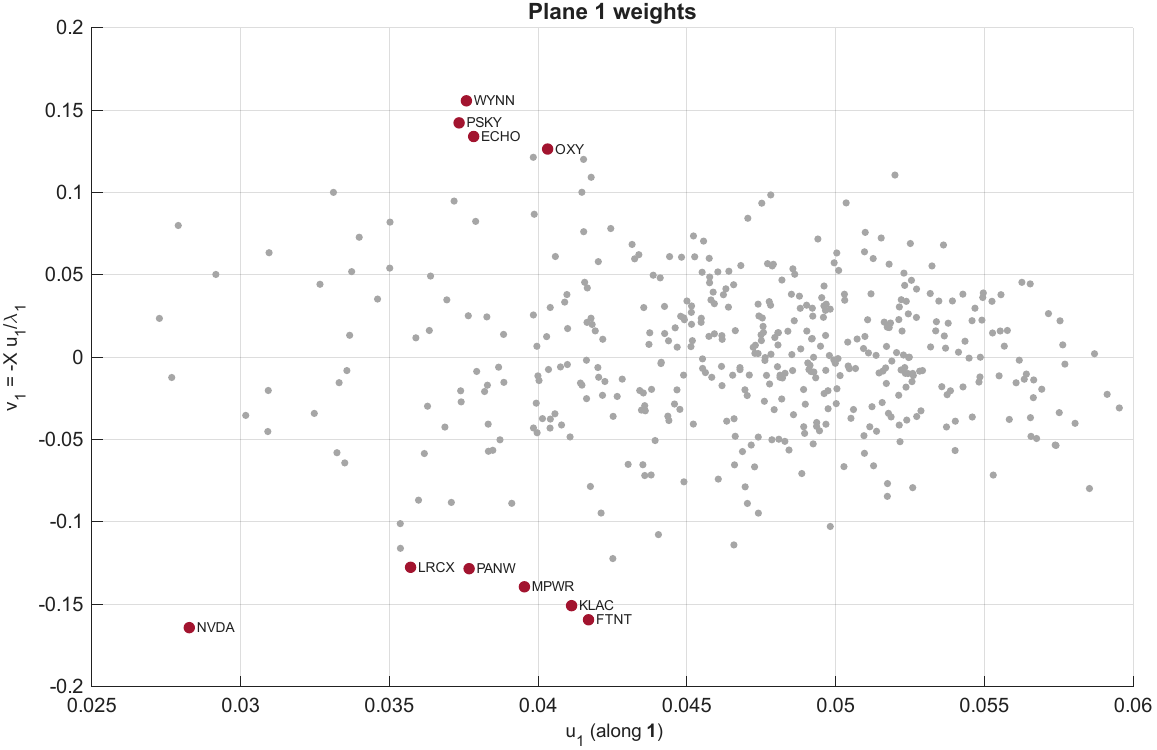}
 \caption{Leading eigenplane: each point is
 a stock, plotted at
 $(\widehat u_{1,\ell},\widehat v_{1,\ell})$.
 The vertical coordinate orders the fitted ranks.}
 \label{fig:SP500_plane1_R2}
\end{figure}

Let us focus on the first four estimated eigenplanes, $\{\widehat{\mathcal V}_j\}_{j=1}^4$, which correspond to the strongest signals. Even among these, some of the weaker signals do not have a clear economic interpretation. Let $\mathbf{1}\in\mathbb{R}^n$ be the all-ones vector, and let $\widehat\Pi_1$ denote the orthogonal projector onto $\widehat{\mathcal V}_1$. We orient the basis of $\widehat{\mathcal V}_1$ by
$$
    \widehat u_1
    =\frac{\widehat\Pi_1\mathbf{1}}{\|\widehat\Pi_1\mathbf{1}\|},
    \qquad
    \widehat v_1
    =-\frac{X_n\widehat u_1}{\lambda_1}.
$$
Thus, $\widehat u_1$ is the unit vector in the plane closest to $\mathbf{1}/\sqrt{n}$. Since the plane is invariant under $X_n$, the vector $\widehat v_1=-X_n\widehat u_1/\lambda_1$ also belongs to it. Moreover, $X_n\widehat u_1$ is orthogonal to $\widehat u_1$ and has norm $\lambda_1$, so $\widehat u_1$ and $\widehat v_1$ form an orthonormal basis. The uniform reference vector treats all stocks symmetrically. Its strong empirical alignment with the leading plane, $\|\widehat\Pi_1\mathbf{1}/\sqrt{n}\|=0.991$, means that $\widehat u_1$ provides a nearly uniform loading direction across stocks.
Figure~\ref{fig:SP500_plane1_R2} displays coordinates of this basis.

This orientation relates $\widehat v_1$ to return ranks and rank momentm strategies. Let $R_{t\ell}$
be the cross-sectional rank of $r_{t\ell}$, assigning rank $n$ to the highest return, and write $\overline R_{\ell}=T^{-1}\sum_tR_{t\ell}$. Then
$$
    (X_n\mathbf{1})_{\ell}
    =\frac{1}{T}\sum_{t=1}^{T}\sum_{k=1}^{n}
      \operatorname{sign}(r_{t\ell}-r_{tk})
    =\frac{1}{T}\sum_{t=1}^{T}(2R_{t\ell}-n-1)=2\overline R_{\ell}-n-1.
$$
Thus, $X_n\mathbf{1}$ is proportional to the demeaned average-rank vector $\overline R$.
So if $\widehat u_1$ is a fitted analogue of $\mathbf{1}/\sqrt{n}$, then $\widehat v_1$ is the (rescaled) fitted analogue of $-X_n\mathbf{1}$, i.e.\ $\widehat v_1$ is approximately a decreasing affine function of average historical rank. In particular, the correlation between $-\widehat v_1$ and
$\overline R$ is $0.99995$. Thus, negative loadings $\widehat v_{1,i}$ identify stocks that
ranked highly more often during the sample, e.g.\ NVIDIA, Fortinet, and KLA Corporation; see Figure~\ref{fig:SP500_plane1_R2}. These loadings connect the first plane to momentum strategies, which buy past winners and sell past losers \citep{efficiency1993returns}. A portfolio based on past observations that buys stocks with negative loadings and shorts those with positive loadings
would use a signal closely related to the rank momentum of \citet{chen2021non}.

\begin{figure}[t]
    \centering
    \begin{subfigure}[b]{0.48\textwidth}
        \centering
        \includegraphics[width=1.3\linewidth]{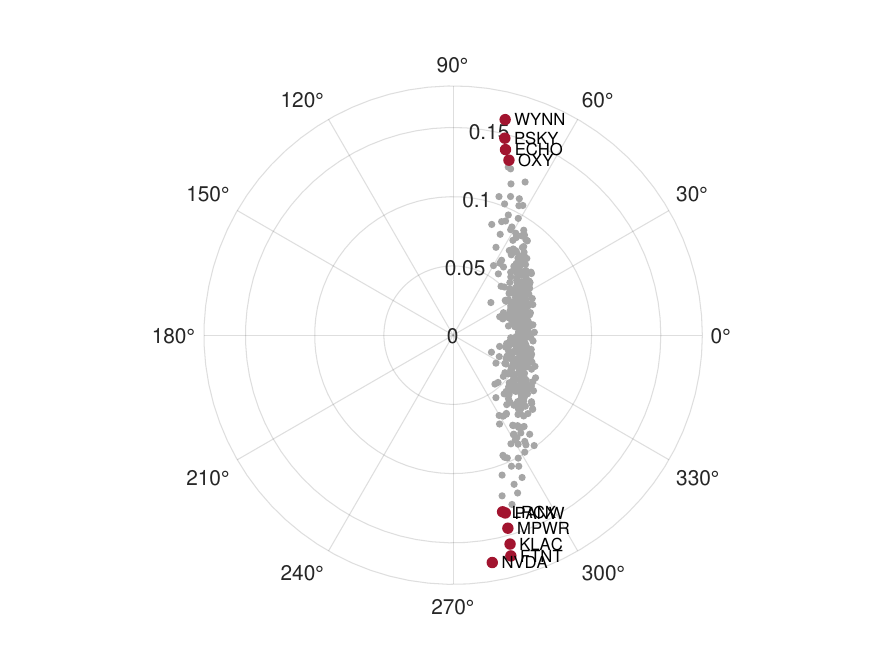}
        \caption{First eigenplane}
        \label{fig:plane1}
    \end{subfigure}
    \hfill
    \begin{subfigure}[b]{0.48\textwidth}
        \centering
        \includegraphics[width=1.3\linewidth]{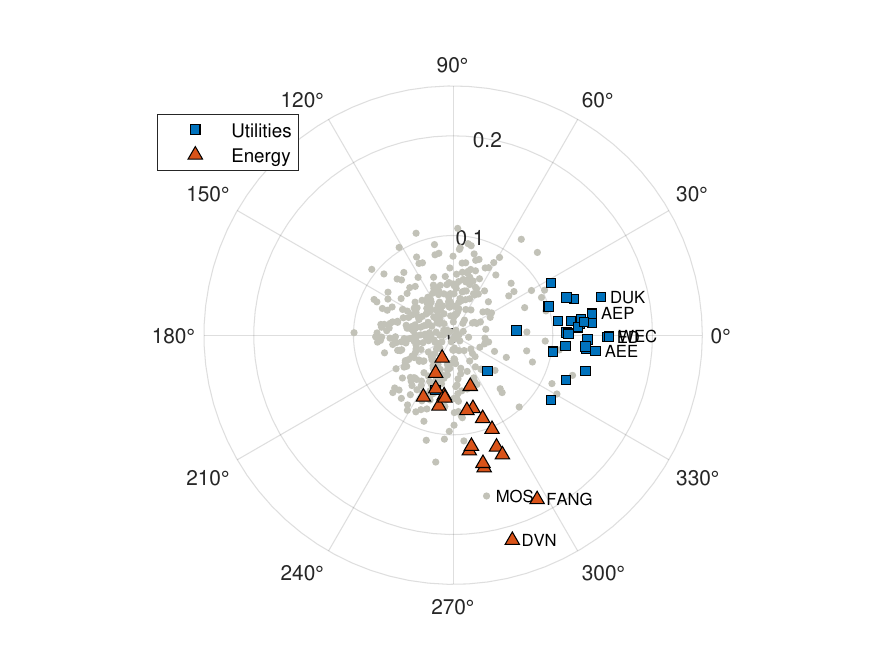}
        \caption{Second eigenplane}
        \label{fig:plane2}
    \end{subfigure}
    \begin{subfigure}[b]{0.48\textwidth}
        \centering
        \includegraphics[width=1.3\linewidth]{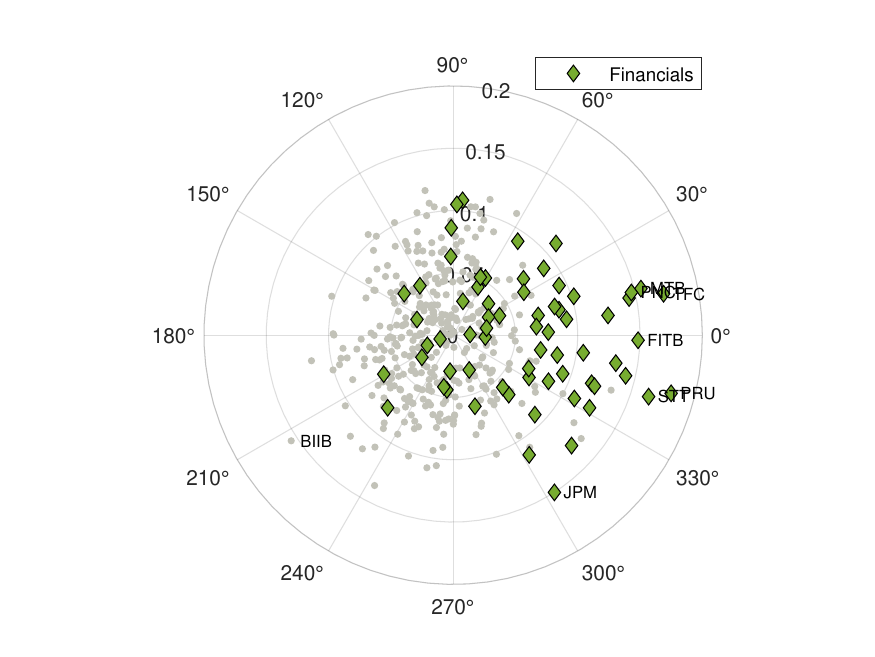}
        \caption{Third eigenplane}
        \label{fig:plane3}
    \end{subfigure}
    \hfill
    \begin{subfigure}[b]{0.48\textwidth}
        \centering
        \includegraphics[width=1.3\linewidth]{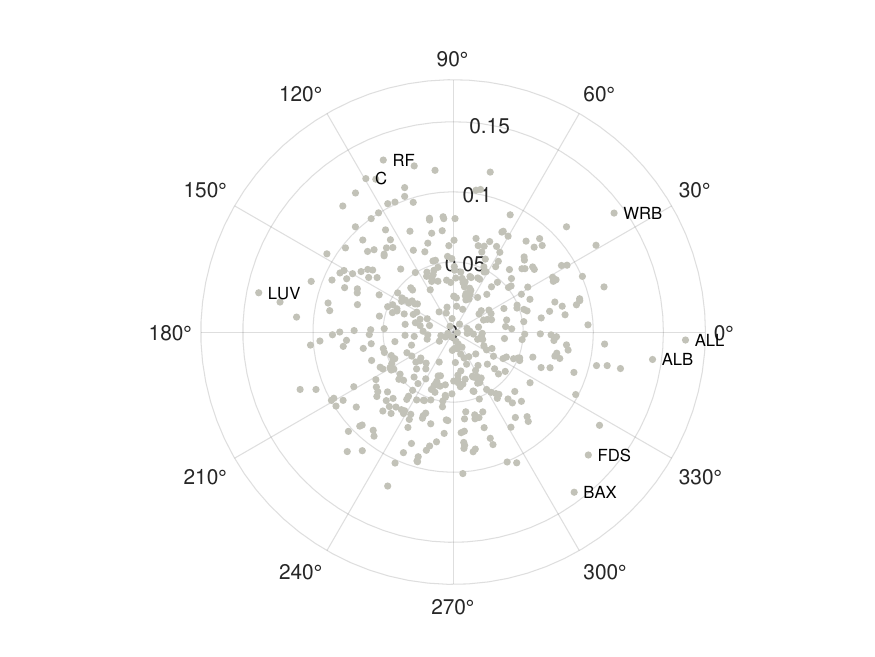}
        \caption{Fourth eigenplane}
        \label{fig:plane4}
    \end{subfigure}
    \caption{S\&P~500: leading eigenplanes in polar coordinates.}
    \label{fig:SP500four_eigenplanes}
\end{figure}

Figure~\ref{fig:SP500four_eigenplanes} illustrates the first four eigenplanes in polar coordinates. We start from a unit eigenvector $\widehat w_j$, $j=1,\ldots,4$ of $\ii X_n$ associated with its $j$th-largest positive eigenvalue $\lambda_j$, and choose the orthonormal basis $\widehat u_j=\sqrt{2}\Re(\widehat w_j)$ and $\widehat v_j=-X_n\widehat u_j/\lambda_j$. Each stock $\ell$ is represented by the polar coordinates of its loading pair $(\widehat u_{j,\ell},\widehat v_{j,\ell})$. Equivalently, the projection of the $\ell$th coordinate vector $e_{\ell}$ onto the plane is
\[
\operatorname{proj}_{\widehat{\mathcal V}_j}e_{\ell}
=
r_{j,\ell}\bigl(\widehat u_j\cos\phi_{j,\ell}+\widehat v_j\sin\phi_{j,\ell}\bigr),
\qquad
r_{j,\ell}=\sqrt{\widehat u_{j,\ell}^{\,2}+\widehat v_{j,\ell}^{\,2}}.
\]
The radii are independent of the orthonormal basis, while rotating the basis shifts all angles by the same constant, preserving their relative differences. 

We next relate the leading eigenplanes shown in polar coordinates in Figure~\ref{fig:SP500four_eigenplanes} to the 11 GICS sectors. Sectoral structure in the eigenvectors of stock-return correlation matrices has previously been documented by \citet{gopikrishnan2001quantifying,plerou2002random}. In our outperformance matrix, the second eigenplane has prominent Utilities and Energy components, while the third has a visible Financials component. 

Utilities and Energy together contain 49 stocks, or $11.0\%$ of the sample, but account for $33.4\%$ of the coordinate shares of $\widehat{\mathcal V}_2$: $21.9\%$ for Utilities and $11.5\%$ for Energy. The largest individual shares belong to Devon Energy and Diamondback Energy in Energy; WEC Energy Group, Consolidated Edison, and Duke Energy in Utilities; and Mosaic in Materials. The canonical correlations between $\widehat{\mathcal V}_2$ and the two-dimensional span of the demeaned Utilities and Energy indicators are $0.66$ and $0.41$, indicating moderate alignment.\footnote{Canonical correlations are the cosines of the canonical angles between the two subspaces, with values closer to one indicating closer alignment.} Using the ten-dimensional span of all 11 demeaned sector indicators gives canonical correlations of $0.78$ and $0.52$.

Financials accounts for $28.9\%$ of the coordinate shares of $\widehat{\mathcal V}_3$, compared with $15.2\%$ of the stocks. The canonical correlation between $\widehat{\mathcal V}_3$ and the span of the demeaned Financials indicator is $0.52$. Seven of the eight stocks with the largest individual shares are in Financials: Prudential Financial, Truist Financial, State Street, M\&T Bank, JPMorgan Chase, Fifth Third Bancorp, and PNC Financial Services Group; the exception is Biogen in Health Care.

The fourth eigenplane in Figure~\ref{fig:plane4} has a less direct sectoral interpretation. Its largest individual shares are dispersed across several sectors: Allstate, W.\ R.\ Berkley, and Regions Financial are in Financials, Albemarle is in Materials, Baxter International is in Health Care, and Southwest Airlines is in Industrials. The estimated limiting canonical angles between the sample and population factor planes are larger for the weaker signals, indicating less accurate recovery. This estimation
error may also make eigenplanes beyond the first few harder to interpret.

A more detailed economic interpretation of the leading eigenplanes remains an open question. In particular, one could compare their loadings with stock characteristics and exposures to factors in the asset pricing ``factor zoo''.

\FloatBarrier
\subsection{Metro-to-metro migration}
\label{sec:census-empirical} 
We next apply the spectral decomposition to migration between metropolitan areas. Migration typically occurs in both directions, and substantial movement need not produce a large net transfer of population. The migration literature therefore distinguishes the volume of movement from its effectiveness in redistributing
population. For the United States, \citet{plane2005migration} use net movement relative to gross exchange to study migration up and down the urban hierarchy. We use the same
net-to-gross normalization, also known as stream effectiveness, at the level of individual pairs of metropolitan areas.

We use the \href{https://www.census.gov/data/tables/2020/demo/geographic-mobility/metro-to-metro-migration.html}{Census Bureau's 2016--2020} estimates of one-year metro-to-metro migration from the American Community Survey (ACS), obtained by pooling and weighting survey responses collected over the five years. Restricting both endpoints to metropolitan areas in the 50 U.S. states, the District of Columbia, and Puerto Rico yields $n=392$ metros. Let $F_{k\ell}$ denote estimated migration from metro $k$ to metro $\ell$. We form the bilateral migration effectiveness matrix $X_n$, with entries
$$
    (X_n)_{k\ell}
    = \frac{F_{k\ell}-F_{\ell k}}{F_{k\ell}+F_{\ell k}},\qquad k\ne\ell,
 \qquad (X_n)_{kk}=0,
    \label{eq:census-matrix}
$$
and set an entry to zero when neither direction has a reported positive flow $F_{k\ell}=F_{\ell k}=0$. The resulting matrix is skew-symmetric. A positive entry indicates net movement
from $k$ to $\ell$, zero represents a balanced exchange or no reported flow, and an absolute value of one indicates that the reported exchange is entirely in one direction. The normalization removes the scale of a given pair's gross flow, so that small-volume links can contribute as much as large-volume links; it does not adjust for population size or the number of migration partners. 

\begin{figure}[t]
        \centering
        \includegraphics[width=0.7\linewidth]{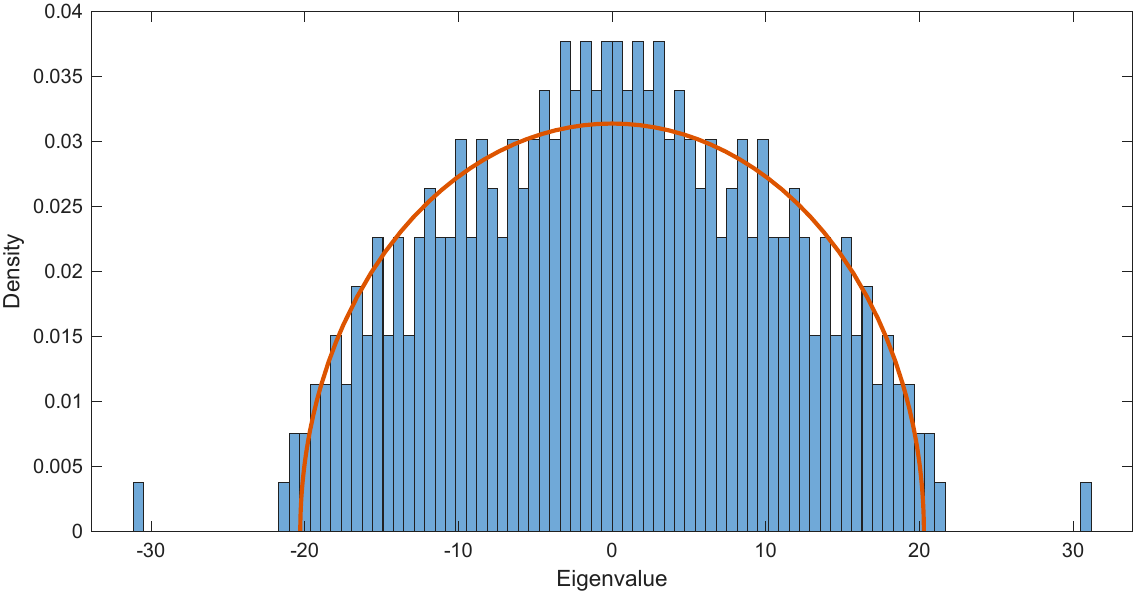}
        \caption{Bilateral migration effectiveness: spectrum of $\ii X_n$ and fitted semicircle.}
        \label{fig:census-spectrum}
\end{figure}

Figure~\ref{fig:census-spectrum} shows that the spectrum of $\ii X_n$ is broadly compatible with a semicircle law. After removing the largest positive and negative eigenvalues,
we fit the semicircle radius as twice the square root of the mean squared remaining eigenvalue. Four positive eigenvalues lie above the fitted cutoff $2\widehat{\sigma}\approx 20.3$. For all four, a sequential Tracy--Widom $F_2$ test rejects the null $\theta_j<\sigma$ at the $5\%$ level. The fourth, $\lambda_4=20.34$, is very close to the cutoff and may correspond to a critical signal, which the test cannot distinguish from a supercritical one. We therefore report results only for the first three signals, with their estimated strengths and the implied accuracy of their factor planes given in Table~\ref{tab:migration_angles}. The leading plane is substantially better recovered than the next two. 

\begin{table}[t]
 \centering
 \begin{tabular}{@{}crrrrr@{}}
 \toprule
 $j$ & $\lambda_j$ & $\widehat\theta_j$ & $\widehat\theta_j/\widehat\sigma$
 & $\widehat{\cos^2\phi_j}$ & $\widehat\phi_j$ (degrees) \\
 \midrule
 1 & 30.80 & 26.99 & 2.66 & 0.86 & 22.1 \\
 2 & 21.33 & 13.96 & 1.38 & 0.47 & 46.6 \\
 3 & 20.85 & 12.82 & 1.26 & 0.37 & 52.3 \\
 \bottomrule
 \end{tabular}
 \caption{Bilateral migration effectiveness: signal strengths and factor-plane accuracy. Each angular estimate applies to the common limit of both canonical angles of Theorem~\ref{thm:canonical_angles_supercritical}.\label{tab:migration_angles}}
\end{table}

We begin the analysis with the leading invariant plane $\widehat{\mathcal V}_1$, associated with the eigenvalue pair $\pm\lambda_1$ of $\ii X_n$. To give the plane an interpretable orientation, let $\mathbf{1}$ be the all-ones vector, and let $\widehat\Pi_1$ denote the orthogonal projector onto $\widehat{\mathcal V}_1$. We choose
$$
    \widehat u_1
    =\frac{\widehat\Pi_1\mathbf{1}}{\|\widehat\Pi_1\mathbf{1}\|},
    \qquad
    \widehat v_1
    =-\frac{X_n\widehat u_1}{\lambda_1}.
$$
Here $\widehat u_1$ is the unit vector in the plane closest to $\mathbf{1}/\sqrt{n}$, and $X_n$ fixes the sign of $\widehat v_1$. The orthonormal basis satisfies $\widehat u_1^{\top}\mathbf{1}=\|\widehat\Pi_1\mathbf{1}\|$ and $\widehat v_1^{\top}\mathbf{1}=0$. We obtain $\|\widehat\Pi_1\mathbf{1}/\sqrt{n}\|=0.82$, so that the estimated basis vector $\widehat{u}_1$ is close to equal weighting $\mathbf{1}/\sqrt{n}$.

This orientation gives $\widehat v_1$ a direct migration interpretation. Let 
$$\tilde{F}_{\ell}=\sum_{k\ne \ell}\frac{F_{k\ell}-F_{\ell k}}{F_{\ell k}+F_{k\ell}},$$ 
using the same zero convention as for $X_n$. That is, $\tilde{F}_{\ell}$ is an unweighted sum
of inward bilateral effectiveness ratios. Since $\tilde{F}=-X_n\mathbf{1}$, its fitted counterpart is
\[
 \widehat F=-X_n\widehat{u}_1=\lambda_1\widehat v_1.
\]
Consequently, positive coordinates of \(\widehat v_1\) correspond to positive weighted inward-imbalance scores, while negative coordinates correspond to negative scores. The weights are the coordinates of \(\widehat u_1\), the unit vector in the leading plane closest to the constant direction.

\begin{figure}[t]
 \centering
 \includegraphics[width=0.8\linewidth]{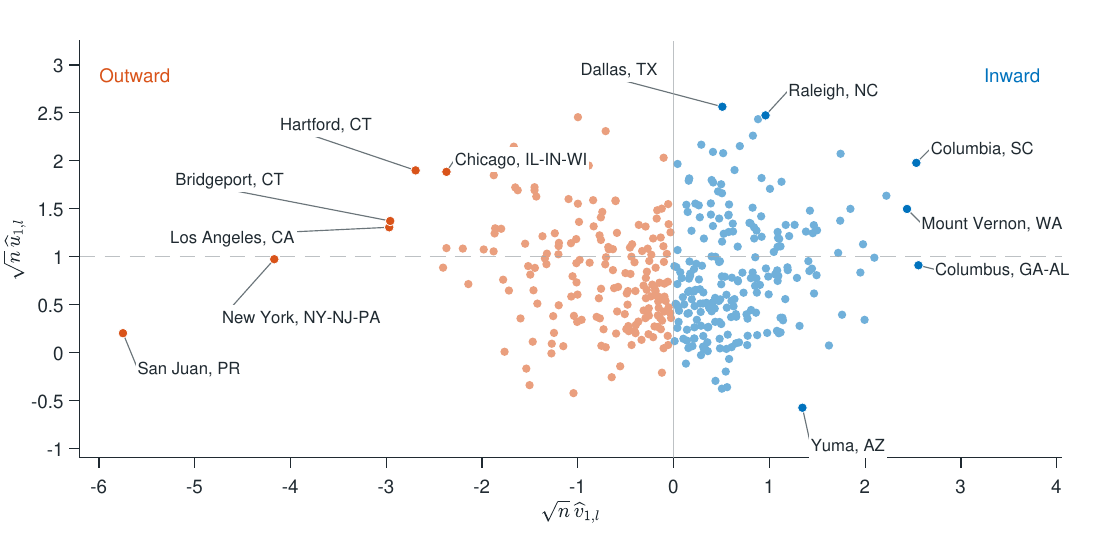}
 \caption{Leading eigenplane: Each point is
 a metropolitan area, plotted at
 $(\sqrt n\,\widehat v_{1,\ell},\sqrt n\,\widehat u_{1,\ell})$.
 The horizontal coordinate orders the fitted inward score.}
 \label{fig:migration_basis}
\end{figure}

Figure~\ref{fig:migration_basis} shows both coordinates of the oriented basis, scaled by $\sqrt n$. Under this scaling, the uniform unit vector $n^{-1/2}\mathbf{1}$ has all coordinates equal to one, as indicated by the horizontal reference line. The horizontal axis shows the signed loading $\sqrt n\widehat v_{1,\ell}$, with positive and negative values corresponding to fitted inward and outward directions, respectively. The vector $\widehat u_1$ is substantially aligned with the uniform direction, although its coordinates exhibit considerable heterogeneity: Dallas and Raleigh have particularly large positive loadings $\widehat u_{1,\ell}$, whereas Yuma has a negative loading. Thus, the ranking by $\widehat v_{1,\ell}$ summarizes the fitted aggregate direction, while the exact sign of each fitted bilateral $(X_n)_{k\ell}$ entry depends on both basis coordinates through
$\lambda_1\left(\widehat u_{1,k}\widehat v_{1,\ell}
-\widehat v_{1,k}\widehat u_{1,\ell}\right).$

\begin{figure}[t]
 \centering
 \includegraphics[width=\linewidth]{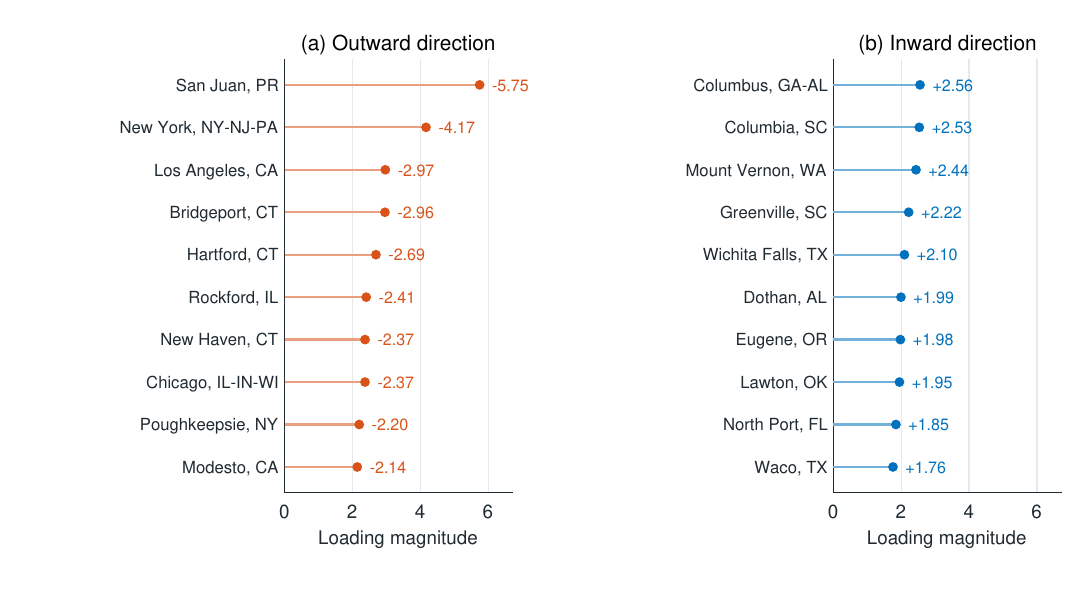}
 \caption{Largest signed loadings of the leading migration direction.
 Each panel reports ten metropolitan areas, ordered by
 $|\sqrt n\,\widehat v_{1,\ell}|$. Horizontal distances use a common scale;
 the printed values retain the signs.}
 \label{fig:migration_tails}
\end{figure}

Figure~\ref{fig:migration_tails} reports the ten largest loadings in each signed direction. San Juan--Bayam\'on--Caguas, PR has the most negative loading,
$\sqrt n\,\widehat v_{1,\ell}=-5.75$, followed by New York--Newark--Jersey City, NY-NJ-PA at $-4.17$ and Los Angeles--Long Beach--Anaheim, CA at $-2.97$. Bridgeport, Hartford, and Chicago also appear in the outward tail. Several major metropolitan areas therefore feature prominently in the fitted outward direction.
The largest positive loadings are Columbus, GA--AL, at $2.56$,
Columbia, SC, at $2.53$, and Mount Vernon--Anacortes, WA, at $2.44$. Most metropolitan areas in the positive tail are in the Southern United States, suggesting a regional pattern in the fitted inward direction.

The second and third eigenplanes, $\widehat{\mathcal V}_2$ and $\widehat{\mathcal V}_3$, have less direct economic interpretations. This may partly reflect their lower estimation accuracy, as suggested by the larger estimated canonical angles in Table~\ref{tab:migration_angles}. Both eigenplanes are geographically dispersed. Medium-sized metros, defined as those with 250,000 to one million residents, account for $50.8\%$ of the coordinate shares of $\widehat{\mathcal V}_2$, compared with their $35.7\%$ share of the metropolitan areas in the sample. Its largest coordinate shares, measured by the basis-invariant quantity $\tfrac1{2}(\widehat u_{2,\ell}^2+\widehat v_{2,\ell}^2)$, occur for Lakeland--Winter Haven, FL (3.54\%), Provo--Orem, UT (2.93\%), Tulsa, OK (1.78\%), Springfield, MO (1.59\%), and San Antonio--New Braunfels, TX (1.51\%). Among the first three, the fitted bilateral directions form the cycle Provo $\to$ Lakeland $\to$ Tulsa $\to$ Provo, which cannot be represented by a single ordering from origin to destination. The observed bilateral effectiveness ratios have the same signs. The gross exchanges on these links are small, illustrating that large proportional imbalances can arise even when the underlying migration volumes are small.

\section{Conclusion}\label{sec:conclusion}

The paper studies spectral estimation in skew-symmetric
factor models. We characterize the limiting eigenvalue
locations and, under Gaussian noise, derive their asymptotic fluctuations. Surprisingly, although the observed data is real, the limiting eigenvalue fluctuations match those of complex Hermitian models. 
We also quantify the accuracy of eigenplane estimation through canonical correlations. For each simple supercritical
strength, both squared canonical correlations have the same
limit below one. Thus, at a fixed signal-to-noise ratio,
the strength can be consistently estimated, while the
estimated plane retains a nonvanishing error.

Applications motivate extensions allowing non-Gaussian noise, unequal variances, and dependence among the upper-triangular entries, as in approximate factor models \citep{bai2008large}. A related problem is then to estimate the noise covariance under suitable structural assumptions, cf.\ covariance thresholding of \citet{fan2013large}.

Extending the fluctuation results beyond Gaussian noise could clarify the role of sparse and dense signal directions. It would also be useful to study whether estimators that exploit sparsity of the factor planes can improve on ordinary spectral estimates, as in sparse principal subspace estimation \citep{cai2013sparse}.

The critical regime also remains to be studied. Deriving the limiting distributions for signal strengths at or approaching $\sigma$ would describe the transition between edge and outlier fluctuations. This would allow uniform inference that remains valid near the separation threshold, as in \citet{bykhovskaya2025weak}.

When all signal strengths are below $\sigma$, the largest eigenvalue has the same limiting fluctuations as under pure noise. Linear spectral statistics may nevertheless contain information about these signals, as they do in classical spiked models \citep{johnstone2020testing}. Comparing the power of tests based on different spectral statistics is therefore a natural next step for skew-symmetric data analysis.

\newpage
\appendix

\section{Preliminaries and general estimates}
\label{app:general_preliminaries}

The appendix collects the preliminary results and proofs. Section~\ref{app:general_preliminaries} introduces the notation and general estimates. Sections~\ref{app:first_order}, \ref{app:fluctuations}, and \ref{app:proof_cca} establish, respectively, the first-order eigenvalue limits, the fluctuation results, and factor recovery. The assumptions are specified locally. The proofs of Theorems~\ref{thm:general_random_skew_outliers} and~\ref{thm:canonical_angles_supercritical} first treat the all-moments case and then extend the conclusions by truncation. Whenever a proof is carried out with $\sigma=1$, the result for general $\sigma>0$ follows by applying the normalized statement to $W_n/\sigma$ and $S_n/\sigma$, whose signal strengths are $\theta_j/\sigma$.

\subsection{Notation and algebraic preliminaries}
\label{app:preliminaries}

We supplement the notation introduced in the main text. Let ${\rm O}(n)$ denote the orthogonal group of $\R^n$. Let $\C^+=\{z\in\C:\Im(z)>0\}$. For a square complex matrix $M$, we write $\Re M=(M+M^*)/2$ for its Hermitian part and $\spec(M)$ for its spectrum. For a rectangular matrix $M$, $\|M\|$ denotes its spectral norm. We write $\Pi_1(M)$ for the orthogonal projection onto the eigenspace associated with the largest eigenvalue of a Hermitian matrix $M$.

The resolvent of a Hermitian matrix $M$ is $(M-zI)^{-1}$, defined for $z\notin\spec(M)$. If $M$ has size $n$ and eigenvalues $\lambda_1(M)\geq\cdots\geq\lambda_n(M)$, its empirical spectral measure is $n^{-1}\sum_{k=1}^n\delta_{\lambda_k(M)}$ and its Stieltjes transform is $n^{-1}\Tr(M-zI)^{-1}$ for $z\in\C^+$. We use the same formula wherever the resolvent is defined.

Let $\mu_{\rm sc}(dx)=\rho_{\rm sc}(x)\,dx$ denote the variance-one semicircle law, where
$$
\rho_{\rm sc}(x)
=\frac{1}{2\pi}\sqrt{4-x^2}\,\mathbf{1}_{[-2,2]}(x),
\qquad
m_{\rm sc}(z)
=\int\frac{\mu_{\rm sc}(dx)}{x-z},
\quad z\in\C\setminus[-2,2].
$$
For deterministic probability measures, $\nu_n\Rightarrow\nu$ denotes weak convergence. A standard complex Gaussian is a random variable $\zeta=(X+\ii Y)/\sqrt2$, where $X,Y\sim\mathcal N(0,1)$ are independent. Finally, $C_b(\R)$ denotes the bounded continuous functions on $\R$.

For the matrices $W_n=\ii\mathcal{E}_n/\sqrt n$ and $H_n=W_n+\ii S_n$ defined in Section~\ref{sec:setting}, we use the eigenvalue notation $\xi_1\geq\cdots\geq\xi_n$ and $\lambda_1\geq\cdots\geq\lambda_n$, respectively. Their resolvents, Stieltjes transforms, and empirical spectral measures are denoted by $G_n,m_n,\mu_n$ and $G_n^{H},m_n^{H},\mu_n^{H}$, respectively, as summarized in Table~\ref{tab:notation-Wn-Wtn}.

\begin{table}[ht]
\centering
\renewcommand{\arraystretch}{1.45}
\fbox{%
\begin{tabular}{c|c|c}
& $W_n$ 
& $H_n$ \\
\hline
eigenvalues 
& $\xi_1 \geq \cdots \geq \xi_n$
& $\lambda_1 \geq \cdots \geq \lambda_n$ \\

resolvent 
& $G_n(z) = (W_n - zI)^{-1}$
& $G_n^{H}(z) = (H_n - zI)^{-1}$ \\

Stieltjes transform 
& $m_n(z) = n^{-1}\operatorname{Tr}G_n(z)$
& $m_n^{H}(z) = n^{-1}\operatorname{Tr}G_n^{H}(z)$ \\

empirical measure 
& $\mu_n$
& $\mu_n^{H}$
\end{tabular}%
}
\caption{Notation for $W_n$ and $H_n$.}
\label{tab:notation-Wn-Wtn}
\end{table}

For the appendix, set \[ m=\#\{j\in[r]:\theta_j>\sigma\}. \] For each $j\in[m]$, write
\be
\la{eq:outlier_roots}
\rho_j=\theta_j+\frac{\sigma^2}{\theta_j}.
\ee
Thus $m$ counts the positive supercritical signal strengths, including multiplicities, and $\rho_j>2\sigma$ is the corresponding limiting positive outlier location. In proofs normalized to $\sigma=1$, this notation becomes $\rho_j=\theta_j+\theta_j^{-1}$.

Using the vectors $w_{j,+}$ and $w_{j,-}$ defined in Section~\ref{sec:setting}, set
$$
U_n=
\bigl[
w_{1,+}\ \cdots\ w_{r,+}\ w_{1,-}\ \cdots\ w_{r,-}
\bigr]\in\C^{n\times2r}
$$
and
$$
\Theta=
\diag(\theta_1,\ldots,\theta_r,
-\theta_1,\ldots,-\theta_r).
$$
Then $U_n^*U_n=I_{2r}$ and
\be\la{eq:finite_rank_representation}
\ii S_n=U_n\Theta U_n^*,
\qquad
H_n=W_n+U_n\Theta U_n^*.
\ee
The following determinant criterion reduces the study of the eigenvalues of $H_n$ outside $\spec(W_n)$ to a determinant equation of size $2r$.

\blem\la{lem:general_det_criterion}
Fix $a\in \C\setminus \spec(W_n)$ and define
$$
M_n(a)=U_n^*G_n(a)U_n\in \C^{2r\times 2r}.
$$
Then
\be\la{eq:general_det_criterion}
a\in \spec(H_n)
\iff
\det\bigl(I_{2r}+\Theta M_n(a)\bigr)=0.
\ee
\elem

\noindent
{\it Proof.}
Let $a\notin \spec(W_n)$. In particular, $W_n-aI$ is invertible. By the matrix determinant lemma,
\begin{align}
\det(H_n-aI)
&=
\det\bigl(W_n+U_n\Theta U_n^*-aI\bigr) \nonumber \\
&=
\det(W_n-aI)\det\bigl(I_{2r}+\Theta U_n^*(W_n-aI)^{-1}U_n\bigr). \la{eq:det_is_real}
\end{align}
Since $\det(W_n-aI)\neq 0$, we obtain \eqref{eq:general_det_criterion}. $\hfill\Box$

\subsection{Standard deterministic and probabilistic tools}
\label{app:standard_tools}

We collect the standard deterministic and probabilistic tools used below. Each result below is stated with its own assumptions.

\blem\la{lem:min-max}
Let $M$ be an $n \times n$ Hermitian matrix with spectrum $\lambda_n(M) \leq \ldots \leq \lambda_1(M)$. Then, for $j=0,\ldots,n-1$,
$$\lambda_{j+1}(M)=\min_{\dim S=n-j}
\ \max_{\substack{x\in S\\ \|x\|=1}}
x^*Mx,$$ and 
$$\lambda_{n-j}(M)=\max_{\dim S=n-j}
\min_{\substack{x\in S\\ \|x\|=1}}
x^*Mx.$$
\elem

\blem[Theorem A.43 in \citealp{bai-silver-book}] \la{lem:rank_ineq}
    Let $Q$ and $R$ be two $n \times n$ Hermitian matrices. Then,
    \begin{equation}
        \|F^Q - F^R\|_{\infty} \leq \frac{1}{n} \mathrm{rank}(Q -R),
    \end{equation} where $F^Q$ and $F^R$ denote the distribution functions of the empirical spectral measures of $Q$ and $R$, respectively. 
\elem

The following notion of a high-probability bound, introduced in a slightly different form by \citet{erdHos2013averaging}, has since been used in a number of works on random matrix theory.

\begin{Def}[Stochastic Domination]\la{def:stochastic_dom}
Let
$$
X = \left(X^{(N)}(u) : N\in\mathbb N,\ u\in U^{(N)}\right),
\qquad
Y = \left(Y^{(N)}(u) : N\in\mathbb N,\ u\in U^{(N)}\right),
$$
be two families of nonnegative random variables, where $U^{(N)}$ is a possibly $N$-dependent parameter set. We write $X\prec Y$ and say that $X$ is stochastically dominated by $Y$, uniformly in $u$, if for all small $\varepsilon>0$ and large $D>0$ we have
$$
\sup_{u\in U^{(N)}}
\P\left[
X^{(N)}(u) > N^\varepsilon Y^{(N)}(u)
\right]
\leq N^{-D}
$$
for large enough $N\geq N_0(\varepsilon,D)$.
\end{Def}

\bDef[Tracy--Widom distribution]\la{def:tracy_widom}
Let $\operatorname{Ai}(\cdot)$ be the Airy function defined by
$$
\operatorname{Ai}(x) =
\frac{1}{2\pi}
\int_{-\infty}^{\infty}
e^{\ii(t+\ii c)^3/3+\ii x(t+\ii c)}\,dt,
$$
where $c>0$ is arbitrary. Consider the Airy kernel
$$
K_{\mathrm{Ai}}(x,y) =
\frac{
\operatorname{Ai}(x)\operatorname{Ai}'(y) -
\operatorname{Ai}'(x)\operatorname{Ai}(y)
}{x-y},
$$
with the diagonal defined by continuity. For $s\in\mathbb R$, define the Fredholm determinant
\begin{align}
F_2(s)
&=
\det\!\left(I-K_{\mathrm{Ai}}\right)_{L^2([s,\infty))} \nonumber\\
&=
\sum_{k=0}^{\infty}
\frac{(-1)^k}{k!}
\int_{[s,\infty)^k}
\det\!\left[
K_{\mathrm{Ai}}(x_i,x_j)
\right]_{i,j=1}^k
\,dx_1\cdots dx_k.
\nonumber
\end{align}
Then $F_2$ is a distribution function, called the Tracy--Widom distribution function for $\beta=2$. We denote by $\TW_2$ a random variable with distribution function $F_2$.
\eDef

\subsection{General resolvent estimates}
\la{app:compressed_resolvent}

We work in the normalized case $\sigma=1$ and establish the spectral and resolvent estimates used below. The following auxiliary assumption strengthens Assumption~\ref{ass:ass1} and imposes finite moments of all orders on the standardized noise entries. It applies to the bounded-entry approximations used in the first-order proofs and to the Gaussian noise used in the fluctuation proofs.

\bAss\la{ass:ass3}
Let $\nu$ be a probability distribution on $\R$, independent of $n$, with mean zero, variance one, and finite moments of all orders. For each $n\geq1$, let $\mathcal E_n\in\R^{n\times n}$ be a real skew-symmetric matrix whose entries strictly above the diagonal are i.i.d.\ with distribution $\nu$.
\eAss

Throughout this subsection, Assumption \ref{ass:ass3} holds. Whenever the perturbation $S_n$, or equivalently $U_n$ and $\Theta$, appears, we additionally assume Assumption \ref{ass:ass2}.

\blem\la{lem:stielt-cv}
Suppose Assumption \ref{ass:ass3} holds. Then $\mu_n\Rightarrow\mu_{\rm sc}$ almost surely. Moreover,
$$
\begin{array}{ll}
\text{(i)}
& \xi_1\xrightarrow{a.s.}2
  \qquad\text{and}\qquad
  \xi_n\xrightarrow{a.s.}-2, \\[2mm]
\text{(ii)}
& \forall x\in\C\setminus[-2,2],\qquad
  m_n(x)\xrightarrow{a.s.}m_{\rm sc}(x).
\end{array}
$$
If Assumption \ref{ass:ass2} also holds, then $\mu_n^{H}\Rightarrow\mu_{\rm sc}$ almost surely.
\elem
\bpr
By \citet[Theorem 2.5]{bai-silver-book}, $\mu_n\Rightarrow\mu_{\rm sc}$ almost surely. Under Assumption \ref{ass:ass2}, the same conclusion holds for $\mu_n^{H}$ by Lemma \ref{lem:rank_ineq}, since $\rank(H_n-W_n)\leq 2r$. Part (i) follows from \citet[Theorem 5.2]{bai-silver-book}. To prove part (ii), consider the probability-one event
$$
\Omega=
\Bigl\{\mu_n\Rightarrow \mu_{\rm sc}\Bigr\}
\cap
\Bigl\{\xi_1\to 2\Bigr\}
\cap
\Bigl\{\xi_n\to -2\Bigr\}.
$$ 
Fix $\omega\in\Omega$ and $x \in \C\setminus[-2,2]$. Let $\delta=\dist\bigl(x,[-2,2]\bigr)>0,$ choose $0<\varepsilon<\delta/2$ and define $I_\varepsilon=[-2-\varepsilon,\,2+\varepsilon]$. Since result (i) holds, there exists $N_0(\omega)$ such that for all $n\ge N_0(\omega)$, $\spec(W_n(\omega))\subset I_\varepsilon.$
Equivalently, $\supp(\mu_n(\omega))\subset I_\varepsilon$ for all $n\ge N_0(\omega).$ Choose $f_x \in C_b(\R)$ satisfying $f_x(t)=(t-x)^{-1}$ for $t \in I_\varepsilon.$ Then, for all $n\ge N_0(\omega)$,
$$
\int_{\R}\frac{1}{t-x}\,\mu_n(\omega)(dt)
=
\int_{I_\varepsilon} f_x(t)\,\mu_n(\omega)(dt).
$$
Combining $\supp(\mu_{\rm sc})=[-2,2]\subset I_\varepsilon$ and $\mu_n(\omega)\Rightarrow \mu_{\rm sc}$ yields
$$
\int_{I_\varepsilon} f_x(t)\,\mu_n(\omega)(dt)
\longrightarrow
\int_{I_\varepsilon} f_x(t)\,\mu_{\rm sc}(dt).
$$
Hence,
$$
m_n(x)
=
\int_{\R}\frac{1}{t-x}\,\mu_n(dt)
\longrightarrow
\int_{\R}\frac{1}{t-x}\,\mu_{\rm sc}(dt)
=
m_{\rm sc}(x)
\qquad \text{a.s.}
$$
\epr

\begin{propo}\la{prop:isotrop}
Suppose Assumption \ref{ass:ass3} holds. Let $v,w\in\C^n$ be deterministic vectors such that $||v||_2=||w||_2=1$ and fix $x\in\C\setminus[-2,2]$. Then,
\be\la{eq:isotrop_law_propo}
w^* G_n(x)v - m_n(x)\,w^* v \xrightarrow[n\to\infty]{\mathrm{a.s.}} 0.
\ee
\end{propo}

\begin{proof}
Fix $x\in\C\setminus[-2,2],\, L >|x|+1$ and $\tau \in (0,1).$ Let $$W_n^{\mathrm{std}} = \ii\mathcal E_n/\sqrt{n-1}, \qquad G_n^{\mathrm{std}}(z)=(W_n^{\mathrm{std}}-zI)^{-1}.$$ The matrix $W_n^{\mathrm{std}}$ satisfies the assumptions of the general class of random matrices introduced by \citet{erdos2013local}. Combining \citet[Theorem 2.3]{erdos2013local} with \citet[Theorem 2.14]{erdos-cov}, for any deterministic unit vectors $v,w\in\C^n$ and uniformly on $\mathcal{D}_n=\mathcal{D}_n(\tau,L)=\{z=E+\ii\eta: |E|\leq L, n^{-1+\tau}\leq\eta\leq L\}$, the following stochastic-domination estimate holds (see Definition \ref{def:stochastic_dom}):
\be
\la{eq:isotropic_vector_bound_sd}
    \left| w^*
        \left(
            G_n^{\mathrm{std}}(z)-m_{\rm sc}(z)I
        \right) v\right|
    \prec
    \sqrt{\frac{\Im m_{\rm sc}(z)}{n\eta}} +
    \frac1{n\eta} \underset{n\to\infty}{\longrightarrow} 0.
\ee
Since $m_{\rm sc}$ is uniformly bounded on $\mathcal D_n$, we have
$$
\sup_{z\in\mathcal D_n}
\left(
\sqrt{\frac{\Im m_{\rm sc}(z)}{n\eta}}
+\frac{1}{n\eta}
\right)
=O(n^{-\tau/2}).
$$
By the lattice--Lipschitz argument of \citet[Remark 2.6]{erdos-cov}, the preceding uniform stochastic domination estimate can be strengthened to hold simultaneously for all $z\in\mathcal D_n$; namely, for every $\varepsilon,D>0$ and all sufficiently large $n$,
$$
\P\left(
\sup_{z\in\mathcal D_n}
\left|w^*\left(G_n^{\rm std}(z)-m_{\rm sc}(z)I\right)v\right|
> n^{\varepsilon-\tau/2}
\right)
\le n^{-D}.
$$
Choosing $\varepsilon=\tau/4$ and $D>1$, the right-hand side is summable in $n$. Hence, by the Borel--Cantelli lemma,
\be\la{eq:stochastic_domination_as}
\sup_{z\in\mathcal D_n}
\left|w^*\left(G_n^{\rm std}(z)-m_{\rm sc}(z)I\right)v\right|
\longrightarrow 0
\qquad\text{a.s.}
\ee

We now prove \eqref{eq:isotrop_law_propo} using this estimate, beginning with the real case. Fix $x\in\R\setminus[-2,2]$ and $\alpha\in(0,1-\tau)$, and set $z_n=x+\ii\eta_n$ with $\eta_n=n^{-\alpha}$. Denote $d=\mathrm{dist}(x, [-2,2])>0.$ By Lemma \ref{lem:stielt-cv}, almost surely, for all large $n$, $\mathrm{dist}(x, \spec(W_n^{\rm std}))\ge\frac{d}{2}.$ Since the spectrum is real and $\Re (z_n)=x$, we obtain $||G_n^{\rm std}(z_n)||\le ||G_n^{\rm std}(x)||\le \frac{2}{d}$ for all large $n$. The resolvent inequality gives 
\be\la{eq:ineq_resolv_ti}
||G_n^{\rm std}(x)-G_n^{\rm std}(z_n)|| \le \eta_n||G_n^{\rm std}(x)||\,||G_n^{\rm std}(z_n)||\le \frac{4\eta_n}{d^2}\longrightarrow0 \quad \text{a.s.}
\ee
Combining \eqref{eq:stochastic_domination_as}, \eqref{eq:ineq_resolv_ti} and the continuity of $m_{\rm sc}$ yields
\be\la{eq:control_errors_1}
w^*(G_n^{\rm std}(x)-m_{\rm sc}(x)I)v \longrightarrow 0 \quad \text{a.s.}
\ee
By Lemma \ref{lem:stielt-cv}, almost surely, $||W_n^{\rm std} - W_n|| = (\sqrt{\frac{n}{n-1}}-1)||W_n|| \to 0$ and, for all large $n$, $\mathrm{dist}(x, \spec(W_n))\ge\frac{d}{2}$. Hence, $||G_n(x)||\le \frac{2}{d}$ for all large $n$. By the resolvent identity,
\be\la{eq:control_errors_2}
\|G_n^{\rm std}(x)-G_n(x)\|
\leq
\frac4{d^2}\|W_n^{\rm std}-W_n\|
\longrightarrow 0
\qquad\text{a.s.}
\ee
Combining \eqref{eq:control_errors_1}, \eqref{eq:control_errors_2} and the fact that $m_n(x)\to m_{\rm sc}(x)$ almost surely by Lemma \ref{lem:stielt-cv} proves the proposition in the real case. 

We now extend to the complex case. First assume that $\Im x>0$. Taking $z_n=x$ in the preceding argument and using the bounds $||G_n^{\rm std}(x)||\le \frac{1}{|\Im x|}$ and $||G_n(x)||\le \frac{1}{|\Im x|}$ completes this case directly. If $\Im x<0$, we apply the previous case to $\overline{x}\in\C^+$ with the vectors $w$ and $v$ interchanged. Since $W_n$ is Hermitian,
$$
G_n(x)^*=G_n(\overline{x}),
\qquad
m_n(\overline{x})=\overline{m_n(x)},
\qquad
m_{\rm sc}(\overline{x})=\overline{m_{\rm sc}(x)}.
$$
Thus,
$$
w^*G_n(x)v-m_n(x)w^*v
=
\overline{
v^*G_n(\overline{x})w
-
m_n(\overline{x})v^*w
}
\longrightarrow 0
\qquad \text{a.s.}
$$
\end{proof}

To apply Lemma \ref{lem:general_det_criterion} asymptotically, we need uniform control of the compressed resolvent $M_n(a)=U_n^*G_n(a)U_n$ away from the bulk. This is provided by the next lemma.

\blem\la{lem:uniform_compressed_resolvent}
Suppose Assumptions \ref{ass:ass3} and \ref{ass:ass2} hold. Let $K\subset \C\setminus[-2,2]$ be compact. Then, almost surely,
\be\la{eq:uniform_Mn_convergence}
\sup_{a\in K}\bigl\|M_n(a)-m_{\textrm sc}(a)I_{2r}\bigr\|\longrightarrow 0.
\ee
Hence, almost surely,
\be\la{eq:uniform_Fn_convergence}
\sup_{a\in K}\Bigl|
\det\bigl(I_{2r}+\Theta M_n(a)\bigr)
-
\det\bigl(I_{2r}+m_{\textrm sc}(a)\Theta\bigr)
\Bigr|
\longrightarrow 0.
\ee
\elem

\begin{proof}
Let $K\subset \C\setminus[-2,2]$ be compact.

Let $D=\{z_\ell:\ell\geq1\}$ be a countable dense subset of $K$. For each fixed $z_\ell\in D$, Proposition \ref{prop:isotrop} applied entrywise gives $M_n(z_\ell)-m_n(z_\ell)I_{2r}\to 0$ almost surely. Moreover, by Lemma \ref{lem:stielt-cv}, $m_n(z_\ell)\to m_{\textrm sc}(z_\ell)$ almost surely. Thus, by a countable intersection argument, there exists a probability-one event on which $M_n(z_\ell)\to m_{\textrm sc}(z_\ell)I_{2r}$ for every $\ell\ge1.$ We now upgrade this pointwise convergence on $D$ to uniform convergence on $K$. Let $ d_K=\mathrm{dist}(K,[-2,2])>0.$ By Lemma \ref{lem:stielt-cv}, almost surely, for all large $n$, $\mathrm{dist}\bigl(K,\spec(W_n)\bigr)\geq \frac{d_K}{2}.$ Therefore, by the resolvent identity, for all large $n$ and all $a,b\in K$,
$$
\|G_n(a)\|\leq \frac{2}{d_K},
\qquad
\|G_n(a)-G_n(b)\|
\leq
|a-b|\,\|G_n(a)\|\,\|G_n(b)\|
\leq
\frac{4}{d_K^2}|a-b|.
$$
Since $U_n$ has orthonormal columns,
$$
\|M_n(a)-M_n(b)\|
\le \|G_n(a)-G_n(b)\|
\leq
\frac{4}{d_K^2}|a-b|.
$$
Thus the family $(M_n)_n$ is uniformly Lipschitz on $K$, for all sufficiently large $n$. Since $m_{\textrm sc}$ is continuous on the compact set $K$, it is uniformly continuous on $K$. Fix $\varepsilon>0$. Choose $\eta>0$ such that, whenever $a,b\in K$ and $|a-b|\leq\eta$, $\|m_{\textrm sc}(a)I_{2r}-m_{\textrm sc}(b)I_{2r}\|\leq \varepsilon.$ Reducing $\eta$ if necessary, we may also assume that $\frac{4}{d_K^2}\eta\leq \varepsilon.$ Since $D$ is dense in $K$ and $K$ is compact, there exist $z_{\ell_1},\ldots,z_{\ell_{N_K}}\in D$ for some integer $N_K\ge1$ such that
$$
K\subset \bigcup_{p=1}^{N_K} B(z_{\ell_p},\eta).
$$
On the probability-one event constructed above, for each $p\in[N_K]$, $M_n(z_{\ell_p})-m_{\textrm sc}(z_{\ell_p})I_{2r}\longrightarrow 0.$ Hence, for all sufficiently large $n$, $\max_{1\leq p\leq N_K}
\bigl\|M_n(z_{\ell_p})-m_{\textrm sc}(z_{\ell_p})I_{2r}\bigr\|
\leq \varepsilon.$ Now let $z\in K$, and choose $p\in[N_K]$ such that $|z-z_{\ell_p}|\leq\eta$. Then, for all sufficiently large $n$,
$$
\begin{aligned}
\bigl\|M_n(z)-m_{\textrm sc}(z)I_{2r}\bigr\|
&\leq
\bigl\|M_n(z)-M_n(z_{\ell_p})\bigr\|
+
\bigl\|M_n(z_{\ell_p})-m_{\textrm sc}(z_{\ell_p})I_{2r}\bigr\| \\
&\hspace{2cm}
+
\bigl\|m_{\textrm sc}(z_{\ell_p})I_{2r}-m_{\textrm sc}(z)I_{2r}\bigr\| \\
&\leq 3\varepsilon.
\end{aligned}
$$
Taking the supremum over $z\in K$ proves \eqref{eq:uniform_Mn_convergence}.

It remains to prove \eqref{eq:uniform_Fn_convergence}. Work on the probability-one event on which \eqref{eq:uniform_Mn_convergence} holds. Since $K$ is compact and $m_{\mathrm{sc}}$ is continuous on $K$, we have $\sup_{a\in K}|m_{\mathrm{sc}}(a)|<\infty$. Thus, uniformly for $a\in K$,
$$
\begin{aligned}
\bigl\|\Theta M_n(a)-m_{\mathrm{sc}}(a)\Theta\bigr\| &\leq \|\Theta\|\,\bigl\|M_n(a)-m_{\mathrm{sc}}(a)I_{2r}\bigr\|\longrightarrow 0.
\end{aligned}
$$
Therefore,
$$
\sup_{a\in K}\bigl\|\bigl(I_{2r}+\Theta M_n(a)\bigr)-\bigl(I_{2r}+m_{\mathrm{sc}}(a)\Theta\bigr)\bigr\|\longrightarrow 0.
$$
 Since the matrices above are uniformly bounded and the dimension $2r$ is fixed, the determinant is Lipschitz on the corresponding bounded set of matrices. Hence there exists $C_K>0$ independent of $n$ such that, for all large $n$ and all $a\in K$,
$$
\Bigl|\det\bigl(I_{2r}+\Theta M_n(a)\bigr)-\det\bigl(I_{2r}+m_{\mathrm{sc}}(a)\Theta\bigr)\Bigr| \leq C_K\bigl\|\Theta M_n(a)-m_{\mathrm{sc}}(a)\Theta\bigr\|.
$$
Taking the supremum over $a\in K$ yields \eqref{eq:uniform_Fn_convergence}.
\end{proof}

The next lemma excludes eigenvalues from compact subsets of $\R\setminus[-2,2]$ on which the limiting determinant does not vanish.

\begin{lem}\la{lem:finite_rank_exclusion}
Suppose Assumptions \ref{ass:ass3} and \ref{ass:ass2} hold. Let $D$ be a fixed real diagonal $2r\times 2r$ matrix and set $H_n(D)=W_n+U_nDU_n^*.$ In particular, $H_n(\Theta)=H_n$. Let $K\subset \R\setminus[-2,2]$ be compact. If $F_D(a)=\det\left(I_{2r}+m_{\rm sc}(a)D\right)$ does not vanish on $K$, then, almost surely, for all large $n$,
$$
\spec(H_n(D))\cap K=\emptyset .
$$
\end{lem}

\begin{proof}
By Lemma \ref{lem:stielt-cv}, almost surely, $K\cap\spec(W_n)=\emptyset$ for all large $n$. Hence, for $a\in K$, Lemma \ref{lem:general_det_criterion} applied with $D$ in place of $\Theta$ implies that $\det\left(I_{2r}+DM_n(a)\right)=0$ whenever $a\in\spec(H_n(D)).$
By Lemma \ref{lem:uniform_compressed_resolvent}, almost surely,
$$
\sup_{a\in K}\left|\det\left(I_{2r}+DM_n(a)\right)-\det\left(I_{2r}+m_{\rm sc}(a)D\right)\right|\longrightarrow0.
$$
Since $F_D$ is continuous and non-vanishing on $K$, $\inf_{a\in K}|F_D(a)|>0$. Therefore the finite-$n$ determinant is non-zero on $K$ for all large $n$, which proves the claim.
\end{proof}

\subsection{Truncation under a finite fourth moment}
\label{app:fourth_moment_truncation}
The following estimate allows us to approximate the noise by bounded
entries while preserving its variance, with an almost-sure control of
the approximation error.
\blem\la{lem:fourth_moment_truncation}
Suppose Assumption~\ref{ass:ass1} holds, and let $\eps$ have the common
distribution of the upper-triangular noise entries. For all sufficiently
large integers $L$, set
$$
\widetilde\eps^{(L)}=\eps\mathbf1_{\{|\eps|\leq L\}}
-\E[\eps\mathbf1_{\{|\eps|\leq L\}}],
\qquad
s_L^2=\E[(\widetilde\eps^{(L)})^2],
\qquad
\eps^{(L)}=\frac{\sigma}{s_L}\widetilde\eps^{(L)},
$$
with $s_L>0$.
Apply this transformation to the entire array
$\{\eps_{k\ell}:1\leq k<\ell<\infty\}$ to define
$\mathcal E_n^{(L)}$, extending by skew-symmetry, and set $W_n^{(L)}=\ii\mathcal E_n^{(L)}/\sqrt n$.
Then the entries of $\mathcal E_n^{(L)}$ strictly above the diagonal
are i.i.d., bounded, centered, and have variance $\sigma^2$.
Moreover, setting
$$
\tau_L=\|\eps-\eps^{(L)}\|_{L^2},
$$
we have $\tau_L\to0$ as $L\to\infty$, and, for each fixed
sufficiently large integer $L$,
\be\la{eq:fourth_moment_truncation_norm}
\|W_n-W_n^{(L)}\|
\xrightarrow[n\to\infty]{\mathrm{a.s.}}2\tau_L.
\ee
These limits hold simultaneously for all such $L$ on an event
of probability one.
\elem
\begin{proof}
Since $\E\eps=0$,
$$
\|\widetilde\eps^{(L)}-\eps\|_{L^4}
\leq2\|\eps\mathbf1_{\{|\eps|>L\}}\|_{L^4}
\longrightarrow0.
$$
Consequently, $s_L\to\sigma>0$ and $\eps^{(L)}\to\eps$ in $L^4$,
which implies $\tau_L\to0$.
For each fixed $L$, the Hermitian matrix
$\ii(\mathcal E_n-\mathcal E_n^{(L)})$ has zero diagonal and
upper-triangular entries
$\ii(\eps_{k\ell}-\eps_{k\ell}^{(L)})$ drawn from one infinite
i.i.d.\ array. These entries are centered, have second absolute
moment $\tau_L^2$, and have a finite fourth moment.
If $\tau_L>0$, the complex Hermitian version of the Bai--Yin theorem,
\citet[Theorem~2.12]{bai1999methodologies}, gives
$$
\lambda_1(W_n-W_n^{(L)})
\xrightarrow[n\to\infty]{\mathrm{a.s.}}2\tau_L.
$$
Since $W_n-W_n^{(L)}$ is $\ii$ times a real skew-symmetric matrix,
its spectrum is symmetric about zero and its norm equals its
largest eigenvalue. This proves
\eqref{eq:fourth_moment_truncation_norm}.
If $\tau_L=0$, the difference matrices vanish almost surely for
every $n$, and the conclusion is immediate.
Finally, the cutoffs form a countable set, so the convergence holds
simultaneously for all sufficiently large integers $L$ on an event
of probability one.
\end{proof}

\section{First-order eigenvalue asymptotics}
\la{app:first_order}

Throughout this section, we impose Assumptions \ref{ass:ass1} and \ref{ass:ass2}. We first prove Theorem \ref{thm:general_random_skew_outliers} under the additional all-moments assumption, using Rouch\'e's theorem and the exclusion lemma, and then extend it by truncation. We next establish the consistency, convergence rate, and asymptotic unbiasedness of the noise variance estimator in Corollary \ref{cor:variance_estimation}.

\subsection{Outlier locations: proof of Theorem \ref{thm:general_random_skew_outliers}}
\label{app:proof_outliers}

We give the proof in the normalized case $\sigma=1$. We first assume additionally that Assumption~\ref{ass:ass3} holds. Throughout this first part of the proof, whenever a limiting argument involves an auxiliary deterministic parameter tending to zero, we implicitly restrict it to a fixed countable sequence, so that all almost-sure statements may be taken on a common probability-one event. Set
$$
F(z)=\det\left(I_{2r}+m_{\rm sc}(z)\Theta\right)
=
\prod_{q=1}^r
\left(1+\theta_qm_{\rm sc}(z)\right)
\left(1-\theta_qm_{\rm sc}(z)\right),
\qquad z\in\C\setminus[-2,2],
$$
and, for $z\notin\spec(W_n)$, define
$$
F_n(z)=\det\left(I_{2r}+\Theta M_n(z)\right).
$$
Let $\vartheta_1>\cdots>\vartheta_s>1$ be the distinct supercritical strengths, with multiplicities $k_1,\ldots,k_s$, and set
$$
p_\ell=\sum_{h<\ell}k_h,
\qquad
\rho_\ell^\star=\vartheta_\ell+\vartheta_\ell^{-1},
\qquad \ell\in[s].
$$
Thus $m=\sum_{\ell=1}^s k_\ell$ and
$\rho_{p_\ell+1}=\cdots=\rho_{p_\ell+k_\ell}=\rho_\ell^\star$.

On $(2,\infty)$, $m_{\rm sc}$ is continuous and strictly increasing, with range $(-1,0)$. Thus the only zeros of $F$ on $(2,\infty)$ are $\rho_1^\star,\ldots,\rho_s^\star$. For each $\ell\in[s]$, we have
$$
m_{\rm sc}(\rho_\ell^\star)=-\vartheta_\ell^{-1},
\qquad
m_{\rm sc}'(\rho_\ell^\star)
=\frac{1}{\vartheta_\ell^2-1}>0.
$$
Hence $1+\vartheta_\ell m_{\rm sc}(z)$ has a simple zero at
$\rho_\ell^\star$. This factor occurs exactly $k_\ell$ times in
the product defining $F$, while all other factors are nonzero
there. Therefore $F$ has a zero of order exactly $k_\ell$ at
$\rho_\ell^\star$.

Fix $L>2+\theta_1$. Since
$\|H_n\|\leq\|W_n\|+\theta_1$ and $\|W_n\|\to2$ almost surely
by Lemma \ref{lem:stielt-cv}, the spectrum of $H_n$ is contained
in $[-L,L]$ for all sufficiently large $n$, almost surely.
If $s\geq1$, choose $\eta>0$ so small that the closed disks
$$
\overline{D_\ell}
=
\{z\in\C:|z-\rho_\ell^\star|\leq\eta\},
\qquad \ell\in[s],
$$
are pairwise disjoint and contained in
$\{z\in\C:2<\Re z<L\}$, and each contains no zero of $F$ other
than $\rho_\ell^\star$. Write
$$
I_\ell=D_\ell\cap\R
=(\rho_\ell^\star-\eta,\rho_\ell^\star+\eta).
$$

We now count the eigenvalues in each $I_\ell$. By
Lemma \ref{lem:stielt-cv}, almost surely, for all sufficiently
large $n$, the disks $\overline{D_\ell}$ are disjoint from
$\spec(W_n)$. Thus $F_n$ is holomorphic on a neighborhood of
each $\overline{D_\ell}$. By
Lemma \ref{lem:uniform_compressed_resolvent},
$$
\sup_{z\in\partial D_\ell}|F_n(z)-F(z)|
\longrightarrow0
\qquad\text{almost surely}.
$$
Since $F$ does not vanish on $\partial D_\ell$, almost surely,
for all sufficiently large $n$,
$$
\sup_{z\in\partial D_\ell}|F_n(z)-F(z)|
<
\min_{z\in\partial D_\ell}|F(z)|.
$$
Rouch\'e's theorem therefore implies that $F_n$ has exactly
$k_\ell$ zeros in $D_\ell$, counted with multiplicity.
Moreover, the determinant identity \eqref{eq:det_is_real} gives
$$
F_n(z)=\frac{\det(H_n-zI)}{\det(W_n-zI)}.
$$
The denominator does not vanish on $\overline{D_\ell}$, so
these zeros are precisely the eigenvalues of $H_n$ in
$D_\ell$, with their algebraic multiplicities. Since $H_n$
is Hermitian, these eigenvalues are real. Consequently,
$I_\ell$ contains exactly $k_\ell$ eigenvalues of $H_n$,
counted with multiplicity.

We next exclude any other eigenvalues above the bulk. If $m=0$, we take $s=0$ and interpret the unions below as empty.
Fix $\varepsilon>0$ small enough that all the intervals
$I_\ell$, $\ell\in[s]$, lie above $2+\varepsilon$, and set
$$
K_\varepsilon
=
[2+\varepsilon,L]\setminus\bigcup_{\ell=1}^s I_\ell.
$$
The function $F$ has no zero on $K_\varepsilon$. Hence, by
Lemma \ref{lem:finite_rank_exclusion}, $H_n$ has no eigenvalue
in $K_\varepsilon$ for all large $n$, almost surely.
Since the intervals $I_1,\ldots,I_s$ are ordered from right to
left and contain $k_1,\ldots,k_s$ eigenvalues, respectively,
we obtain
$$
\lambda_{p_\ell+a}\in I_\ell,
\qquad \ell\in[s],\quad a\in[k_\ell],
\qquad
\lambda_{m+1}\leq2+\varepsilon
$$
for all large $n$. Since $\eta$ is arbitrary, this gives
$\lambda_j\to\rho_j$ almost surely for every $j\in[m]$.

It remains to prove that the next eigenvalues stick to the
edge. The preceding upper bound holds for every sufficiently
small $\varepsilon>0$. On the other hand, Lemma \ref{lem:stielt-cv} guarantees that
$\mu_n^{H}\Rightarrow\mu_{\rm sc}$ almost surely.
Since $\mu_{\rm sc}((2-\varepsilon,\infty))>0$, for every
fixed $k\geq1$, almost surely, for all large $n$, $H_n$ has
at least $m+k$ eigenvalues in $(2-\varepsilon,\infty)$.
Thus
$$
2-\varepsilon
\leq\lambda_{m+k}
\leq\lambda_{m+1}
\leq2+\varepsilon.
$$
Letting $\varepsilon\downarrow0$, we obtain
$\lambda_{m+k}\to2$ almost surely for every fixed $k\geq1$. This proves the theorem under Assumption~\ref{ass:ass3}.

We now retain only Assumption~\ref{ass:ass1}. Use the truncation from Lemma~\ref{lem:fourth_moment_truncation}, and write
$$
H_n^{(L)}=W_n^{(L)}+\ii S_n,
\qquad
d_{n,L}=\|H_n-H_n^{(L)}\|
=\|W_n-W_n^{(L)}\|.
$$
For each fixed sufficiently large integer $L$, the truncated noise
satisfies Assumption~\ref{ass:ass3}. The preceding part of the proof
therefore gives the required almost-sure limits for the ordered
eigenvalues $\lambda_1^{(L)}\geq\cdots\geq\lambda_n^{(L)}$
of $H_n^{(L)}$.
These limits are independent of $L$, since the truncation preserves
the noise variance and leaves the signal unchanged.
Since the cutoffs and fixed eigenvalue indices form countable sets,
we may work on one event of probability one on which all these
limits and \eqref{eq:fourth_moment_truncation_norm} hold.
For any index $q$ occurring in \eqref{eq:first_order_limss}, denote
its stated limit by $\ell_q$. Weyl’s inequality gives
$$
|\lambda_q-\ell_q|
\leq |\lambda_q-\lambda_q^{(L)}|
+|\lambda_q^{(L)}-\ell_q|
\leq d_{n,L}+|\lambda_q^{(L)}-\ell_q|.
$$
Applying Lemma \ref{lem:fourth_moment_truncation} and taking the limit superior as $n\to\infty$ with $L$ fixed gives
$$
\limsup_{n\to\infty}|\lambda_q-\ell_q|\leq2\tau_L.
$$
Letting $L\to\infty$ through integers proves the almost-sure
convergence. This completes the proof. $\hfill\Box$

\subsection{Noise variance estimation: proof of Corollary \ref{cor:variance_estimation}}
\label{app:proof_variance_estimation}

We work here with a general $\sigma>0$. Set
$$
V_n=\frac1n\Tr(H_n^2)
=\frac1n\sum_{j=1}^n\lambda_j^2.
$$
Since $H_n=\ii(\mathcal E_n/\sqrt n+S_n)$,
$$
V_n=\frac2n\sum_{1\leq a<b\leq n}
\left(\frac{\eps_{ab}}{\sqrt n}+(S_n)_{ab}\right)^2.
$$
The entries $\eps_{ab}$ are independent, centered, have variance
$\sigma^2$ and a finite fourth moment, and $S_n$ is deterministic.
Moreover,
$$
\sum_{a<b}(S_n)_{ab}^2
=\frac12\Tr((\ii S_n)^2)
=\sum_{j=1}^r\theta_j^2.
$$
It follows that
\be\la{eq:expect_sigma_hat}
\E V_n
=\frac{n-1}{n}\sigma^2
+\frac2n\sum_{j=1}^r\theta_j^2
\longrightarrow\sigma^2.
\ee
Using independence and the inequality
$\operatorname{Var}(X+Y)\leq 2\operatorname{Var}(X)+2\operatorname{Var}(Y)$ gives
$$
\begin{aligned}
\operatorname{Var}(V_n)
&=\frac{4}{n^2}\sum_{a<b}
\operatorname{Var}\left(
\frac{\eps_{ab}^2}{n}
+\frac{2(S_n)_{ab}\eps_{ab}}{\sqrt n}
\right)\\
&\leq
\frac{8}{n^4}\sum_{a<b}\left(\E[\eps_{12}^4]-\sigma^4\right)
+\frac{32\sigma^2}{n^3}\sum_{a<b}(S_n)_{ab}^2\\
&=
\frac{4(n-1)}{n^3}\left(\E[\eps_{12}^4]-\sigma^4\right)
+\frac{32\sigma^2}{n^3}\sum_{j=1}^r\theta_j^2
=O(n^{-2}).
\end{aligned}
$$
Chebyshev’s inequality and \eqref{eq:expect_sigma_hat} give
\be\la{eq:bigO_bound}
V_n-\sigma^2
=(V_n-\E V_n)+(\E V_n-\sigma^2)
=O_\P(n^{-1}).
\ee
Moreover, for every $t>0$,
$$
\sum_n\P(|V_n-\E V_n|>t)
\leq\frac1{t^2}\sum_n\operatorname{Var}(V_n)<\infty.
$$
The Borel--Cantelli lemma, applied to $t=1,1/2,\ldots$, and
\eqref{eq:expect_sigma_hat} therefore imply
$V_n\to\sigma^2$ almost surely.
By Theorem~\ref{thm:general_random_skew_outliers},
$\|H_n\|$ is bounded almost surely for all sufficiently large $n$.
Trimming the first and last $k_n$ eigenvalues removes the $2k_n$
largest squared eigenvalues. Therefore,
$$
0\leq\widehat\sigma_n^2\leq V_n,
\qquad
0\leq V_n-\widehat\sigma_n^2
\leq\frac{2k_n}{n}\|H_n\|^2.
$$
Since $k_n/n^{1/3}\to0$ almost surely,
$$
0\leq n^{2/3}(V_n-\widehat\sigma_n^2)
\leq2\frac{k_n}{n^{1/3}}\|H_n\|^2
\xrightarrow{\mathrm{a.s.}}0.
$$
Together with $V_n\to\sigma^2$ almost surely, this proves
$\widehat\sigma_n^2\to\sigma^2$ almost surely. Using \eqref{eq:bigO_bound} gives
$$
n^{2/3}(\widehat\sigma_n^2-\sigma^2)
=n^{2/3}(V_n-\sigma^2)
-n^{2/3}(V_n-\widehat\sigma_n^2)
\overset{\P}{\longrightarrow}0,
$$
which proves the required rate.
It remains to prove asymptotic unbiasedness. By Fatou’s lemma,
nonnegativity, and \eqref{eq:expect_sigma_hat},$$
\sigma^2
=\E\left[\liminf_{n\to\infty}\widehat\sigma_n^2\right]
\leq\liminf_{n\to\infty}\E\widehat\sigma_n^2
\leq\limsup_{n\to\infty}\E\widehat\sigma_n^2
\leq\lim_{n\to\infty}\E V_n
=\sigma^2.
$$
This proves the expectation convergence and concludes the corollary.
$\hfill\Box$


\section{Eigenvalue fluctuations}
\la{app:fluctuations}

This section is dedicated to the proof of Theorem~\ref{thm:eigenvalue_fluctuations_ext}. We first prove the joint outlier limits in Section~\ref{app:proof_gaussian_fluctuations}, and then establish the soft-edge limits in Section~\ref{app:proof_tw}. Finally, Section~\ref{app:proof_joint_fluctuations} proves their joint convergence and asymptotic independence.

\subsection{Supercritical outlier fluctuations}
\label{app:proof_gaussian_fluctuations}

We prove the outlier cluster limits used in Theorem~\ref{thm:eigenvalue_fluctuations_ext}, allowing critical strengths among the remaining spikes.

\bTh\la{thm:supercritical_gaussian_fluctuations}
Suppose Assumptions~\ref{ass:ass1} and~\ref{ass:ass2} hold with $\eps_{ij}\sim\mathcal N(0,\sigma^2)$, and assume that $m\geq1$. Use the supercritical block parameters
$s,\vartheta_\ell,k_\ell,p_\ell$ defined in
Theorem~\ref{thm:eigenvalue_fluctuations_ext}, and set
\[
\rho_\ell^\star
=\vartheta_\ell+\frac{\sigma^2}{\vartheta_\ell},
\qquad
a_\ell=1-\frac{\sigma^2}{\vartheta_\ell^2},
\qquad \ell\in[s].
\]
Then, jointly over $\ell=1,\ldots,s$,
\[
\bigl(\sqrt n(\lambda_{p_\ell+b}-\rho_\ell^\star)\bigr)_{b=1}^{k_\ell}
\Longrightarrow
\sigma\sqrt{a_\ell}\,(\gamma_{\ell,b})_{b=1}^{k_\ell},
\]
where $(\gamma_{\ell,b})_{b=1}^{k_\ell}$ is the decreasingly ordered
eigenvalue vector of an independent standard $\mathrm{GUE}(k_\ell)$
matrix. The corresponding negative-outlier fluctuations are exactly
the negatives of their positive partners, since
$\lambda_{n-j+1}=-\lambda_j$.
\eTh

We work with $\sigma=1$ for the auxiliary results and the proof below. After the auxiliary estimates, we prove a matrix central limit theorem for the compressed resolvents associated with the distinct supercritical strengths (Lemma~\ref{lem:joint_gaussian_resolvent_fluctuations}). We then linearize each entire outlier cluster (Lemma~\ref{lem:outlier_linearization}). Combining these results proves Theorem~\ref{thm:supercritical_gaussian_fluctuations}.

\blem\la{lem:real_gaussian_block_qf}
Let $q\geq1$ be fixed. For $n>q$, let $T_n$ be a random Hermitian
$(n-q)\times(n-q)$ matrix, independent of
$$
\Gamma_n=(\gamma_{1,n}\ \cdots\ \gamma_{q,n})
\in\R^{(n-q)\times q},
$$
where the columns are independent standard Gaussian vectors in $\R^{n-q}$. Assume that $\|T_n\|=O_{\P}(1)$. Then
$$
\left\|
\frac1n\Gamma_n^\top T_n\Gamma_n
-\frac1n\Tr(T_n)I_q
\right\|=O_{\P}(n^{-1/2}).
$$
\elem

\begin{proof}
We first keep $T_n=T$ fixed and write $\E_\Gamma$ for integration over the Gaussian columns. Let $\gamma$ be a standard real Gaussian vector in $\R^{n-q}$. For any coordinate indices $i,j,k,l\in[n-q]$, independence of its coordinates, together with their second and fourth moments, gives
$$
\E_\Gamma[(\gamma_i\gamma_j-\delta_{ij})
(\gamma_k\gamma_l-\delta_{kl})]
=\delta_{ik}\delta_{jl}+\delta_{il}\delta_{jk}.
$$
Consequently,
$$
\begin{aligned}
\E_\Gamma|\gamma^\top T\gamma-\Tr T|^2
&=\sum_{i,j,k,l}T_{ij}\overline{T_{kl}}
(\delta_{ik}\delta_{jl}+\delta_{il}\delta_{jk})\\
&=\sum_{i,j}|T_{ij}|^2
+\sum_{i,j}T_{ij}\overline{T_{ji}}\\
&=\Tr(T^2)+\Tr(TT^\top)
\leq2\Tr(T^2).
\end{aligned}
$$
Here we used $T=T^*$ and
$|\Tr(TT^\top)|\leq\sum_{i,j}|T_{ij}|^2=\Tr(T^2)$.
For two distinct columns, independence similarly gives
$$
\E_\Gamma|\gamma_{i,n}^\top T\gamma_{j,n}|^2=\Tr(T^2),
\qquad i\ne j.
$$
Set
$$
D_n=\frac1n\Gamma_n^\top T_n\Gamma_n-\frac1n\Tr(T_n)I_q.
$$
Summing the absolute second moments over its $q^2$ entries gives, with
$T_n$ fixed,
$$
\E_\Gamma\|D_n\|_{\rm F}^2
\leq\frac{2q+q(q-1)}{n^2}\Tr(T_n^2)
=\frac{q(q+1)}{n^2}\Tr(T_n^2).
$$
For $L,M>0$, split according to the event $\{\|T_n\|\leq L\}$. On this event, $\Tr(T_n^2)\leq(n-q)L^2$. Using $\|D_n\|\leq\|D_n\|_{\rm F}$, the tower property and Markov's inequality conditional on $T_n$, we obtain
$$
\begin{aligned}
\P(\sqrt n\|D_n\|>M)
&\leq\P(\|T_n\|>L)
 +\E\left[\one_{\{\|T_n\|\leq L\}}
     \P(\sqrt n\|D_n\|_{\rm F}>M\mid T_n)\right]\\
&\leq\P(\|T_n\|>L)+\frac{q(q+1)L^2}{M^2}.
\end{aligned}
$$
Given $\varepsilon>0$, tightness of $\|T_n\|$ allows us to choose $L$ so that the first term is at most $\varepsilon$ for all large $n$. We then choose $M$ so that the second term is at most $\varepsilon$. This proves $\|D_n\|=O_{\P}(n^{-1/2})$.
\end{proof}

\blem\la{lem:uniform_derivative_compressed_appendix}
Suppose Assumptions \ref{ass:ass3} and \ref{ass:ass2} hold with $\sigma=1$. If $K\subset\R\setminus[-2,2]$ is compact, then almost surely,
$$
\sup_{a\in K}
\left\|
M'_n(a)-m_{\rm sc}'(a)I_{2r}
\right\|
\longrightarrow 0.
$$
\elem

\begin{proof}
Choose $h_0>0$ so small that $K_{h_0}\!=\!\{a+t\!: a\in K,\, |t|\leq h_0\}$ is still compactly contained in $\R\setminus[-2,2]$. By Lemma \ref{lem:uniform_compressed_resolvent},
$$
\sup_{a\in K_{h_0}}\|M_n(a)-m_{\rm sc}(a)I_{2r}\|\longrightarrow0
$$
almost surely. Let $d=\dist(K_{h_0},[-2,2])>0.$ By Lemma \ref{lem:stielt-cv}, almost surely, for all sufficiently large $n$, $\sup_{a\in K_{h_0}}\|G_n(a)\| \leq \frac{2}{d}.$  Moreover, for $0<|h|\leq h_0$, the resolvent identity gives
\be\la{eq:derivative_compressed_resolvent_identity}
\frac{G_n(a+h)-G_n(a)}{h}-G_n(a)^2
=
hG_n(a+h)G_n(a)^2.
\ee
Thus, since $G'_n(a)=G_n(a)^2$ and $U_n$ has orthonormal columns, there exists $C_1>0$ such that, uniformly for $a\in K$,
$$
\left\|
\frac{M_n(a+h)-M_n(a)}{h}
-
M'_n(a)
\right\|
\leq C_1|h|.
$$
The same deterministic inequality holds for $m_{\rm sc}$, for some $C_2>C_1$, since it is twice continuously differentiable on $K_{h_0}$. Hence
$$
\sup_{a\in K}
\left\|
M'_n(a)-m_{\rm sc}'(a)I_{2r}
\right\|
\leq
2C_2|h|
+
\frac{2}{|h|}
\sup_{b\in K_{h_0}}\|M_n(b)-m_{\rm sc}(b)I_{2r}\|.
$$
Taking first $n\to\infty$ and then $h\to0$ proves the claim.
\end{proof}

\blem\la{lem:minor_trace_estimate}
Suppose Assumption \ref{ass:ass3} holds. Let $q\geq1$ be fixed. For $n>q$, let $W_n^\circ$ be the principal minor obtained from $W_n$ by deleting the first $q$ rows and columns. For $\rho>2$, set
$$
G_n^\circ(\rho)=(W_n^\circ-\rho I_{n-q})^{-1},
\qquad m_n^\circ(\rho)=\frac1n\Tr G_n^\circ(\rho).
$$
Then
$$
m_n^\circ(\rho)-m_{\rm sc}(\rho)=o_{\P}(n^{-1/2}).
$$
Moreover,
$$
\|G_n^\circ(\rho)\|=O_{\P}(1),
\qquad
\frac1n\Tr\bigl((G_n^\circ(\rho))^2\bigr)
\longrightarrow m_{\rm sc}'(\rho)
\quad\text{in probability}.
$$
These conclusions hold simultaneously for any fixed finite collection of
points in $(2,\infty)$.
\elem

\begin{proof}
We first prove the full-trace estimate
$m_n(\rho)-m_{\rm sc}(\rho)=o_{\P}(n^{-1/2})$.
Choose $\eta>0$ such that $2+\eta<\rho$, and set
$$
\Omega_n=\{\spec(W_n)\subset[-2-\eta,2+\eta]\}.
$$
By Lemma \ref{lem:stielt-cv}, $\P(\Omega_n)\to1$.
Fix $\varepsilon\in(0,1/6)$ and set
$$
\eta_n=n^{-1/3},
\qquad
z_n=\rho+\ii\eta_n.
$$
Let $W_n^{\rm std}$ and $G_n^{\rm std}$ be as in the proof of
Proposition \ref{prop:isotrop}, and write
$$
c_n=\sqrt{\frac{n}{n-1}},
\qquad
m_n^{\rm std}(z)=\frac1n\Tr G_n^{\rm std}(z).
$$
Since $W_n^{\rm std}=c_nW_n$, we have
$$
G_n(z)=c_nG_n^{\rm std}(c_nz),
\qquad
m_n(z)=c_nm_n^{\rm std}(c_nz).
$$
The averaged local semicircle law of
\citet[Theorem 2.3, equation (2.19)]{erdos2013local},
applied to $W_n^{\rm std}$, gives
$$
\left|m_n^{\rm std}(c_nz_n)-m_{\rm sc}(c_nz_n)\right|
\prec\frac1{n\eta_n}=n^{-2/3}.
$$
Moreover, $m_{\rm sc}$ and its derivative are bounded on a
fixed complex neighborhood of $\rho>2$. Since
$c_n-1=O(n^{-1})$ and $z_n\to\rho$, it follows that
$$
\left|c_nm_{\rm sc}(c_nz_n)-m_{\rm sc}(z_n)\right|
=O(n^{-1}).
$$
Consequently, by definition of stochastic domination,
$$
\begin{aligned}
m_n(z_n)-m_{\rm sc}(z_n)
={}&
c_n\left(m_n^{\rm std}(c_nz_n)-m_{\rm sc}(c_nz_n)\right)\\
&+c_nm_{\rm sc}(c_nz_n)-m_{\rm sc}(z_n)
=O_{\P}(n^{-2/3+\varepsilon}).
\end{aligned}
$$
We now pass to the real axis. For every real $a\neq0$,
$$
\left|\frac1a-\Re\frac1{a-\ii\eta_n}\right|
=
\frac{\eta_n^2}{|a|(a^2+\eta_n^2)}
\leq\frac{\eta_n^2}{|a|^3}.
$$
On $\Omega_n$, all eigenvalues $\xi_k$ satisfy
$|\xi_k-\rho|\geq\rho-2-\eta>0$.
Applying the preceding inequality to each spectral
denominator gives
$$
\left|m_n(\rho)-\Re m_n(z_n)\right|
\leq\frac{\eta_n^2}{(\rho-2-\eta)^3}.
$$
Similarly, since $\mu_{\rm sc}$ is supported on $[-2,2]$,
$$
\left|m_{\rm sc}(\rho)-\Re m_{\rm sc}(z_n)\right|
\leq\frac{\eta_n^2}{(\rho-2)^3}.
$$
Thus, on $\Omega_n$,
$$
\left|m_n(\rho)-m_{\rm sc}(\rho)\right|
\leq
\left|m_n(z_n)-m_{\rm sc}(z_n)\right|+C\eta_n^2,
$$
where $C>0$ is independent of $n$. Since $\P(\Omega_n)\to1$
and $\eta_n^2=n^{-2/3}$, we obtain
$$
m_n(\rho)-m_{\rm sc}(\rho)
=O_{\P}(n^{-2/3+\varepsilon})
=o_{\P}(n^{-1/2}),
$$
where the last equality follows from $\varepsilon<1/6$.

We next compare the full trace and the minor trace. Let $\xi_1^\circ\geq\cdots\geq\xi_{n-q}^\circ$ be the eigenvalues of $W_n^\circ$. Cauchy's interlacing theorem gives
$$
\xi_k\geq\xi_k^\circ\geq\xi_{k+q},
\qquad k\in[n-q].
$$
On $\Omega_n$, both spectra lie in $[-2-\eta,2+\eta]$. Set $g(x)=(x-\rho)^{-1}$ for $x<\rho$. The function $g$ is negative and decreasing on this interval, so
$$
\sum_{k=q+1}^n g(\xi_k)
\geq\sum_{k=1}^{n-q}g(\xi_k^\circ)
\geq\sum_{k=1}^{n-q}g(\xi_k).
$$
Consequently, on $\Omega_n$,
$$
\begin{aligned}
0\leq\Tr G_n^\circ(\rho)-\Tr G_n(\rho)
&=\sum_{k=1}^{n-q}g(\xi_k^\circ) - \sum_{k=1}^n g(\xi_k)\\
&\leq-\sum_{k=1}^q g(\xi_k)
\leq\frac{q}{\rho-2-\eta}.
\end{aligned}
$$
Since $q$ is fixed, division by $n$ gives
$m_n^\circ(\rho)-m_n(\rho)=O_{\P}(n^{-1})$. Combining this with the full-trace estimate proves the first assertion of the lemma.

Moreover, on $\Omega_n$, $\|G_n^\circ(\rho)\|\leq(\rho-2-\eta)^{-1}.$ Since $\P(\Omega_n)\to1$, this proves $\|G_n^\circ(\rho)\|=O_\P(1)$. To prove the squared-trace limit, write
$$
\mu_n^\circ=\frac1{n-q}\sum_{k=1}^{n-q}\delta_{\xi_k^\circ}.
$$
The matrix $\widetilde W_{n-q}=\sqrt{n/(n-q)}\,W_n^\circ$ has standardized upper-triangular noise satisfying Assumption~\ref{ass:ass3}, with dimension $n-q$. Hence its empirical spectral measure converges weakly to $\mu_{\rm sc}$ almost surely, by the semicircle law used in Lemma \ref{lem:stielt-cv}. Since $W_n^\circ=\sqrt{(n-q)/n}\,\widetilde W_{n-q}$ and $\sqrt{(n-q)/n}\to1$, the same almost-sure weak convergence holds for $\mu_n^\circ$. Now choose a bounded continuous function $f$ on $\R$ equal to $(x-\rho)^{-2}$ on
$[-2-\eta,2+\eta]$. On $\Omega_n$,
$$
\frac1n\Tr\bigl((G_n^\circ(\rho))^2\bigr)
=\frac{n-q}{n}\int f\,d\mu_n^\circ.
$$
The right-hand side converges almost surely to $\int f\,d\mu_{\rm sc}=m_{\rm sc}'(\rho)$. Since $\P(\Omega_n)\to1$, the left-hand side converges to the same limit in probability. Finally, applying the estimates at each point of a fixed finite collection and using a union bound proves the simultaneous claim.
\end{proof}

\blem\la{lem:gaussian_quadratic_form_appendix}
Let $s,q,k_1,\ldots,k_s\geq1$ be fixed integers. For $n>q$, let $\mathcal F_n$ be a sigma-field and let
$T_{1,n},\ldots,T_{s,n}$ be $\mathcal F_n$-measurable Hermitian
$(n-q)\times(n-q)$ matrices. Independently of $\mathcal F_n$, let
$Z_{\ell,n}\in\C^{(n-q)\times k_\ell}$, $\ell\in[s]$, be mutually
independent matrices with independent standard complex Gaussian entries.
Suppose that
\[
\max_{\ell\in[s]}\|T_{\ell,n}\|=O_{\P}(1),
\qquad
\frac1n\Tr(T_{\ell,n}^2)\longrightarrow v_\ell
\quad\text{in probability},
\]
where $v_1,\ldots,v_s\geq0$ are deterministic. Set
\[
Q_{\ell,n}
=\frac{Z_{\ell,n}^*T_{\ell,n}Z_{\ell,n}
       -\Tr(T_{\ell,n})I_{k_\ell}}{\sqrt n}.
\]
Then
\[
(Q_{\ell,n})_{\ell=1}^s
\Longrightarrow(\sqrt{v_\ell}\,\mathcal G_\ell)_{\ell=1}^s,
\]
where $\mathcal G_1,\ldots,\mathcal G_s$ are independent standard
$\mathrm{GUE}(k_1),\ldots,\mathrm{GUE}(k_s)$ matrices. More precisely,
conditionally on $\mathcal F_n$, for every
fixed collection of Hermitian matrices $B_\ell\in\C^{k_\ell\times k_\ell}$,
\be\la{eq:matrix_qf_conditional_cf}
\E\left[\left.
 e^{\ii\sum_{\ell=1}^s\Tr(B_\ell Q_{\ell,n})}
 \right|\mathcal F_n\right]
\longrightarrow
\exp\left(-\frac12\sum_{\ell=1}^s
v_\ell\Tr(B_\ell^2)\right)
\ee
in probability and in $L^1$.
\elem

\begin{proof}
Fix deterministic Hermitian matrices
$B_\ell\in\C^{k_\ell\times k_\ell}$, $\ell\in[s]$, and write
$\mathbf B=(B_1,\ldots,B_s)$.
For $n>q$, condition on $\mathcal F_n$. Under this conditional law,
the matrices $T_{\ell,n}$, $\ell\in[s]$, are fixed, while the
matrices $Z_{\ell,n}$ remain mutually independent with independent
standard complex Gaussian entries. For each $\ell\in[s]$, choose unitary diagonalizations
\[
T_{\ell,n}=V_{\ell,n}\Xi_{\ell,n}V_{\ell,n}^*,
\qquad
\Xi_{\ell,n}
=\diag(\tau_{\ell,1},\ldots,\tau_{\ell,n-q}),
\]
and
\[
B_\ell=U_\ell D_\ell U_\ell^*,
\qquad
D_\ell=\diag(\beta_{\ell,1},\ldots,\beta_{\ell,k_\ell}).
\]
Set
\[
\widetilde Z_{\ell,n}
:=V_{\ell,n}^*Z_{\ell,n}U_\ell
=(\zeta_{\ell,a,b})_{\substack{1\leq a\leq n-q\\
                              1\leq b\leq k_\ell}}.
\]
Left and right multiplication by fixed unitary matrices preserve
the standard complex Gaussian matrix law. Thus the entries of each
$\widetilde Z_{\ell,n}$ are again independent standard complex
Gaussian variables. Since these transformations act separately
on the matrices $Z_{\ell,n}$, this independence also holds
across $\ell$, conditionally on $\mathcal F_n$. By cyclicity of the trace,
\[
\Tr(B_\ell Z_{\ell,n}^*T_{\ell,n}Z_{\ell,n})
=
\Tr(D_\ell\widetilde Z_{\ell,n}^*
          \Xi_{\ell,n}\widetilde Z_{\ell,n})
=
\sum_{a=1}^{n-q}\sum_{b=1}^{k_\ell}
\tau_{\ell,a}\beta_{\ell,b}|\zeta_{\ell,a,b}|^2.
\]
Moreover,
\[
\Tr(T_{\ell,n})\Tr(B_\ell)
=
\sum_{a=1}^{n-q}\sum_{b=1}^{k_\ell}
\tau_{\ell,a}\beta_{\ell,b}.
\]
Consequently, for every $\ell\in[s]$,
\[
\Tr(B_\ell Q_{\ell,n})
=
\frac1{\sqrt n}
\sum_{a=1}^{n-q}\sum_{b=1}^{k_\ell}
\tau_{\ell,a}\beta_{\ell,b}
\bigl(|\zeta_{\ell,a,b}|^2-1\bigr).
\]

For a standard complex Gaussian variable $\zeta$, the variable
$|\zeta|^2$ is exponential with mean one, and hence
\[
\E e^{\ii u(|\zeta|^2-1)}
=\frac{e^{-\ii u}}{1-\ii u},
\qquad u\in\R.
\]
Using the conditional independence established above, we obtain
\[
\begin{aligned}
\varphi_n(\mathbf B\mid\mathcal F_n)
&:=\E\left[\left.
\exp\left(\ii\sum_{\ell=1}^s\Tr(B_\ell Q_{\ell,n})\right)
\right|\mathcal F_n\right]\\
&=\prod_{\ell=1}^s\prod_{a=1}^{n-q}\prod_{b=1}^{k_\ell}
\frac{\exp(-\ii\tau_{\ell,a}\beta_{\ell,b}/\sqrt n)}
{1-\ii\tau_{\ell,a}\beta_{\ell,b}/\sqrt n}.
\end{aligned}
\]
Write
$\varphi_n(\mathbf B\mid\mathcal F_n)=\exp(\mathcal L_n(\mathbf B))$,
where
\[
\mathcal L_n(\mathbf B)
:=
\sum_{\ell=1}^s\sum_{a=1}^{n-q}\sum_{b=1}^{k_\ell}
\left[
-\frac{\ii\tau_{\ell,a}\beta_{\ell,b}}{\sqrt n}
-\log\left(
1-\frac{\ii\tau_{\ell,a}\beta_{\ell,b}}{\sqrt n}
\right)
\right].
\]
Here $\log$ denotes the principal branch. Define the remainder
$\mathcal R_n(\mathbf B)$ by
\be\la{eq:joint_qf_log_expansion}
\mathcal L_n(\mathbf B)
=
-\frac1{2n}\sum_{\ell=1}^s
\Tr(T_{\ell,n}^2)\Tr(B_\ell^2)
+\mathcal R_n(\mathbf B).
\ee

Fix $L>0$. On the event
$\{\max_{\ell\in[s]}\|T_{\ell,n}\|\leq L\}$, we have
$|\tau_{\ell,a}\beta_{\ell,b}|/\sqrt n\leq1/2$
for every $\ell,a,b$ and all sufficiently large $n$.
Applying
\[
\left|-u-\log(1-u)-\frac{u^2}{2}\right|
\leq\frac23|u|^3,
\qquad u\in\C,\quad |u|\leq\frac12,
\]
with $u=\ii\tau_{\ell,a}\beta_{\ell,b}/\sqrt n$ gives
\[
\begin{aligned}
|\mathcal R_n(\mathbf B)|
&\leq
\frac{2}{3n^{3/2}}
\sum_{\ell=1}^s
\left(\sum_{a=1}^{n-q}|\tau_{\ell,a}|^3\right)
\left(\sum_{b=1}^{k_\ell}|\beta_{\ell,b}|^3\right)\\
&\leq
\frac{2L^3}{3\sqrt n}
\sum_{\ell=1}^s\sum_{b=1}^{k_\ell}|\beta_{\ell,b}|^3.
\end{aligned}
\]

Since $\max_{\ell\in[s]}\|T_{\ell,n}\|=O_\P(1)$, for every
$\delta>0$ we may choose $L>0$ such that
\[
\limsup_{n\to\infty}
\P\left(\max_{\ell\in[s]}\|T_{\ell,n}\|>L\right)
\leq\delta.
\]
For every $\varepsilon>0$, the preceding remainder bound then yields
\[
\limsup_{n\to\infty}
\P\bigl(|\mathcal R_n(\mathbf B)|>\varepsilon\bigr)
\leq\delta.
\]
Letting $\delta\downarrow0$ proves
$\mathcal R_n(\mathbf B)=o_\P(1)$.

Combining \eqref{eq:joint_qf_log_expansion} with the assumed limits
of $n^{-1}\Tr(T_{\ell,n}^2)$, we obtain
\[
\mathcal L_n(\mathbf B)
\longrightarrow
-\frac12\sum_{\ell=1}^s v_\ell\Tr(B_\ell^2)
\qquad\text{in probability}.
\]
By continuity of the exponential map,
\[
\varphi_n(\mathbf B\mid\mathcal F_n)
\longrightarrow
\psi(\mathbf B)
:=
\exp\left(-\frac12\sum_{\ell=1}^s
v_\ell\Tr(B_\ell^2)\right)
\qquad\text{in probability}.
\]
Since conditional characteristic functions have absolute value
at most one, this convergence also holds in $L^1$, proving
\eqref{eq:matrix_qf_conditional_cf}. Taking expectations and using
the tower property gives
\[
\E\exp\left(\ii\sum_{\ell=1}^s\Tr(B_\ell Q_{\ell,n})\right)
\longrightarrow\psi(\mathbf B).
\]
For any deterministic Hermitian matrix $B\in\C^{k\times k}$ and any standard $\mathrm{GUE}(k)$ matrix $\mathcal G$, recall that
\be\la{eq:charac_gue_formula}
\E e^{\ii\Tr(B\mathcal G)}
=\exp\left(-\frac12\Tr(B^2)\right).
\ee
It follows that $\psi(\mathbf B)$ is the joint characteristic function of $(\sqrt{v_\ell}\,\mathcal G_\ell)_{\ell=1}^s$, where $\mathcal G_1,\ldots,\mathcal G_s$ are independent standard GUE matrices with $\mathcal G_\ell\sim\mathrm{GUE}(k_\ell)$ for each $\ell\in[s]$. Applying L\'evy's continuity theorem proves the claimed joint convergence.
\end{proof}

With these auxiliary estimates in hand, we now establish the two main ingredients specific to the supercritical outlier fluctuations: the joint resolvent central limit theorem and the outlier linearization.

\blem\la{lem:joint_gaussian_resolvent_fluctuations}
Suppose Assumptions~\ref{ass:ass1} and~\ref{ass:ass2} hold with $\sigma=1$ and $\eps_{ij}\sim\mathcal N(0,1)$, and assume that $m\geq1$.
Let $\vartheta_1>\cdots>\vartheta_s>1$ be the distinct supercritical
strengths, with multiplicities $k_1,\ldots,k_s$, so that
$m=\sum_{\ell=1}^s k_\ell$. For each $\ell\in[s]$, set
\[
p_\ell=\sum_{h<\ell}k_h,\qquad
\rho_\ell^\star=\vartheta_\ell+\vartheta_\ell^{-1},\qquad
\mathcal J_\ell=\{p_\ell+1,\ldots,p_\ell+k_\ell\},
\]
and let $V_{\ell,n}=(w_{j,+})_{j\in\mathcal J_\ell} \in \C^{n \times k_\ell}$. Define
\[
R_{\ell,n}
:=\sqrt n\left[
V_{\ell,n}^*G_n(\rho_\ell^\star)V_{\ell,n}
-m_{\rm sc}(\rho_\ell^\star)I_{k_\ell}
\right].
\]
Then
\be\la{eq:joint_clt_compressed_resolvent}
(R_{\ell,n})_{\ell=1}^s
\Longrightarrow
\left(
\frac{\mathcal G_\ell}
{\vartheta_\ell\sqrt{\vartheta_\ell^2-1}}
\right)_{\ell=1}^s,
\ee
where $\mathcal G_1,\ldots,\mathcal G_s$ are independent standard
GUE matrices with $\mathcal G_\ell\sim\mathrm{GUE}(k_\ell)$
for each $\ell\in[s]$.
\elem

\begin{proof}
Set $q=2m$. Choose one deterministic orthogonal matrix $O_n$ such that
$O_ne_{2j-1}=u_j$ and $O_ne_{2j}=v_j$ for every $j\in[m]$, and set
\[
\omega_j=\frac{e_{2j-1}-\ii e_{2j}}{\sqrt2},
\qquad
\Omega_\ell=(\omega_j)_{j\in\mathcal J_\ell}.
\]
Then $w_{j,+}=O_n\omega_j$ and $V_{\ell,n}=O_n\Omega_\ell$.
Gaussian orthogonal conjugation invariance gives
$O_n^\top W_nO_n\stackrel{d}{=}W_n$.
Applying this same rotation to all the compressed resolvents
therefore gives the equality of joint laws
\be\la{eq:transfer_laws}
\bigl(V_{\ell,n}^*G_n(\rho_\ell^\star)V_{\ell,n}\bigr)_{\ell=1}^s
\stackrel{d}{=}
\bigl(\Omega_\ell^*G_n(\rho_\ell^\star)\Omega_\ell\bigr)_{\ell=1}^s.
\ee
Thus it suffices to prove the joint limit in these canonical
coordinates. For $n>q$, decompose
\[
W_n=
\begin{pmatrix}
J_n&C_n^*\\
C_n&W_n^\circ
\end{pmatrix},
\qquad
C_n=-\frac{\ii}{\sqrt n}\Gamma_n,
\]
where $J_n\in\C^{q\times q}$, $W_n^\circ$ is obtained by deleting
the first $q$ rows and columns, and
\[
(\Gamma_n)_{ak}=\eps_{k,q+a},
\qquad a\in[n-q],\quad k\in[q].
\]
The matrices $J_n$, $\Gamma_n$ and $W_n^\circ$ use disjoint upper-triangular entries of the noise and are mutually independent. The entries of $\Gamma_n$ are independent standard real Gaussians. In the following block calculations, we identify each $\omega_j$ with its restriction to $\C^q$, since it is supported on its first $q$ coordinates. Correspondingly, we identify $\Omega_\ell$ with its restriction to $\C^{q\times k_\ell}$ by keeping its $q$ first rows. In particular, \[ \Omega_\ell^*\Omega_\ell=I_{k_\ell}, \qquad \|\Omega_\ell\|=1. \] For each $\ell\in[s]$, set
\[
T_{\ell,n}
=(W_n^\circ-\rho_\ell^\star I_{n-q})^{-1},
\qquad
m_{\ell,n}^\circ=\frac1n\Tr(T_{\ell,n}).
\]
Lemma~\ref{lem:minor_trace_estimate}, applied with this $q$ at the
fixed points $\rho_1^\star,\ldots,\rho_s^\star>2$, gives
\be\la{eq:joint_minor_resolvent_estimates}
\begin{gathered}
\max_{\ell\in[s]}\|T_{\ell,n}\|=O_\P(1),
\qquad
\max_{\ell\in[s]}
|m_{\ell,n}^\circ-m_{\rm sc}(\rho_\ell^\star)|
=o_\P(n^{-1/2}),\\[2pt]
\max_{\ell\in[s]}
\left|
\frac1n\Tr(T_{\ell,n}^2)
-\frac1{\vartheta_\ell^2-1}
\right|
\longrightarrow0
\qquad\text{in probability}.
\end{gathered}
\ee
Here we used
$m_{\rm sc}'(\rho_\ell^\star)=(\vartheta_\ell^2-1)^{-1}$.
Since each $T_{\ell,n}$ is Hermitian and independent of $\Gamma_n$,
Lemma~\ref{lem:real_gaussian_block_qf} gives
\[
\left\|C_n^*T_{\ell,n}C_n-m_{\ell,n}^\circ I_q\right\|
=
\left\|
\frac1n\Gamma_n^\top T_{\ell,n}\Gamma_n
-\frac1n\Tr(T_{\ell,n})I_q
\right\|=O_\P(n^{-1/2}).
\]
The maximum over $\ell\in[s]$ has the same order by a finite union bound. Define, for each $\ell\in[s]$,
\[
\Delta_{\ell,n}
=
J_n-\bigl(C_n^*T_{\ell,n}C_n-m_{\ell,n}^\circ I_q\bigr),
\qquad
d_{\ell,n}=\rho_\ell^\star+m_{\ell,n}^\circ.
\]
Since $q$ is fixed, $\sqrt nJ_n$ has a fixed Gaussian distribution,
so $\|J_n\|=O_\P(n^{-1/2})$. Consequently,
\[
\max_{\ell\in[s]}\|\Delta_{\ell,n}\|
=O_\P(n^{-1/2}).
\]
Also,
$\rho_\ell^\star+m_{\rm sc}(\rho_\ell^\star)=\vartheta_\ell$,
and hence
\[
\max_{\ell\in[s]}|d_{\ell,n}-\vartheta_\ell|
=
\max_{\ell\in[s]}
|m_{\ell,n}^\circ-m_{\rm sc}(\rho_\ell^\star)|
=o_\P(n^{-1/2}).
\]

Set $\vartheta_*=\min_{\ell\in[s]}\vartheta_\ell>1$ and define
\[
\mathcal A_n^\star
=
\left\{
\max_{\ell\in[s]}|d_{\ell,n}-\vartheta_\ell|
\leq\frac{\vartheta_*}{2}
\right\}
\cap
\left\{
\max_{\ell\in[s]}\|\Delta_{\ell,n}\|
\leq\frac{\vartheta_*}{4}
\right\}.
\]
The preceding estimates imply $\P(\mathcal A_n^\star)\to1$.
On this event, for every $\ell\in[s]$,
\[
|d_{\ell,n}|
\geq\vartheta_\ell-|d_{\ell,n}-\vartheta_\ell|
\geq\frac{\vartheta_*}{2},
\]
and therefore
\[
\|d_{\ell,n}^{-1}\Delta_{\ell,n}\|
\leq
\frac{\vartheta_*/4}{\vartheta_*/2}
=\frac12.
\]

By the Schur complement formula, the $q\times q$ upper-left
block of $G_n(\rho_\ell^\star)$ is
\[
 G_{\ell,n}
=
\bigl(J_n-\rho_\ell^\star I_q-C_n^*T_{\ell,n}C_n\bigr)^{-1}
=
-(d_{\ell,n}I_q-\Delta_{\ell,n})^{-1}.
\]
Set $X_{\ell,n}=d_{\ell,n}^{-1}\Delta_{\ell,n}$.
On $\mathcal A_n^\star$, we have $\|X_{\ell,n}\|\leq1/2$, so
$\|(I_q-X_{\ell,n})^{-1}\|\leq2$.
The identity
\[
(I_q-X_{\ell,n})^{-1}
=
I_q+X_{\ell,n}
+X_{\ell,n}^2(I_q-X_{\ell,n})^{-1}
\]
therefore gives
\be\la{eq:remainder_joint}
 G_{\ell,n}
=
-d_{\ell,n}^{-1}I_q
-d_{\ell,n}^{-2}\Delta_{\ell,n}
+\mathcal S_{\ell,n},
\ee
where, on this event,
\[
\mathcal S_{\ell,n}
=
-d_{\ell,n}^{-3}\Delta_{\ell,n}^2
(I_q-X_{\ell,n})^{-1}.
\]
Consequently,
\[
\|\mathcal S_{\ell,n}\|
\leq
2|d_{\ell,n}|^{-3}\|\Delta_{\ell,n}\|^2.
\]
Since $\P(\mathcal A_n^\star)\to1$, this yields
$\max_{\ell\in[s]}\|\mathcal S_{\ell,n}\|=O_\P(n^{-1})$. Compressing \eqref{eq:remainder_joint} by $\Omega_\ell$ gives,
on $\mathcal A_n^\star$,
\[
\Omega_\ell^* G_{\ell,n}\Omega_\ell
=
-d_{\ell,n}^{-1}I_{k_\ell}
-d_{\ell,n}^{-2}\Omega_\ell^*\Delta_{\ell,n}\Omega_\ell
+\mathfrak e_{\ell,n},
\]
where
\[
\mathfrak e_{\ell,n}
:=\Omega_\ell^*\mathcal S_{\ell,n}\Omega_\ell,
\qquad
\max_{\ell\in[s]}\|\mathfrak e_{\ell,n}\|=O_\P(n^{-1}).
\]

Let $\gamma_k$ denote column $k$ of $\Gamma_n$, and define
\[
z_{j,n}
:=-\frac{\gamma_{2j}+\ii\gamma_{2j-1}}{\sqrt2},
\qquad j\in[m],
\]
and
\[
Z_{\ell,n}=(z_{j,n})_{j\in\mathcal J_\ell},
\qquad \ell\in[s].
\]
The real and imaginary parts of each $z_{j,n}$ have independent
$\mathcal N(0,1/2)$ coordinates. Distinct $j$ use disjoint pairs
of columns of $\Gamma_n$. Thus the matrices $Z_{\ell,n}$ are
mutually independent matrices with independent standard complex
Gaussian entries. Their entire family is independent of
$(J_n,W_n^\circ)$. For each $\ell\in[s]$, set
\[
A_{\ell,n}:=\sqrt n\,\Omega_\ell^*J_n\Omega_\ell,
\qquad
Q_{\ell,n}
:=
\frac{
Z_{\ell,n}^*T_{\ell,n}Z_{\ell,n}
-\Tr(T_{\ell,n})I_{k_\ell}
}{\sqrt n}.
\]
The matrices $A_{\ell,n}$ are independent standard
$\mathrm{GUE}(k_\ell)$ matrices, exactly for every $n$.
To verify this, fix $\ell\in[s]$. For $a\in[k_\ell]$, let
$j=p_\ell+a$. Direct calculation gives
\[
(A_{\ell,n})_{aa}
=
\sqrt n\,\omega_j^*J_n\omega_j
=\eps_{2j-1,2j}.
\]
For $1\leq a<b\leq k_\ell$, let $j=p_\ell+a$ and $h=p_\ell+b$.
Expanding $\sqrt n\,\omega_j^*J_n\omega_h$ gives
\[
(A_{\ell,n})_{ab}
=
\frac12\left[
\eps_{2j-1,2h}-\eps_{2j,2h-1}
+\ii\bigl(\eps_{2j-1,2h-1}+\eps_{2j,2h}\bigr)
\right].
\]
Thus each diagonal entry is standard real Gaussian, while each
upper-triangular entry has independent real and imaginary parts
with distribution $\mathcal N(0,1/2)$.
All these real Gaussian variables use disjoint noise coordinates,
also across distinct blocks $\ell$.
Since $A_{\ell,n}$ is Hermitian, this proves the asserted
independent GUE laws. Moreover, the entire family
$(A_{\ell,n})_{\ell=1}^s$ is independent of
$(\Gamma_n,W_n^\circ)$, and hence of
$\bigl((Z_{\ell,n})_{\ell=1}^s,W_n^\circ\bigr)$.

Direct multiplication gives, for every $j\in[m]$,
\[
C_n\omega_j
=
-\frac{\gamma_{2j}+\ii\gamma_{2j-1}}{\sqrt{2n}}
=
\frac{z_{j,n}}{\sqrt n}.
\]
Therefore $C_n\Omega_\ell=Z_{\ell,n}/\sqrt n$, and the definition
of $\Delta_{\ell,n}$ yields
\[
\begin{aligned}
\sqrt n\,\Omega_\ell^*\Delta_{\ell,n}\Omega_\ell
&=
\sqrt n\,\Omega_\ell^*J_n\Omega_\ell
-\frac{
Z_{\ell,n}^*T_{\ell,n}Z_{\ell,n}
-\Tr(T_{\ell,n})I_{k_\ell}
}{\sqrt n}\\
&=A_{\ell,n}-Q_{\ell,n}.
\end{aligned}
\]
In particular,
\[
\max_{\ell\in[s]}\|A_{\ell,n}-Q_{\ell,n}\|
\leq
\sqrt n\,\max_{\ell\in[s]}\|\Delta_{\ell,n}\|
=O_\P(1).
\]
Since $m_{\rm sc}(\rho_\ell^\star)=-\vartheta_\ell^{-1}$,
\[
-d_{\ell,n}^{-1}-m_{\rm sc}(\rho_\ell^\star)
=
\frac{d_{\ell,n}-\vartheta_\ell}
{d_{\ell,n}\vartheta_\ell},
\qquad
\max_{\ell\in[s]}
|d_{\ell,n}^{-2}-\vartheta_\ell^{-2}|
=o_\P(n^{-1/2}).
\]
Define $r_{\ell,n}$ by
\be\la{eq:joint_schur_fluctuation}
R_{\ell,n}
=
-\vartheta_\ell^{-2}(A_{\ell,n}-Q_{\ell,n})
+r_{\ell,n}.
\ee
On $\mathcal A_n^\star$, the compressed inverse expansion gives
\[
r_{\ell,n}
=
\sqrt n\,
\frac{d_{\ell,n}-\vartheta_\ell}
{d_{\ell,n}\vartheta_\ell}I_{k_\ell}
+
(\vartheta_\ell^{-2}-d_{\ell,n}^{-2})
(A_{\ell,n}-Q_{\ell,n})
+\sqrt n\,\mathfrak e_{\ell,n}.
\]
The maxima of the operator norms of these three terms have orders
$o_\P(1)$, $o_\P(n^{-1/2})$ and $O_\P(n^{-1/2})$,
respectively. Since $\P(\mathcal A_n^\star)\to1$, it follows that
\[
\max_{\ell\in[s]}\|r_{\ell,n}\|
=o_\P(1).
\]
It remains to establish the joint limit of the leading matrices.
Let $\mathcal F_n=\sigma(W_n^\circ)$.
The matrices $T_{\ell,n}$, $\ell\in[s]$, are Hermitian and
$\mathcal F_n$-measurable. The entire family
$(Z_{\ell,n})_{\ell=1}^s$ is independent of $\mathcal F_n$ and
consists of mutually independent standard complex Gaussian
matrices. The norm bounds and squared-trace limits required by
Lemma~\ref{lem:gaussian_quadratic_form_appendix} are precisely
\eqref{eq:joint_minor_resolvent_estimates}, with
\[
q=2m,
\qquad
v_\ell=\frac1{\vartheta_\ell^2-1},
\qquad \ell\in[s].
\]
Consequently,
\[
(Q_{\ell,n})_{\ell=1}^s
\Longrightarrow
\left(
\frac{\mathcal G_\ell}{\sqrt{\vartheta_\ell^2-1}}
\right)_{\ell=1}^s.
\]
More precisely, for every fixed collection of deterministic
Hermitian matrices $B_\ell\in\C^{k_\ell\times k_\ell}$,
$\ell\in[s]$, that lemma gives
\be\la{eq:expectation_limit_cf}
\E\left[\left.
\exp\left(\ii\sum_{\ell=1}^s\Tr(B_\ell Q_{\ell,n})\right)
\right|\mathcal F_n\right]
\longrightarrow
\exp\left(
-\frac12\sum_{\ell=1}^s
\frac{\Tr(B_\ell^2)}{\vartheta_\ell^2-1}
\right)
\ee
in probability and in $L^1$. The family $(A_{\ell,n})_{\ell=1}^s$ is independent of
$\bigl((Q_{\ell,n})_{\ell=1}^s,\mathcal F_n\bigr)$.
Using its exact independent GUE laws therefore gives
\be\la{eq:joint_conditional_leading_cf}
\begin{aligned}
&\E\left[\left.
\exp\left(
\ii\sum_{\ell=1}^s
\Tr\bigl(B_\ell(A_{\ell,n}-Q_{\ell,n})\bigr)
\right)
\right|\mathcal F_n\right]\\
&\qquad=
\exp\left(-\frac12\sum_{\ell=1}^s\Tr(B_\ell^2)\right)
\E\left[\left.
\exp\left(-\ii\sum_{\ell=1}^s\Tr(B_\ell Q_{\ell,n})\right)
\right|\mathcal F_n\right]\\
&\qquad\longrightarrow
\exp\left(
-\frac12\sum_{\ell=1}^s
\frac{\vartheta_\ell^2}{\vartheta_\ell^2-1}
\Tr(B_\ell^2)
\right).
\end{aligned}
\ee
This convergence holds in probability and also in $L^1$,
since the conditional characteristic functions have absolute
value at most one. Taking expectations and using the tower
property gives the same deterministic limit for the unconditional
characteristic functions. Moreover, the limiting function is the joint characteristic function of
independent GUE matrices scaled by
$\vartheta_\ell/\sqrt{\vartheta_\ell^2-1}$.
L\'evy's continuity theorem therefore gives
\[
(A_{\ell,n}-Q_{\ell,n})_{\ell=1}^s
\Longrightarrow
\left(
\frac{\vartheta_\ell}{\sqrt{\vartheta_\ell^2-1}}
\mathcal G_\ell
\right)_{\ell=1}^s.
\]
Multiplying the $\ell$th matrix by $-\vartheta_\ell^{-2}$
gives the limiting matrix
\[
-\frac{\mathcal G_\ell}
{\vartheta_\ell\sqrt{\vartheta_\ell^2-1}}.
\]
Thus, since
$\max_{\ell\in[s]}\| r_{\ell,n}\|=o_\P(1)$,
Slutsky's theorem applied to
\eqref{eq:joint_schur_fluctuation} gives
\eqref{eq:joint_clt_compressed_resolvent} in canonical coordinates.
The initial equality of joint laws \eqref{eq:transfer_laws} transfers this convergence to the original signal directions and completes the proof.
\end{proof}

\blem\la{lem:outlier_linearization}
Suppose Assumptions~\ref{ass:ass1} and~\ref{ass:ass2} hold with $\sigma=1$ and $\eps_{ij}\sim\mathcal N(0,1)$. With the block notation of the previous lemma, for each supercritical strength $\vartheta_\ell$,
\be\la{eq:outlier_linearization}
\max_{1\leq a\leq k_\ell}
\left|
\sqrt n(\lambda_{p_\ell+a}-\rho_\ell^\star)
-\lambda_a\!\left(-\bigl(\vartheta_\ell^2-1\bigr)R_{\ell,n}\right)
\right|\longrightarrow0
\qquad\text{in probability}.
\ee
These approximations hold simultaneously over all supercritical blocks. The remaining strengths may have arbitrary multiplicities and may equal $1$.
\elem

\begin{proof}
We use the notation of Lemma~\ref{lem:joint_gaussian_resolvent_fluctuations}. Fix $\ell\in[s]$ and abbreviate
\[
\vartheta=\vartheta_\ell,\qquad
k=k_\ell,\qquad
p=p_\ell,\qquad
\rho=\rho_\ell^\star,\qquad
V=V_{\ell,n},\qquad
R_n=R_{\ell,n}.
\]
Since $\theta\mapsto\theta+\theta^{-1}$ is strictly increasing on
$(1,\infty)$, the distinct supercritical strengths have distinct
limiting outlier locations. We may therefore choose $\delta>0$
such that
\[
I=[\rho-\delta,\rho+\delta]\subset(2,\infty)
\]
contains no other limiting outlier location.
By the all-moments case established in Subsection~\ref{app:proof_outliers}, almost surely, for all sufficiently large $n$, exactly the eigenvalues $\lambda_{p+1},\ldots,\lambda_{p+k}$ lie in $I$, counted with multiplicity. Moreover, Lemma~\ref{lem:stielt-cv} gives
$\|W_n\|\to2$ almost surely. Thus $I$ is eventually disjoint from
$\spec(W_n)$, almost surely. Recall that
\[
H_n=W_n+U_n\Theta U_n^*,
\qquad
M_n(x)=U_n^*G_n(x)U_n.
\]
Permute the columns of $U_n$ and the diagonal entries of $\Theta$
by the same permutation so that the first $k$ columns of $U_n$
are the columns of $V$, in their original order. We retain the
notation $U_n$, $\Theta$ and $M_n$ for the resulting matrices.
This leaves $H_n$ unchanged. For $x\in I$, consider the Hermitian matrix
\[
F_n(x)=\Theta^{-1}+M_n(x)
=
\begin{pmatrix}
\vartheta^{-1}I_k+V^*G_n(x)V & B_n(x)^*\\
B_n(x) & D_n(x)
\end{pmatrix},
\]
where
\[
B_n(x)\in\C^{(2r-k)\times k},
\qquad
D_n(x)\in\C^{(2r-k)\times(2r-k)}.
\]
By Lemma~\ref{lem:uniform_compressed_resolvent},
\begin{equation}\label{eq:uniform_secular_matrix_convergence}
\sup_{x\in I}
\left\|
F_n(x)-\bigl(\Theta^{-1}+m_{\rm sc}(x)I_{2r}\bigr)
\right\|
\longrightarrow0
\qquad\text{almost surely}.
\end{equation}
Thus, using $m_{\rm sc}(\rho)=-\vartheta^{-1}$, the diagonal
entries of the limiting lower-right block at $x=\rho$ are
\[
\theta_j^{-1}-\vartheta^{-1}
\quad(j\in[r]\setminus\mathcal J_\ell),
\qquad
-\theta_j^{-1}-\vartheta^{-1}
\quad(j\in[r]).
\]
In particular, they are all nonzero.
After shrinking $\delta$ if necessary, this limiting block is
therefore uniformly invertible on $I$. Consequently, almost
surely, $D_n(x)$ is invertible for every $x\in I$ and all
sufficiently large $n$, with
\be\la{eq:norm_inv_dn}
\sup_{x\in I}\|D_n(x)^{-1}\|=O(1).
\ee
Since the limiting matrix is diagonal, the uniform convergence
in \eqref{eq:uniform_secular_matrix_convergence} also gives
\be\la{eq:bn_to_zero_unif}
\sup_{x\in I}\|B_n(x)\|\longrightarrow0
\qquad\text{almost surely}.
\ee
Define the Hermitian Schur complement
\[
S_n(x)
=
\vartheta^{-1}I_k+V^*G_n(x)V-\mathcal C_n(x),
\qquad
\mathcal C_n(x):=B_n(x)^*D_n(x)^{-1}B_n(x).
\]
We first estimate its value at $\rho$.
Fix $\varepsilon\in(0,1/4)$ and set
$z_n=\rho+\ii n^{-1/2}$.
Let $W_n^{\rm std}$ and $G_n^{\rm std}$ be as in the proof of
Proposition~\ref{prop:isotrop}. For any two distinct columns
$u,v$ of $U_n$, we have $u^*v=0$. Since $\rho>2$,
$\Im m_{\rm sc}(z_n)=O(n^{-1/2})$.
Applying \eqref{eq:isotropic_vector_bound_sd} with any fixed
$\tau\in(0,1/2)$ and $L>\rho+1$ therefore gives
\[
|u^*G_n^{\rm std}(z_n)v|
\prec
\sqrt{\frac{\Im m_{\rm sc}(z_n)}{n^{1/2}}}
+\frac1{n^{1/2}}
=O(n^{-1/2}).
\]
By the definition of stochastic domination, it follows that
\be\la{eq:approx_z_n}
u^*G_n^{\rm std}(z_n)v
=O_\P(n^{-1/2+\varepsilon}).
\ee
We now pass to the real resolvent with the original
normalization. Since
$W_n^{\rm std}=\sqrt{n/(n-1)}\,W_n$, Lemma~\ref{lem:stielt-cv}
gives
\[
\|W_n^{\rm std}-W_n\|=O_\P(n^{-1}),
\qquad
\|G_n(\rho)\|+\|G_n^{\rm std}(z_n)\|=O_\P(1).
\]
Indeed, the spectra of $W_n$ and $W_n^{\rm std}$ remain at a
fixed positive distance from $\rho$ almost surely for all
sufficiently large $n$. The resolvent identity yields
\begin{align}
\|G_n(\rho)-G_n^{\rm std}(z_n)\|
&\leq
\|G_n(\rho)\|\,\|G_n^{\rm std}(z_n)\|
\left(\|W_n^{\rm std}-W_n\|+n^{-1/2}\right) \nonumber\\
&=O_\P(n^{-1/2})\la{eq:approx_z_n_two}.
\end{align}
Consequently, combining \eqref{eq:approx_z_n} and \eqref{eq:approx_z_n_two}, every entry of $B_n(\rho)$ is
$O_\P(n^{-1/2+\varepsilon})$. Since its dimensions are fixed,
\[
\|B_n(\rho)\|=O_\P(n^{-1/2+\varepsilon}).
\]
Together with $\|D_n(\rho)^{-1}\|=O_\P(1)$, this gives
\[
\|\mathcal C_n(\rho)\|
\leq
\|B_n(\rho)\|^2\|D_n(\rho)^{-1}\|
=O_\P(n^{-1+2\varepsilon})
=o_\P(n^{-1/2}),
\]
where the last equality follows from $\varepsilon<1/4$.
Since $m_{\rm sc}(\rho)=-\vartheta^{-1}$, we have
\[
S_n(\rho)=\frac{R_n}{\sqrt n}-\mathcal C_n(\rho).
\]
Lemma~\ref{lem:joint_gaussian_resolvent_fluctuations} gives
$\|R_n\|=O_\P(1)$. Therefore,
\be\la{eq:block_schur_at_outlier}
\sqrt n\,S_n(\rho)=R_n+o_\P(1),
\qquad
\|S_n(\rho)\|=O_\P(n^{-1/2}).
\ee
We next control the derivative of $S_n$ uniformly on $I$.
For all sufficiently large $n$, the matrices involved are
continuously differentiable on $I$. Since $G_n'(x)=G_n(x)^2$,
\[
S_n'(x)=V^*G_n(x)^2V-\mathcal C_n'(x).
\]
The first term is the corresponding diagonal block of $M_n'(x)$.
Hence Lemma~\ref{lem:uniform_derivative_compressed_appendix} gives
\[
\sup_{x\in I}
\|V^*G_n(x)^2V-m_{\rm sc}'(x)I_k\|
\longrightarrow0
\qquad\text{almost surely}.
\]
Moreover,
\[
\begin{aligned}
\mathcal C_n'(x)
={}&
B_n'(x)^*D_n(x)^{-1}B_n(x)
+B_n(x)^*D_n(x)^{-1}B_n'(x)\\
&-
B_n(x)^*D_n(x)^{-1}D_n'(x)
D_n(x)^{-1}B_n(x).
\end{aligned}
\]
The same derivative lemma applied to the other blocks implies
\[
\sup_{x\in I}\|B_n'(x)\|=O(1),
\qquad
\sup_{x\in I}\|D_n'(x)\|=O(1)
\qquad\text{almost surely}.
\]
Every term in $\mathcal C_n'(x)$ contains a factor $B_n(x)$
or $B_n(x)^*$, whose norm tends to zero uniformly on $I$ from \eqref{eq:bn_to_zero_unif},
while all other factors are uniformly bounded.
Consequently,
\[
\sup_{x\in I}\|\mathcal C_n'(x)\|
\longrightarrow0
\qquad\text{almost surely}.
\]
Combining these estimates gives
\be\la{eq:block_schur_derivative}
\sup_{x\in I}
\|S_n'(x)-m_{\rm sc}'(x)I_k\|
\longrightarrow0
\qquad\text{almost surely}.
\ee
We now show that each ordered eigenvalue of $S_n(\cdot)$ has
a unique zero in $I$, by establishing strict monotonicity
and opposite signs at the endpoints. Since $m_{\rm sc}'$ is continuous and strictly positive on $I$,
the deterministic constant
\[
c:=\frac12\min_{x\in I}m_{\rm sc}'(x)
\]
is positive. It follows from \eqref{eq:block_schur_derivative}
that, almost surely, for all sufficiently large $n$,
\[
S_n'(x)\succeq cI_k
\qquad\text{for every }x\in I.
\]
Also, combining \eqref{eq:uniform_secular_matrix_convergence}, \eqref{eq:norm_inv_dn} and \eqref{eq:bn_to_zero_unif} implies $S_n(x)\rightarrow
\bigl(\vartheta^{-1}+m_{\rm sc}(x)\bigr)I_k$ uniformly on $I$, almost surely.
The scalar function $x\mapsto\vartheta^{-1}+m_{\rm sc}(x)$ is strictly
increasing on $I$ and vanishes at $\rho$. Therefore, almost surely,
for all sufficiently large $n$,
\be\la{eq:sign_endpoints}
S_n(\rho-\delta)\prec0,
\qquad
S_n(\rho+\delta)\succ0.
\ee
We now work on the event where the preceding spectral, invertibility and monotonicity properties hold; its probability tends to one. For $x\in I$, let $\kappa_{1,n}(x)\geq\cdots\geq\kappa_{k,n}(x)$ be the ordered eigenvalues of $S_n(x)$. These functions are
continuous by Weyl's inequality. For $x<y$ in $I$,
\[
S_n(y)-S_n(x)
=
\int_x^y S_n'(t)\,dt
\succeq c(y-x)I_k.
\]
Lemma \ref{lem:min-max} consequently gives, for every $b\in[k]$,
\be\la{eq:mono_eigval_schur}
\kappa_{b,n}(y)-\kappa_{b,n}(x)\geq c(y-x).
\ee
Thus each $\kappa_{b,n}$ is strictly increasing. The signs at the endpoints given in \eqref{eq:sign_endpoints} imply that it has a unique zero $\alpha_{b,n}\in I$. Since $\kappa_{b,n}(x)\geq\kappa_{b+1,n}(x)$, these zeros satisfy $\alpha_{1,n}\leq\cdots\leq\alpha_{k,n}$.

We now identify these zeros with the eigenvalues of $H_n$ in $I$.
For each $x\in I$, a direct calculation shows that the map
$y\mapsto-G_n(x)U_ny$ is an isomorphism from $\ker F_n(x)$
onto $\ker(H_n-xI_n)$, with inverse $h\mapsto\Theta U_n^*h$. Since $D_n(x)$ is invertible, Schur complementation also gives
an isomorphism
\[
\ker S_n(x)\longrightarrow\ker F_n(x),
\qquad
v\longmapsto
\begin{pmatrix}
v\\
-D_n(x)^{-1}B_n(x)v
\end{pmatrix}.
\]
It follows that $\dim\ker S_n(x)=\dim\ker(H_n-xI_n).$ Because both matrices are Hermitian, these dimensions are the
multiplicities of their zero eigenvalues. Equivalently, the
number of indices $b$ for which $\alpha_{b,n}=x$ equals the
multiplicity of $x$ as an eigenvalue of $H_n$.
The roots $\alpha_{1,n},\ldots,\alpha_{k,n}$ therefore coincide,
as a multiset, with $\lambda_{p+1},\ldots,\lambda_{p+k}$. More precisely,
\be\la{eq:block_ordered_roots}
\lambda_{p+a}=\alpha_{k-a+1,n},
\qquad a\in[k].
\ee
Since $\kappa_{b,n}(\alpha_{b,n})=0$, applying \eqref{eq:mono_eigval_schur} between
$\rho$ and $\alpha_{b,n}$ gives
\[
c|\alpha_{b,n}-\rho|
\leq|\kappa_{b,n}(\rho)|
\leq\|S_n(\rho)\|.
\]
Thus, by \eqref{eq:block_schur_at_outlier},
\[
\max_{b\in[k]}|\alpha_{b,n}-\rho|
\leq c^{-1}\|S_n(\rho)\|
=O_\P(n^{-1/2}).
\]

It remains to expand $S_n$ at these random roots.
Set $d:=m_{\rm sc}'(\rho)=(\vartheta^2-1)^{-1}.$ For each $b\in[k]$, the integral identity gives
\be\la{eq:linearization_integral_identity}
S_n(\alpha_{b,n})
=
S_n(\rho)
+
(\alpha_{b,n}-\rho)
\int_0^1
S_n'\bigl(\rho+t(\alpha_{b,n}-\rho)\bigr)\,dt.
\ee
Define
\[
\eta_n
:=
\max_{b\in[k]}\sup_{0\leq t\leq1}
\left\|
S_n'\bigl(\rho+t(\alpha_{b,n}-\rho)\bigr)-dI_k
\right\|.
\]
By the triangle inequality,
\[
\begin{aligned}
\eta_n
\leq{}&
\sup_{x\in I}
\|S_n'(x)-m_{\rm sc}'(x)I_k\|\\
&+
\max_{b\in[k]}\sup_{0\leq t\leq1}
\left|
m_{\rm sc}'\bigl(\rho+t(\alpha_{b,n}-\rho)\bigr)
-m_{\rm sc}'(\rho)
\right|.
\end{aligned}
\]
The first term tends to zero almost surely by
\eqref{eq:block_schur_derivative}.
The second tends to zero in probability by continuity of
$m_{\rm sc}'$ and
$\max_b|\alpha_{b,n}-\rho|=O_\P(n^{-1/2})$.
Hence $\eta_n=o_\P(1)$. Multiplying \eqref{eq:linearization_integral_identity} by
$\sqrt n$ and separating the leading terms gives
\[
\sqrt n\,S_n(\alpha_{b,n})
=
R_n+d\sqrt n(\alpha_{b,n}-\rho)I_k+\mathfrak e_{b,n},
\]
where we define the remainder by
\[
\begin{aligned}
\mathfrak e_{b,n}
:={}&\sqrt n\,S_n(\rho)-R_n\\
&+\sqrt n(\alpha_{b,n}-\rho)
\int_0^1
\left[
S_n'\bigl(\rho+t(\alpha_{b,n}-\rho)\bigr)-dI_k
\right]dt.
\end{aligned}
\]
By the definition of $\eta_n$, the norm of the integral is at
most $\eta_n$, uniformly over $b\in[k]$. Using
\eqref{eq:block_schur_at_outlier},
$\max_b|\alpha_{b,n}-\rho|=O_\P(n^{-1/2})$, and
$\eta_n=o_\P(1)$, we obtain
\[
\begin{aligned}
\max_{b\in[k]}\|\mathfrak e_{b,n}\|
&\leq
\|\sqrt n\,S_n(\rho)-R_n\|
+
\sqrt n\max_{b\in[k]}|\alpha_{b,n}-\rho|\,\eta_n\\
&=
o_\P(1)+O_\P(1)o_\P(1)
=o_\P(1).
\end{aligned}
\]
The $b$-th largest eigenvalue of
$\sqrt n\,S_n(\alpha_{b,n})$ is zero.
The $b$th largest eigenvalue of
$R_n+d\sqrt n(\alpha_{b,n}-\rho)I_k$ is
$\lambda_b(R_n)+d\sqrt n(\alpha_{b,n}-\rho)$.
Weyl's inequality therefore yields
\[
\left|
\lambda_b(R_n)+d\sqrt n(\alpha_{b,n}-\rho)
\right|
\leq\|\mathfrak e_{b,n}\|.
\]
Consequently,
\[
\max_{b\in[k]}
\left|
\sqrt n(\alpha_{b,n}-\rho)+d^{-1}\lambda_b(R_n)
\right|
\longrightarrow0
\qquad\text{in probability}.
\]
Finally, since $d>0$, multiplication by $-d^{-1}$ reverses the
order of the eigenvalues:
\[
\lambda_a(-d^{-1}R_n)
=
-d^{-1}\lambda_{k-a+1}(R_n),
\qquad a\in[k].
\]
Combining this identity with \eqref{eq:block_ordered_roots}
and $d^{-1}=\vartheta^2-1$ proves
\eqref{eq:outlier_linearization} for the chosen block.
There are finitely many blocks, so a finite union bound gives
the simultaneous conclusion.
\end{proof}

\medskip

We can now combine Lemmas \ref{lem:joint_gaussian_resolvent_fluctuations} and \ref{lem:outlier_linearization} to conclude the proof of Theorem \ref{thm:supercritical_gaussian_fluctuations}.

{\it Proof of Theorem~\ref{thm:supercritical_gaussian_fluctuations}.}

We give the proof in the normalized case $\sigma=1$. By
Lemma~\ref{lem:joint_gaussian_resolvent_fluctuations}, jointly,
\[
\bigl(- (\vartheta_\ell^2-1)R_{\ell,n}\bigr)_{\ell=1}^s
\Longrightarrow
\bigl(\sqrt{1-\vartheta_\ell^{-2}}\,\mathcal G_\ell\bigr)_{\ell=1}^s,
\]
where the $\mathcal G_\ell$ are independent standard GUE matrices. The ordered-eigenvalue maps are continuous by Weyl's inequality. Lemma~\ref{lem:outlier_linearization}, the continuous mapping theorem, and Slutsky's theorem therefore give the asserted joint cluster limits.
$\hfill\Box$

\subsection{Soft-edge fluctuations}
\la{app:proof_tw}

We establish the Tracy--Widom edge limit used in Theorem \ref{thm:eigenvalue_fluctuations_ext}. The theorem below holds without a simplicity assumption on the supercritical signal strengths.

\bTh\la{thm:tw_deformations}
Suppose Assumptions \ref{ass:ass1} and \ref{ass:ass2} hold with $\eps_{ij}\sim\mathcal N(0,\sigma^2)$. Assume, moreover, that either $m=r$ or $\theta_{m+1} < \sigma$, where $m=\#\{j\in[r]:\theta_j>\sigma\}$. Then
\be\la{eq:tw_thm_formula}
    \frac{n^{2/3}}{\sigma}
    \left(\lambda_{m+1}-2\sigma\right)
    \Longrightarrow \TW_2
    \qquad \text{and} \qquad
    \frac{n^{2/3}}{\sigma}
    \left(\lambda_{n-m}+2\sigma\right)
    \Longrightarrow -\TW_2.
\ee
The cases $m=0$ and $m=r$ correspond, respectively, to all spikes being subcritical and all positive spikes being supercritical.
\eTh

For the proof below, we work under the hypotheses of Theorem \ref{thm:tw_deformations} in the normalized case $\sigma=1$. The auxiliary results are stated with their own assumptions. We first establish the Tracy--Widom limit for the unperturbed spectral edge and the resolvent and eigenvector estimates needed for the fully subcritical regime, in which $\theta_1<1$. We then prove Theorem \ref{thm:tw_deformations} when $m=0$ and extend the argument to $m\geq1$ using finite-rank spectral inequalities and a projection argument.

\bpropo\la{propo:tw_no_deformations}
Under Assumption \ref{ass:ass1} with $\sigma=1$ and $\eps_{ij}\sim\mathcal N(0,1)$,
$$
    n^{2/3}\left(\xi_1-2\right)
    \Longrightarrow
    \TW_2.
$$
\epropo

\begin{proof}
Since $\mathcal E_n$ is real skew-symmetric, the nonzero eigenvalues of $\ii\mathcal E_n=\sqrt n\,W_n$ come in opposite pairs. More precisely, if $n=2p$, then
$$
    \spec(\ii\mathcal E_n)=
    \{\pm\ell_{1,n},\dots, \pm\ell_{p,n}\},
$$
whereas if $n=2p+1$, then
$$
    \spec(\ii\mathcal E_n) =
    \{\pm\ell_{1,n},\dots, \pm\ell_{p,n}, 0\},
$$
where, in both cases, $\ell_{1,n}>\cdots>\ell_{p,n}>0.$ By \citet[Exercise 1.3.5]{forrester2010log}, 
the joint density of the positive eigenvalues $(\ell_{1,n},\dots,\ell_{p,n})$ is, up to a normalizing constant,
\be
    \prod_{j=1}^{p} e^{-\ell_{j,n}^2/2}
    \prod_{1\leq j<k\leq p}
    \left(\ell_{j,n}^2-\ell_{k,n}^2\right)^2,
    \qquad n=2p,
    \la{eq:even-positive-density}
\ee
and
\be
    \prod_{j=1}^{p} \ell_{j,n}^2 e^{-\ell_{j,n}^2/2}
    \prod_{1\leq j<k\leq p}
    \left(\ell_{j,n}^2-\ell_{k,n}^2\right)^2,
    \qquad n=2p+1.
    \la{eq:odd-positive-density}
\ee

Define $X_{j,n}=\ell_{j,n}^2/2$ for $j \in [p].$ Then $d\ell_{j,n}=(2X_{j,n})^{-1/2}dX_{j,n}$. Therefore, by \eqref{eq:even-positive-density} and \eqref{eq:odd-positive-density}, $(X_{1,n},\dots,X_{p,n})$ has joint density proportional to
$$
    \prod_{1\leq j<k\leq p}(X_{j,n}-X_{k,n})^2
    \prod_{j=1}^{p} X_{j,n}^{\gamma_n}e^{-X_{j,n}},
$$
where $\gamma_n=-1/2$ for even $n$ and $\gamma_n=1/2$ for odd $n$. Thus, along each parity subsequence, $(X_{1,n},\ldots,X_{p,n})$ has the joint eigenvalue density of the standard, unscaled Laguerre Unitary Ensemble (LUE) of size $p$ and parameter $\gamma_n>-1$. Let $\lambda^{\mathrm L}_{p,\gamma}$ denote the largest eigenvalue of the standard, unscaled LUE of size $p\ge1$ and parameter $\gamma>-1$. In the notation above, $\lambda^{\mathrm L}_{p,\gamma_n}$ has the same distribution as $X_{1,n}$. Denote $P(p,\gamma,t) = \mathbb P\left(\lambda^{\mathrm L}_{p,\gamma}\leq t\right)$ for $t \in \R.$ For every fixed $\gamma>-1$ and every fixed $t\in \mathbb R$, \citet[Corollary 3.1]{bornemann2026asymptotic} implies
$$
    P\left( p,\gamma,
        4p+2\gamma+2(2p)^{1/3}t
    \right)
    \underset{p\to\infty}{\longrightarrow} F_2(t),
$$
where $F_2$ is the distribution function of the $\TW_2$ law. Applying this result with $\gamma=-1/2$ along the even subsequence $n=2p$, and with $\gamma=1/2$ along the odd subsequence $n=2p+1$, gives
\be\la{eq:tw_interm_term}
    Z_n = \frac{
        X_{1,n}-(4p+2\gamma_n)
    }{
        2(2p)^{1/3}
    }
    \Longrightarrow
    \TW_2 .
\ee
Since $4p+2\gamma_n=2n-1$, we have $X_{1,n}=2n-1+2(2p)^{1/3}Z_n$. Using $\xi_1=\sqrt{2X_{1,n}/n}$ and rationalizing gives
$$
\begin{aligned}
    n^{2/3}(\xi_1-2)
    &=
    \frac{2n^{-1/3}(X_{1,n}-2n)}{\xi_1+2}\\
    &=
    \frac{4(2p/n)^{1/3}}{\xi_1+2}Z_n
    -
    \frac{2n^{-1/3}}{\xi_1+2}.
\end{aligned}
$$
Now $2p/n\to 1$, $\xi_1\to 2$ in probability by Lemma \ref{lem:stielt-cv}, and $Z_n\Longrightarrow \TW_2$ by \eqref{eq:tw_interm_term}. Hence, by Slutsky's theorem, $n^{2/3}\left(\xi_1-2\right)
    \Rightarrow
    \TW_2.$
\end{proof}

\begin{lem}
\la{lem:edge_resolvent_compression}
Suppose Assumptions \ref{ass:ass3} and \ref{ass:ass2} hold with $\sigma=1$. For $t\in\R$, set $E_n(t)=2+n^{-2/3}t.$ Then, for every fixed $x\in\R$, every $\varepsilon>0$, and every $\delta>0$,
\be
\la{eq:edge_resolvent_compression}
    \mathbb P\left(
        \xi_1\leq E_n(x-\varepsilon),
        \quad
        \sup_{E\geq E_n(x)}
        \left\|
            U_n^*(EI-W_n)^{-1}U_n
        \right\|
        >1+\delta
    \right)
    \longrightarrow 0.
\ee
\end{lem}

\begin{proof}
Fix $x\in\R$, $\varepsilon>0$, and $\delta>0$. Let $\mathcal A_n=
    \left\{
        \xi_1\leq E_n(x-\varepsilon)
    \right\}.$ Set $E_0=E_n(x)$ and, for $E>\xi_1$, define
$$
    K_n(E)=U_n^*(EI-W_n)^{-1}U_n=-M_n(E).
$$
The matrix $K_n(E)$ is positive semidefinite and satisfies
$$
    K_n'(E)
    =
    -U_n^*(EI-W_n)^{-2}U_n
    \preceq0.
$$
Consequently, on $\mathcal A_n$,
$$
    0\preceq K_n(E)\preceq K_n(E_0),
    \qquad E\geq E_0,
$$
and hence
\be\la{eq:endpoint_identity}
    \sup_{E\geq E_n(x)}
    \left\|U_n^*(EI-W_n)^{-1}U_n\right\|
    =
    \|K_n(E_0)\|.
\ee

Choose $\kappa>0$ small enough so that
\be
\la{eq:rho_choice}
    (1+\kappa^2)\left(1+\frac{\delta}{2}\right)
    \leq 1+\delta.
\ee
Set $\eta_n=\kappa\varepsilon n^{-2/3}$ and
$z_n=E_0+\ii\eta_n$. On the event $\mathcal A_n$, for every
eigenvalue $\xi_k$, we have
$$
    E_0-\xi_k
    \geq E_0-\xi_1
    \geq E_n(x)-E_n(x-\varepsilon)
    =
    \varepsilon n^{-2/3}.
$$
Therefore, $0<\eta_n/(E_0-\xi_k)\leq\kappa$. Moreover, for every
$a>0$ such that $\eta_n/a\leq\kappa$, we have
$$
    \frac1a
    \leq
    (1+\kappa^2)\frac{a}{a^2+\eta_n^2}
    =
    (1+\kappa^2)\Re\frac{1}{a+\ii\eta_n}.
$$
Applying this with $a=E_0-\xi_k$ and compressing to the columns
of $U_n$ gives, on $\mathcal A_n$,
$$
    0\preceq K_n(E_0)
    \preceq
    -(1+\kappa^2)\Re M_n(z_n).
$$
Hence,
\begin{align}
    \|K_n(E_0)\|
    &\leq
    (1+\kappa^2)\|\Re M_n(z_n)\| \notag\\
    &\leq
    (1+\kappa^2)\|M_n(z_n)\|.
    \label{eq:real_resolvent_regularized}
\end{align}
We now apply the local law at the single spectral parameter
$z_n$. Let $W_n^{\rm std}$ and $G_n^{\rm std}$ be as in the
proof of Proposition \ref{prop:isotrop}, and set
$c_n=\sqrt{n/(n-1)}$. Since $W_n^{\rm std}=c_nW_n$, we have
\be\la{eq:scaling_identity}
    G_n(z_n)=c_nG_n^{\rm std}(c_nz_n).
\ee
Fix $\tau\in(0,1/3)$. For all large $n$,
$c_nz_n\in\mathcal D_n(\tau,3)$ and
$n\Im(c_nz_n)\asymp n^{1/3}$. Since $m_{\rm sc}$ is uniformly
bounded on this domain, the right-hand side of
\eqref{eq:isotropic_vector_bound_sd} is $O(n^{-1/6})$ at
$c_nz_n$. Applying this estimate to every pair of columns
of $U_n$, whose number is fixed, gives
$$
    \left\|
        U_n^*
        \left(
            G_n^{\rm std}(c_nz_n)-m_{\rm sc}(c_nz_n)I
        \right)
        U_n
    \right\|
    \longrightarrow0
    \qquad\text{in probability}.
$$
Moreover, $c_n\to1$, $c_nz_n\to2$, and continuity of
$m_{\rm sc}$ up to the real axis gives
$m_{\rm sc}(c_nz_n)\to-1$. Together with
$U_n^*U_n=I_{2r}$ and \eqref{eq:scaling_identity}, this yields
\be
\la{eq:compressed_complex_resolvent}
    \|M_n(z_n)+I_{2r}\|
    \longrightarrow0
    \qquad\text{in probability}.
\ee
Using \eqref{eq:endpoint_identity}, \eqref{eq:rho_choice}, and
\eqref{eq:real_resolvent_regularized}, we obtain
$$
\begin{aligned}
    &\mathbb P\left(
        \mathcal A_n,\quad
        \sup_{E\geq E_n(x)}
        \left\|U_n^*(EI-W_n)^{-1}U_n\right\|
        >1+\delta
    \right)\\
    &\qquad\leq
    \mathbb P\left(
        \|M_n(z_n)\|>1+\frac{\delta}{2}
    \right)
    \longrightarrow0,
\end{aligned}
$$
where the convergence follows from
\eqref{eq:compressed_complex_resolvent}.
This proves \eqref{eq:edge_resolvent_compression}.
\end{proof}

\begin{lem}
\la{lem:edge_overlap}
Suppose Assumptions \ref{ass:ass1} and \ref{ass:ass2} hold with $\sigma=1$ and $\eps_{ij}\sim\mathcal N(0,1)$. Let $q_{1,n}$ be a unit eigenvector of $W_n$ associated with $\xi_1$. Then
$$
    n^{2/3}
    \left|
        q_{1,n}^*U_n\Theta U_n^*q_{1,n}
    \right|
    \longrightarrow 0
    \qquad \text{in probability.}
$$
\end{lem}
\begin{proof}
Let $\Pi_{1,n}=\Pi_{1}(W_n)=q_{1,n}q_{1,n}^*$ be the spectral projection associated with the largest eigenvalue $\xi_1$. Since the entries of $\mathcal E_n$ above the diagonal are nondegenerate Gaussian random variables, the largest positive eigenvalue of $W_n$ is simple almost surely, so $\Pi_{1,n}$ is well-defined almost surely.

We first claim that $\E \Pi_{1,n}=I_n/n.$ Indeed, by Gaussian orthogonal invariance, $OW_nO^\top \stackrel{d}{=} W_n,$ for all $O\in {\rm O}(n).$ Moreover, spectral projections are equivariant under orthogonal conjugation, $\Pi_{1}(OW_nO^\top)=O\Pi_{1}(W_n)O^\top.$ Taking expectations and using the equality in distribution gives $O\E\Pi_{1,n}O^\top=\E\Pi_{1,n}$ for all $O\in {\rm O}(n).$ Therefore $\E\Pi_{1,n}$ commutes with every orthogonal matrix, which implies that $\E\Pi_{1,n}$ is a scalar multiple of the identity. Since $\operatorname{Tr}\Pi_{1,n}=1$, we get $\operatorname{Tr}\E\Pi_{1,n}=1$, which proves the claim. Hence, 
$$
\begin{aligned}
    \E\|U_n^*q_{1,n}\|^2
    =
    \operatorname{Tr}
    \left(
        U_n^*(\E \Pi_{1,n})U_n
    \right) 
    =
    \frac1n
    \operatorname{Tr}(U_n^*U_n) 
    =
    \frac{2r}{n}.
\end{aligned}
$$

Moreover, $\left|
        q_{1,n}^*U_n\Theta U_n^*q_{1,n}
    \right|
    \leq
    \|\Theta\|\,
    \|U_n^*q_{1,n}\|^2$
where $\|\Theta\|=\theta_1 $. Hence, $$\E
    \left|
         q_{1,n}^*U_n\Theta U_n^*q_{1,n}
    \right|
    \leq
    \frac{2\theta_1 r}{n}.$$ The conclusion follows from Markov’s inequality.
\end{proof}

\medskip

We can now combine the unperturbed Tracy--Widom limit with the preceding perturbation estimates to prove Theorem \ref{thm:tw_deformations}.

{\it Proof of Theorem \ref{thm:tw_deformations}.}

We give the proof in the normalized case $\sigma=1$. Let $q$ be a unit eigenvector of $W_n$ associated with $\xi_1$. We suppress its dependence on $n$ for notational simplicity. Set $Q_n=\ii S_n=U_n\Theta U_n^*$. We first assume $m=0$, so that $\|\Theta\|=\theta_1<1$. Fix $x\in\R$ and $\varepsilon>0$. Choose $\delta>0$ such that $\theta_1(1+\delta)<1$. For $t\in\R$, let $E_n(t)=2+n^{-2/3}t$. Define the events $\mathcal A_n=
    \left\{
        \xi_1\leq E_n(x-\varepsilon)
    \right\}$ and
$$
    \mathcal K_n =
    \left\{
        \sup_{E\geq E_n(x)}
        \left\|
            U_n^*(EI-W_n)^{-1}U_n
        \right\|
        \leq 1+\delta
    \right\}.
$$
We claim that, for all large $n$,
\be
\la{eq:event_inclusion_lower}
    \mathcal A_n\cap \mathcal K_n
    \subset
    \left\{
        \lambda_1\leq E_n(x)
    \right\}.
\ee
Indeed, assume that $\mathcal A_n\cap \mathcal K_n$ holds and suppose, for contradiction, that $\lambda_1>E_n(x)$. Since $\mathcal A_n$ holds, we have $\lambda_1>E_n(x)>E_n(x-\varepsilon)\geq \xi_1.$ Therefore, by Lemma \ref{lem:general_det_criterion}, $\det( I_{2r} + \Theta M_n(\lambda_1))=0.$ It follows that $1 \leq \| \Theta M_n(\lambda_1)\|.$ On the other hand, since $\lambda_1>E_n(x)$ and $\mathcal K_n$ holds, $\|\Theta M_n(\lambda_1)\| \leq
    \theta_1(1+\delta)
    <1.$ This contradiction proves
\eqref{eq:event_inclusion_lower}.

Taking probabilities in \eqref{eq:event_inclusion_lower}, we get $\mathbb P(\lambda_1\leq E_n(x))\geq
    \mathbb P(\mathcal A_n\cap \mathcal K_n) =
    \mathbb P(\mathcal A_n)-\mathbb P(\mathcal A_n\cap \mathcal K_n^c).$
By Lemma \ref{lem:edge_resolvent_compression}, $\mathbb P(\mathcal A_n\cap \mathcal K_n^c)=o(1).$ Therefore, $\mathbb P(
        \lambda_1\leq E_n(x))\geq
    \mathbb P(
        \xi_1\leq E_n(x-\varepsilon))-o(1).$
Moreover, Proposition \ref{propo:tw_no_deformations} implies $\mathbb P\left(
        \xi_1\leq E_n(x-\varepsilon)
    \right) \to F_2(x-\varepsilon).$ Hence,
\be
\la{eq:liminf_subcritical}
    \liminf_{n\to\infty}
    \mathbb P(
        \lambda_1\leq E_n(x))\geq
    F_2(x-\varepsilon).
\ee
We next establish an upper bound for the limit superior of $\mathbb P(\lambda_1\leq E_n(x))$. By the variational characterization of the largest eigenvalue, $\lambda_1
    \geq q^*H_nq
    =\xi_1+q^*Q_nq.$ On the event
$\{\xi_1>E_n(x+\varepsilon),\, \lambda_1\leq E_n(x)\}$, this implies
$$
    q^*Q_nq
    \leq\lambda_1-\xi_1
    <E_n(x)-E_n(x+\varepsilon)
    =-\varepsilon n^{-2/3}.
$$
Consequently, by Lemma \ref{lem:edge_overlap},
$$
    \mathbb P\left(
        \xi_1>E_n(x+\varepsilon),\,
        \lambda_1\leq E_n(x)
    \right)\leq
    \mathbb P\left(
        |q^*Q_nq|>\varepsilon n^{-2/3}
    \right)
    =o(1).
$$
Therefore, $\mathbb P( \lambda_1\leq E_n(x)) \leq \mathbb P(\xi_1\leq E_n(x+\varepsilon)) +o(1).$ Proposition \ref{propo:tw_no_deformations} implies $\mathbb P(\xi_1\leq E_n(x+\varepsilon)) \to F_2(x+\varepsilon).$ Hence,
\be
\la{eq:limsup_subcritical}
    \limsup_{n\to\infty}
    \mathbb P(
        \lambda_1\leq E_n(x))\leq
    F_2(x+\varepsilon).
\ee
Combining \eqref{eq:liminf_subcritical} and \eqref{eq:limsup_subcritical} and letting $\varepsilon\to0$, we obtain $\mathbb P(\lambda_1\leq E_n(x))\to F_2(x)$ by continuity of $F_2$. This proves the positive-edge assertion when $m=0$.

Assume now that $1\leq m\leq r$. Define
$$
    B_n=W_n+
    \sum_{j=m+1}^r\theta_j
    \left(w_{j,+}w_{j,+}^*-w_{j,-}w_{j,-}^*\right),
    \qquad b_n=\lambda_1(B_n),
$$
where an empty sum is understood to be zero, and set
$$
    Q_{+,n}=\sum_{j=1}^m\theta_jw_{j,+}w_{j,+}^*,
    \qquad
    Q_{-,n}=\sum_{j=1}^m\theta_jw_{j,-}w_{j,-}^*.
$$
Thus $H_n=B_n+Q_{+,n}-Q_{-,n}$, where both $Q_{+,n}$ and $Q_{-,n}$ are positive semidefinite and have rank $m$. Since $H_n\preceq B_n+Q_{+,n}$ and $\dim\ker Q_{+,n}=n-m$, Lemma \ref{lem:min-max} yields
\begin{align}
    \lambda_{m+1}
    &\leq\lambda_{m+1}(B_n+Q_{+,n}) \notag\\
    &\leq
    \max_{\substack{x\in\ker Q_{+,n}\\ \|x\|=1}}
    x^*(B_n+Q_{+,n})x \notag\\
    &=
    \max_{\substack{x\in\ker Q_{+,n}\\ \|x\|=1}}
    x^*B_nx
    \leq\lambda_1(B_n)=b_n.
    \la{eq:ineq_b_n}
\end{align}
Since the deformation in $B_n$ contains only subcritical spike pairs, the subcritical case proved above (or Proposition \ref{propo:tw_no_deformations} when $m=r$, in which case $B_n=W_n$) gives
\be\la{eq:tw_b_n}
    n^{2/3}(b_n-2)\Longrightarrow\TW_2.
\ee

We next derive a lower bound for $\lambda_{m+1}$ using the top eigenvector $q$. The proof of Lemma \ref{lem:edge_overlap} establishes $\E(qq^*)=I_n/n$. Hence,
$$
    \E|q^*Q_nq|\leq\frac{2r\theta_1}{n},
    \qquad
    \E\|Q_nq\|^2
    =\frac{\Tr(\Theta^2)}{n}
    =\frac{2}{n}\sum_{j=1}^r\theta_j^2.
$$
By Markov's inequality, both $|q^*Q_nq|$ and $\|Q_nq\|^2$ are $O_{\P}(n^{-1})$.

Theorem \ref{thm:general_random_skew_outliers} identifies the first $m$ positive outliers, counting multiplicities. Together with Proposition \ref{propo:tw_no_deformations}, it gives
$$
    \lambda_m-\xi_1
    \longrightarrow\rho_m-2
    =\frac{(\theta_m-1)^2}{\theta_m}>0
    \qquad\text{in probability}.
$$
Choose $c\in(0,\rho_m-2)$. Let $\widehat w_{1,n},\ldots,\widehat w_{m,n}$ be orthonormal eigenvectors of $H_n$ associated with $\lambda_1,\ldots,\lambda_m$, and set
$$
    d_{j,n}=\lambda_j-\xi_1,
    \qquad
    \beta_{j,n}=\widehat w_{j,n}^*q,
    \qquad
    a_n=\sum_{j=1}^m|\beta_{j,n}|^2.
$$
With probability tending to one, $d_{j,n}\geq c$ for every $j\in[m]$. The eigenvalue equations imply
\be\la{eq:outlier_edge_overlap_identity}
d_{j,n}\beta_{j,n}=\widehat w_{j,n}^*Q_nq,
\qquad j\in[m].
\ee
On the event $\{\min_{j\in[m]}d_{j,n}\geq c\}$, orthonormality gives
$$
    a_n\leq c^{-2}
    \sum_{j=1}^m|\widehat w_{j,n}^*Q_nq|^2
    \leq c^{-2}\|Q_nq\|^2.
$$
In particular, $a_n=O_{\P}(n^{-1})$. On the event $\{\min_{j\in[m]}d_{j,n}\geq c,\ a_n\leq1/2\}$, which has probability tending to one, let
$$
    \widetilde q_n=
    \frac{q-\sum_{j=1}^m\beta_{j,n}\widehat w_{j,n}}
         {\sqrt{1-a_n}}.
$$
This is a unit vector orthogonal to the first $m$ eigenvectors of $H_n$. Hence $\lambda_{m+1}\geq \widetilde q_n^*H_n\widetilde q_n$. Using \eqref{eq:outlier_edge_overlap_identity}, we obtain
$$
\begin{aligned}
    \widetilde q_n^*H_n\widetilde q_n
    &=\frac{q^*H_nq
        -\sum_{j=1}^m\lambda_j|\beta_{j,n}|^2}
        {1-a_n}\\
    &=\xi_1+
      \frac{q^*Q_nq-
        \sum_{j=1}^m
        |\widehat w_{j,n}^*Q_nq|^2/d_{j,n}}
        {1-a_n}\\
    &\geq\xi_1-2|q^*Q_nq|
        -\frac{2}{c}\|Q_nq\|^2.
\end{aligned}
$$
The two error terms in the last line are $O_{\P}(n^{-1})$, hence $o_{\P}(n^{-2/3})$. As in the case $m=0$, the preceding lower bound, valid with probability tending to one, implies that, for every fixed $x\in\R$ and $\varepsilon>0$,
$$
    \mathbb P\left(
        \xi_1>E_n(x+\varepsilon),\,
        \lambda_{m+1}\leq E_n(x)
    \right)=o(1).
$$
Combining this estimate with \eqref{eq:ineq_b_n} gives
$$
\begin{aligned}
    \mathbb P(b_n\leq E_n(x))
    &\leq\mathbb P(\lambda_{m+1}\leq E_n(x))\\
    &\leq\mathbb P(\xi_1\leq E_n(x+\varepsilon))+o(1).
\end{aligned}
$$
Using \eqref{eq:tw_b_n} and Proposition \ref{propo:tw_no_deformations}, we obtain
$$
\begin{aligned}
    F_2(x)
    &\leq\liminf_{n\to\infty}
      \mathbb P(\lambda_{m+1}\leq E_n(x))\\
    &\leq\limsup_{n\to\infty}
      \mathbb P(\lambda_{m+1}\leq E_n(x))
      \leq F_2(x+\varepsilon).
\end{aligned}
$$
Letting $\varepsilon\to0$ proves the positive-edge assertion. Finally, exact spectral symmetry gives $\lambda_{n-m}=-\lambda_{m+1}$. This proves the negative-edge assertion for every $0\leq m\leq r$ and completes the proof.
$\hfill\Box$

\subsection{Joint outlier and soft-edge fluctuations:
proof of Theorem~\ref{thm:eigenvalue_fluctuations_ext}}
\la{app:proof_joint_fluctuations}

We now combine the preceding fluctuation results and prove
asymptotic independence by conditioning on the common noise minor
used in the outlier fluctuation argument.

When $m=0$, the conclusion follows from Theorem~\ref{thm:tw_deformations} and spectral symmetry. For $m\geq1$, the outlier assertion follows from Theorem~\ref{thm:supercritical_gaussian_fluctuations}. It remains to prove the joint assertion under the additional assumption that either $m=r$ or $\theta_{m+1}<\sigma$. We therefore assume $m\geq1$ and work in the normalized case $\sigma=1$.
Recall that $\vartheta_1>\cdots>\vartheta_s>1$ are the distinct
supercritical strengths, with multiplicities $k_1,\ldots,k_s$,
and that
\[
m=\sum_{\ell=1}^s k_\ell,\qquad
p_\ell=\sum_{h<\ell}k_h,\qquad
\rho_\ell^\star=\vartheta_\ell+\vartheta_\ell^{-1},\qquad
\mathcal J_\ell=\{p_\ell+1,\ldots,p_\ell+k_\ell\}.
\]

Choose the deterministic orthogonal matrix $O_n$ used in the
proof of Lemma~\ref{lem:joint_gaussian_resolvent_fluctuations},
so that $O_ne_{2j-1}=u_j$ and $O_ne_{2j}=v_j$ for every
$j\in[m]$.
Replacing $(W_n,S_n)$ by
$(O_n^\top W_nO_n,O_n^\top S_nO_n)$ preserves the spectra
of both $W_n$ and $H_n$ and, by Gaussian orthogonal invariance,
preserves the noise law. We may therefore assume that
$u_j=e_{2j-1}$ and $v_j=e_{2j}$ for every $j\in[m]$.
All the random variables below are constructed from this same
rotated model.

Set $N_n=n-2m$ and let $W_n^\circ$ be the principal minor of
$W_n$ obtained by deleting the first $2m$ rows and columns,
as in the proof of
Lemma~\ref{lem:joint_gaussian_resolvent_fluctuations}.
Define
\[
T_n=n^{2/3}(\lambda_{m+1}-2),
\qquad
T_n^\circ=n^{2/3}\bigl(\lambda_1(W_n^\circ)-2\bigr).
\]
We first show that
\be\la{eq:joint_edge_minor_comparison}
T_n-T_n^\circ\longrightarrow0
\qquad\text{in probability}.
\ee

The matrix
$\widetilde W_n^\circ=\sqrt{n/N_n}\,W_n^\circ$
has the normalized Gaussian skew-symmetric noise law in
dimension $N_n$. By Proposition~\ref{propo:tw_no_deformations},
\[
N_n^{2/3}
\bigl(\lambda_1(\widetilde W_n^\circ)-2\bigr)
\Longrightarrow\TW_2.
\]
Moreover,
\[
T_n^\circ
=
\left(\frac{n}{N_n}\right)^{1/6}
N_n^{2/3}
\bigl(\lambda_1(\widetilde W_n^\circ)-2\bigr) +
2n^{2/3}\left(\sqrt{\frac{N_n}{n}}-1\right).
\]
Since $m$ is fixed, the first coefficient tends to one and
the last term is $O(n^{-1/3})$. Thus
$T_n^\circ\Longrightarrow\TW_2$.

Let $q_n$ be a unit eigenvector of $W_n$ associated with
$\xi_1$, and fix
$c\in(0,\rho_s^\star-2)$.
The projection argument in the proof of
Theorem~\ref{thm:tw_deformations} gives, on an event of
probability tending to one,
\[
\lambda_{m+1}\geq\xi_1-\mathfrak e_n,
\qquad
\mathfrak e_n
:=
2|q_n^*(\ii S_n)q_n|
+\frac{2}{c}\|(\ii S_n)q_n\|^2
=O_\P(n^{-1}).
\]
By eigenvalue interlacing,
$\lambda_1(W_n^\circ)\leq\xi_1$. Consequently, on the same event,
\[
0\leq(T_n^\circ-T_n)_+
\leq n^{2/3}\mathfrak e_n
=O_\P(n^{-1/3}),
\]
where $x_+=\max\{x,0\}$. Hence
\be\la{eq:tw2_pos_part}
(T_n^\circ-T_n)_+\longrightarrow0
\qquad\text{in probability}.
\ee

On the other hand, Theorem~\ref{thm:tw_deformations} gives
$T_n\Longrightarrow\TW_2$.
Since both coordinates converge in distribution, the pairs
$(T_n^\circ,T_n)$ are tight.
Let $(S,S')$ be any subsequential joint limit.
Then $S$ and $S'$ both have distribution $\TW_2$.
The continuous mapping theorem and \eqref{eq:tw2_pos_part}
give $(S-S')_+=0$ almost surely, so $S\leq S'$ almost surely.
Since $\arctan$ is bounded and strictly increasing,
\[
\arctan(S')-\arctan(S)\geq0
\quad\text{almost surely},
\qquad
\E[\arctan(S')-\arctan(S)]=0.
\]
Thus $S=S'$ almost surely.
Every subsequence therefore has a further subsequence along
which the continuous mapping theorem gives
$T_n-T_n^\circ\Longrightarrow S'-S=0$.
By the subsequence criterion, the full sequence converges in
distribution to zero. Since the limit is deterministic, this
is equivalent to convergence in probability, proving
\eqref{eq:joint_edge_minor_comparison}.

We use the matrices $A_{\ell,n}$ and $Q_{\ell,n}$ from the
proof of Lemma~\ref{lem:joint_gaussian_resolvent_fluctuations},
constructed from this same rotated model and the same minor
$W_n^\circ$. Set $\mathcal F_n=\sigma(W_n^\circ)$ and, for
each $\ell\in[s]$, define
\[
a_\ell=1-\vartheta_\ell^{-2},
\qquad
L_{\ell,n}=a_\ell(A_{\ell,n}-Q_{\ell,n}).
\]
We first relate the eigenvalues of $L_{\ell,n}$ to the outlier
fluctuations.
Equation~\eqref{eq:joint_schur_fluctuation} gives
\[
R_{\ell,n}
=
-\vartheta_\ell^{-2}(A_{\ell,n}-Q_{\ell,n})
+ r_{\ell,n},
\qquad
\max_{\ell\in[s]}\|r_{\ell,n}\|=o_\P(1).
\]
Since
$(\vartheta_\ell^2-1)\vartheta_\ell^{-2}=a_\ell$,
it follows that
\[
-(\vartheta_\ell^2-1)R_{\ell,n}
=
L_{\ell,n}
-(\vartheta_\ell^2-1) r_{\ell,n}.
\]
Weyl's inequality therefore gives
\[
\begin{aligned}
\max_{1\leq b\leq k_\ell}
\bigl|
\lambda_b\bigl(-(\vartheta_\ell^2-1)R_{\ell,n}\bigr)
-\lambda_b(L_{\ell,n})
\bigr|
\leq
(\vartheta_\ell^2-1)\| r_{\ell,n}\|.
\end{aligned}
\]
Combining this estimate with
\eqref{eq:outlier_linearization}, and using that the number of
blocks is fixed, yields
\be\la{eq:joint_outlier_leading_terms}
\max_{\substack{\ell\in[s]\\1\leq b\leq k_\ell}}
\left|
\sqrt n(\lambda_{p_\ell+b}-\rho_\ell^\star)
-\lambda_b(L_{\ell,n})
\right|
\longrightarrow0
\qquad\text{in probability}.
\ee
Fix deterministic Hermitian matrices
$B_\ell\in\C^{k_\ell\times k_\ell}$, $\ell\in[s]$, and write
$\mathbf B=(B_1,\ldots,B_s)$. Define
\[
\begin{aligned}
\Phi_n(\mathbf B)
&:=
\E\left[\left.
\exp\left(\ii\sum_{\ell=1}^s\Tr(B_\ell L_{\ell,n})\right)
\right|\mathcal F_n\right],\\
\psi(\mathbf B)
&:=
\exp\left(
-\frac12\sum_{\ell=1}^s a_\ell\Tr(B_\ell^2)
\right).
\end{aligned}
\]
Since
\[
a_\ell^2\frac{\vartheta_\ell^2}{\vartheta_\ell^2-1}
=a_\ell,
\]
applying \eqref{eq:joint_conditional_leading_cf} with
$B_\ell$ replaced by $a_\ell B_\ell$ gives
\be\la{eq:joint_conditional_cf}
\E|\Phi_n(\mathbf B)-\psi(\mathbf B)|
\longrightarrow0.
\ee
To identify the joint limit, let
$\mathcal G_1,\ldots,\mathcal G_s$ be independent standard
GUE matrices with
$\mathcal G_\ell\sim\mathrm{GUE}(k_\ell)$, and let
$T\sim\TW_2$ be independent of this entire family.
The function $\psi(\mathbf B)$ is the joint characteristic
function of
$(\sqrt{a_\ell}\,\mathcal G_\ell)_{\ell=1}^s$.
Since $T_n^\circ$ is $\mathcal F_n$-measurable, the tower
property and \eqref{eq:joint_conditional_cf} give, for every
$u\in\R$,
\[
\begin{aligned}
&\left|
\E\exp\left(
\ii\sum_{\ell=1}^s\Tr(B_\ell L_{\ell,n})+\ii uT_n^\circ
\right)
-\psi(\mathbf B)\E e^{\ii uT_n^\circ}
\right|\\
&\qquad=
\left|
\E\left[
e^{\ii uT_n^\circ}
\bigl(\Phi_n(\mathbf B)-\psi(\mathbf B)\bigr)
\right]
\right|\\
&\qquad\leq
\E|\Phi_n(\mathbf B)-\psi(\mathbf B)|
\longrightarrow0.
\end{aligned}
\]
Moreover, $T_n^\circ\Longrightarrow T$, so
$\E e^{\ii uT_n^\circ}\to\E e^{\ii uT}$.
Consequently,
\[
\E\exp\left(
\ii\sum_{\ell=1}^s\Tr(B_\ell L_{\ell,n})+\ii uT_n^\circ
\right)
\longrightarrow
\psi(\mathbf B)\E e^{\ii uT}.
\]
The right-hand side is the joint characteristic function of
$\bigl((\sqrt{a_\ell}\,\mathcal G_\ell)_{\ell=1}^s,T\bigr)$.
L\'evy's continuity theorem therefore gives
\[
\bigl((L_{\ell,n})_{\ell=1}^s,T_n^\circ\bigr)
\Longrightarrow
\bigl((\sqrt{a_\ell}\,\mathcal G_\ell)_{\ell=1}^s,T\bigr).
\]

For each $\ell\in[s]$, let
$\gamma_{\ell,1}\geq\cdots\geq\gamma_{\ell,k_\ell}$
be the ordered eigenvalues of $\mathcal G_\ell$.
The ordered-eigenvalue map on Hermitian matrices is continuous
by Weyl's inequality. Since $a_\ell>0$, the continuous mapping
theorem gives
\[
\left(
\bigl((\lambda_b(L_{\ell,n}))_{b=1}^{k_\ell}\bigr)_{\ell=1}^s,
T_n^\circ
\right)
\Longrightarrow
\left(
\bigl((\sqrt{a_\ell}\,\gamma_{\ell,b})_{b=1}^{k_\ell}\bigr)_{\ell=1}^s,
T
\right).
\]
Equations~\eqref{eq:joint_outlier_leading_terms} and \eqref{eq:joint_edge_minor_comparison}, together with Slutsky's theorem, now establish the joint convergence asserted in Theorem~\ref{thm:eigenvalue_fluctuations_ext} for $\sigma=1$. The lower-edge and negative-outlier conclusions follow from the exact identities $\lambda_{n-j+1}=-\lambda_j$ and $\lambda_{n-m}=-\lambda_{m+1}$.
\hfill$\Box$

\section{Factor recovery: proof of Theorem \ref{thm:canonical_angles_supercritical}}
\la{app:proof_cca}

We first prove Theorem~\ref{thm:canonical_angles_supercritical} under
the additional assumption that $\mathcal E_n/\sigma$ satisfies Assumption~\ref{ass:ass3}, while retaining Assumption~\ref{ass:ass2}. We then extend the conclusion under Assumption~\ref{ass:ass1} by truncation. Since division by $\sigma>0$ does not change the eigenspaces, it suffices to work with $\sigma=1$.

Fix $j\in[m]$ such that $\theta_j$ has multiplicity one.
Let $e_{j,+}$ and $e_{j,-}$ denote, respectively, the $j$-th
and $(r+j)$-th vectors of the canonical basis of
$\mathbb C^{2r}$. The corresponding diagonal entries of
$\Theta$ are $\theta_j$ and $-\theta_j$.
Since $\theta\mapsto\theta+\theta^{-1}$ is strictly increasing
on $(1,\infty)$, the outlier location $\rho_j$ is distinct
from $\rho_k$ for every $k\in[m]\setminus\{j\}$.
By the all-moments case of Theorem~\ref{thm:general_random_skew_outliers} established in the first part of Subsection~\ref{app:proof_outliers}, almost surely, $\lambda_j\to\rho_j>2$ and $\lambda_j$ is simple for all sufficiently large $n$.
For such $n$, let $\widehat w_{j,n}$ be a normalized
eigenvector of $H_n$ associated with $\lambda_j$, and let
$\widehat\Pi_{j,n}^+=\widehat w_{j,n}\widehat w_{j,n}^*$
be the corresponding spectral projection.

Choose a compact interval $I\subset(2,\infty)$ containing
$\rho_j$ in its interior. By Lemma \ref{lem:stielt-cv},
almost surely, $I$ is disjoint from $\spec(W_n)$ for all
sufficiently large $n$. Since $\lambda_j\to\rho_j$,
we also have $\lambda_j\in I$ for all sufficiently large $n$.
Lemmas \ref{lem:uniform_compressed_resolvent} and
\ref{lem:uniform_derivative_compressed_appendix} therefore give
\be\la{eq:cca_resolvent_limits}
M_n(\lambda_j)\longrightarrow-\theta_j^{-1}I_{2r},
\qquad
M_n'(\lambda_j)\longrightarrow m_{\rm sc}'(\rho_j)I_{2r}
\qquad\text{a.s.}
\ee
Set $y_n=U_n^*\widehat w_{j,n}$, so that $\|y_n\|\leq1$.
The eigenvector equation gives
\be\la{eq:cca_eigenvector_equations}
\widehat w_{j,n}=-G_n(\lambda_j)U_n\Theta y_n,
\qquad
\bigl(I_{2r}+M_n(\lambda_j)\Theta\bigr)y_n=0.
\ee
Combining \eqref{eq:cca_resolvent_limits} and
\eqref{eq:cca_eigenvector_equations}, and using $\|y_n\|\leq1$,
we obtain
$$
(I_{2r}-\theta_j^{-1}\Theta)y_n
=
-\bigl(M_n(\lambda_j)+\theta_j^{-1}I_{2r}\bigr)\Theta y_n
\longrightarrow0
\qquad\text{a.s.}
$$
Since $I_{2r}-\theta_j^{-1}\Theta$ is diagonal and has exactly
one zero entry, at the coordinate $e_{j,+}$, it follows that
\be\la{eq:cca_coordinate_alignment}
y_n-(e_{j,+}^*y_n)e_{j,+}\longrightarrow0
\qquad\text{a.s.}
\ee

Moreover, $G_n(\lambda_j)$ is Hermitian and
$M_n'(\lambda_j)=U_n^*G_n(\lambda_j)^2U_n$.
Taking squared norms in the first identity of
\eqref{eq:cca_eigenvector_equations} therefore gives
\be\la{eq:cca_normalization}
1=y_n^*\Theta M_n'(\lambda_j)\Theta y_n.
\ee
By \eqref{eq:cca_resolvent_limits} and $\|y_n\|\leq1$,
$$
\begin{aligned}
&\left|
y_n^*\Theta
\bigl(M_n'(\lambda_j)-m_{\rm sc}'(\rho_j)I_{2r}\bigr)
\Theta y_n
\right|\\
&\qquad\leq
\|\Theta\|^2
\left\|M_n'(\lambda_j)-m_{\rm sc}'(\rho_j)I_{2r}\right\|
\longrightarrow0
\qquad\text{a.s.}
\end{aligned}
$$
Thus the right-hand side of \eqref{eq:cca_normalization}
equals $m_{\rm sc}'(\rho_j)\|\Theta y_n\|^2+o(1)$ almost surely.
Since $\Theta$ is diagonal and $\Theta e_{j,+}=\theta_j e_{j,+}$,
\eqref{eq:cca_coordinate_alignment} also gives
$$
\begin{aligned}
\|\Theta y_n\|^2
&=
\theta_j^2|e_{j,+}^*y_n|^2
+\left\|\Theta\bigl(y_n-(e_{j,+}^*y_n)e_{j,+}\bigr)\right\|^2\\
&=
\theta_j^2|e_{j,+}^*y_n|^2+o(1)
\qquad\text{a.s.}
\end{aligned}
$$
Substituting these two estimates into \eqref{eq:cca_normalization}
yields, almost surely,
$$
1=\theta_j^2m_{\rm sc}'(\rho_j)|e_{j,+}^*y_n|^2+o(1).
$$
Since $m_{\rm sc}'(\rho_j)=(\theta_j^2-1)^{-1}$, the preceding estimates give
$|e_{j,+}^*y_n|^2\to1-\theta_j^{-2}$ almost surely.
Combining this with \eqref{eq:cca_coordinate_alignment}
and $\|y_n\|\leq1$, we obtain
$$
\begin{aligned}
U_n^*\widehat\Pi_{j,n}^+U_n
&=y_ny_n^*\\
&=|e_{j,+}^*y_n|^2e_{j,+}e_{j,+}^{\top}+o(1)\\
&\longrightarrow
(1-\theta_j^{-2})e_{j,+}e_{j,+}^{\top}
\qquad\text{a.s.}
\end{aligned}
$$
We now consider the negative outlier. Since
$\overline{H_n}=-H_n$, the vector
$\overline{\widehat w_{j,n}}$ is a normalized eigenvector
associated with the eigenvalue $-\lambda_j$.
Let
$\widehat\Pi_{j,n}^-=
\overline{\widehat w_{j,n}}\widehat w_{j,n}^{\top}$
be the corresponding spectral projection. Set
$$
P=\begin{pmatrix}0&I_r\\ I_r&0\end{pmatrix},
\qquad\text{so that}\qquad
\overline{U_n}=U_nP.
$$
Then
$$
U_n^*\widehat\Pi_{j,n}^-U_n
=P\,\overline{U_n^*\widehat\Pi_{j,n}^+U_n}\,P
\longrightarrow
(1-\theta_j^{-2})e_{j,-}e_{j,-}^{\top}
\qquad\text{a.s.}
$$
Therefore, setting
$\widehat\Pi_{j,n}=\widehat\Pi_{j,n}^+
+\widehat\Pi_{j,n}^-,$ we obtain
\be\la{eq:compressed_projection_limit}
    U_n^*\widehat\Pi_{j,n}U_n
    \longrightarrow
    (1-\theta_j^{-2})
    \left(e_{j,+}e_{j,+}^{\top}
    +e_{j,-}e_{j,-}^{\top}\right)
    \qquad\text{a.s.}
\ee
It remains to translate this complex statement into a real statement on the spike planes. Write $\widehat w_{j,n}=a_{j,n}+\ii b_{j,n}$ for $a_{j,n},b_{j,n}\in\mathbb R^n.$ Since $\widehat w_{j,n}$ and $\overline{\widehat w_{j,n}}$ are eigenvectors associated with the distinct eigenvalues $\lambda_j$ and $-\lambda_j$, they are orthogonal. Hence, $\|a_{j,n}\|^2-\|b_{j,n}\|^2-2\ii\,a_{j,n}^{\top}b_{j,n}=0.$ Together with $\|\widehat w_{j,n}\|=1$, this gives $\|a_{j,n}\|=\|b_{j,n}\|=(\sqrt{2})^{-1}$ and $a_{j,n}^{\top}b_{j,n}=0.$ Thus, the vectors $\widehat u_{j,n}=\sqrt 2\,a_{j,n}$ and $\widehat v_{j,n}=-\sqrt 2\,b_{j,n}$ form an orthonormal pair in $\mathbb R^n$. Moreover, $\widehat w_{j,n}=
    \frac{1}{\sqrt 2}
    \left(\widehat u_{j,n}-\ii\widehat v_{j,n}\right),$ and therefore
\be\la{eq:proj_overlap_matrix}
    \widehat\Pi_{j,n}
    =
    \widehat w_{j,n}\widehat w_{j,n}^{*}
    +
    \overline{\widehat w_{j,n}}\widehat w_{j,n}^{\top}
    =
    \widehat u_{j,n}\widehat u_{j,n}^{\top}
    +
    \widehat v_{j,n}\widehat v_{j,n}^{\top}
    =
    \widehat A_{j,n}\widehat A_{j,n}^{\top},
\ee
where $\widehat A_{j,n}=
    \begin{pmatrix}
        \widehat u_{j,n} & \widehat v_{j,n}
    \end{pmatrix}.$ 
Set
$$
A_{j}=\begin{pmatrix}u_{j} & v_{j}\end{pmatrix},
\qquad
B_{j}=\begin{pmatrix}w_{j,+} & w_{j,-}\end{pmatrix},
\qquad
C=\frac{1}{\sqrt2}
\begin{pmatrix}
1 & 1\\
-\ii & \ii
\end{pmatrix}.
$$
Then $B_{j}=A_{j}C$, and $C$ is unitary. Since $A_{j}$ is real and $A_{j}=B_{j}C^*$, we have
\be\la{eq:interm_form_cca}
A_{j}^{\top}\widehat\Pi_{j,n}A_{j}=
A_{j}^{*}\widehat\Pi_{j,n}A_{j}=
C B_{j}^{*}\widehat\Pi_{j,n}B_{j} C^*.
\ee 
Moreover, $B_{j}=U_nR_j$ with $R_j=\begin{pmatrix}e_{j,+} & e_{j,-}\end{pmatrix}\in\mathbb C^{2r\times 2}.$
Combining this with \eqref{eq:compressed_projection_limit}, we obtain
$$
B_{j}^{*}\widehat\Pi_{j,n}B_{j}
=
R_j^*U_n^*\widehat\Pi_{j,n}U_nR_j
\longrightarrow (1-\theta_j^{-2}) I_2 \qquad \text{a.s.}
$$
From \eqref{eq:proj_overlap_matrix} and \eqref{eq:interm_form_cca}, we conclude that
$$
A_{j}^{\top}\widehat A_{j,n}\widehat A_{j,n}^{\top}A_{j}
\longrightarrow
\left(1-\theta_j^{-2}\right)I_2 \qquad \text{a.s.}
$$
Since the canonical correlations between the population plane
$\operatorname{Span}_{\mathbb R}(u_j,v_j)$ and the estimated plane
$\operatorname{Span}_{\mathbb R}(\widehat u_{j,n},\widehat v_{j,n})$
are the singular values of $A_j^\top\widehat A_{j,n}$, their squares
are the eigenvalues of
$A_j^\top\widehat A_{j,n}\widehat A_{j,n}^\top A_j$.
This proves the theorem under Assumption~\ref{ass:ass3}.

We now retain only Assumption~\ref{ass:ass1}. Use the truncation from
Lemma~\ref{lem:fourth_moment_truncation}, and write
$$
H_n^{(L)}=W_n^{(L)}+\ii S_n,
\qquad
d_{n,L}=\|H_n-H_n^{(L)}\|=\|W_n-W_n^{(L)}\|.
$$
Let $\widehat\Pi_{j,n}^{(L)}$ be the real-plane projector associated
with the $j$th eigenvalue $\lambda_j^{(L)}$ of $H_n^{(L)}$.
For both matrices, use the convention of
Section~\ref{sec:asy_eigenvectors} on exceptional events.
For each fixed sufficiently large integer $L$, the all-moments case
proved above applies to $H_n^{(L)}$, with the original variance.
We work on one event of probability one on which
\eqref{eq:fourth_moment_truncation_norm}, the eigenvalue limits of
Theorem~\ref{thm:general_random_skew_outliers}, and the bounded-entry
eigenplane limits hold for every such $L$.
This is possible because the cutoffs form a countable set.
Since $\theta_j$ is simple and supercritical, the eigenvalue limits
imply that $\lambda_j$ is positive and simple for all sufficiently
large $n$. The same limits for $H_n^{(L)}$ show that there is a
constant $g>0$, independent of $L$, such that, for each fixed
sufficiently large integer $L$ and all sufficiently large $n$,
$$
\lambda_j^{(L)}>g,
\qquad \min_{q\ne j}|\lambda_j^{(L)}-\lambda_q^{(L)}|>g.
$$
Fix such an $L$ with $2\tau_L<g/4$. By
\eqref{eq:fourth_moment_truncation_norm}, $d_{n,L}<g/4$ for all
sufficiently large $n$.
Let $\widehat\Pi_{j,n}^{+,(L)}$ be the rank-one spectral projector
of $H_n^{(L)}$ associated with $\lambda_j^{(L)}$, and set
$Q_{n,L}=I_n-\widehat\Pi_{j,n}^{+,(L)}$.
The eigenvector equation gives
$$
(H_n^{(L)}-\lambda_jI_n)Q_{n,L}\widehat w_{j,n}
=-Q_{n,L}(H_n-H_n^{(L)})\widehat w_{j,n}.
$$
The restriction of $H_n^{(L)}-\lambda_jI_n$ to
the range of $Q_{n,L}$ has eigenvalues
$\lambda_q^{(L)}-\lambda_j$, $q\ne j$, whose absolute values
are at least $g-d_{n,L}$ by Weyl’s inequality. Thus
$$
\|Q_{n,L}\widehat w_{j,n}\|
\leq\frac{d_{n,L}}{g-d_{n,L}}
\leq\frac{2d_{n,L}}g.
$$
The rank-one projection identity yields
$$
\|\widehat\Pi_{j,n}^{+}-\widehat\Pi_{j,n}^{+,(L)}\|
=\|Q_{n,L}\widehat w_{j,n}\|.
$$
Since the negative-outlier projectors are the complex conjugates
of the positive-outlier projectors,
\eqref{eq:proj_overlap_matrix} therefore gives
$$
\|\widehat\Pi_{j,n}-\widehat\Pi_{j,n}^{(L)}\|
\leq\frac{4d_{n,L}}g.
$$
For each fixed sufficiently large integer $L$, the all-moments case proved above gives, on the same probability-one event,
$$
A_j^\top\widehat\Pi_{j,n}^{(L)}A_j
\longrightarrow(1-\theta_j^{-2})I_2.
$$
Since $\|A_j\|=1$, the triangle inequality gives
$$
\|A_j^\top\widehat\Pi_{j,n}A_j-(1-\theta_j^{-2})I_2\|
\leq\|\widehat\Pi_{j,n}-\widehat\Pi_{j,n}^{(L)}\| + \|A_j^\top\widehat\Pi_{j,n}^{(L)}A_j
-(1-\theta_j^{-2})I_2\|.
$$
Using \eqref{eq:fourth_moment_truncation_norm} and taking the limit superior as $n\to\infty$ gives
$$
\limsup_{n\to\infty}
\|A_j^\top\widehat\Pi_{j,n}A_j-(1-\theta_j^{-2})I_2\|
\leq\frac{8\tau_L}{g}.
$$
Letting $L\to\infty$ through integers proves the required
almost-sure convergence. The limits of the squared canonical
correlations follow by continuity of the eigenvalues of this
$2\times2$ matrix. This completes the proof.
$\hfill\Box$

\bibliographystyle{abbrvnat}
\bibliography{skew_biblio}

\end{document}